\documentclass[letterpaper,journal]{IEEEtran}

\usepackage{cite}
\usepackage{amsmath,amssymb,amsfonts}
\usepackage{graphicx}
\usepackage{textcomp}
\usepackage{stfloats}
\usepackage{xcolor}
\usepackage{array}
\usepackage{url}
\usepackage{verbatim}
\usepackage{balance}
\def\BibTeX{{\rm B\kern-.05em{\sc i\kern-.025em b}\kern-.08em
	T\kern-.1667em\lower.7ex\hbox{E}\kern-.125emX}}

\usepackage{subfigure}
\usepackage{epstopdf}

\usepackage{algorithmic}
\usepackage[ruled,vlined]{algorithm2e}

\usepackage{multirow}

\begin{document}

\title{Construction and Dynamic Update of Channel Gain Maps via 3D Gaussian Splatting}

\author{Yilong Chen, Yuan Guo, Juncong Zhou, Jie Xu, \IEEEmembership{Fellow, IEEE}, and Rui Zhang, \IEEEmembership{Fellow, IEEE}
	\thanks{Y. Chen, Y. Guo, J. Zhou, and J. Xu are with School of Science and Engineering, Shenzhen Future Network of Intelligence Institute (FNii-Shenzhen), and Guangdong Provincial Key Laboratory of Future Networks of Intelligence, The Chinese University of Hong Kong (Shenzhen), Shenzhen 518172, China (e-mail: chenyilong@cuhk.edu.cn; guoyuan@cuhk.edu.cn; juncongzhou@link.cuhk.edu.cn; xujie@cuhk.edu.cn).}
	\thanks{R. Zhang is with the Department of Electrical and Computer Engineering, National University of Singapore, Singapore 117583 (e-mail: elezhang@nus.edu.sg).}
	\thanks{J. Xu is the corresponding author.}
}

\maketitle

\begin{abstract}
	Channel knowledge maps (CKMs) have emerged as a promising technique for providing scene-specific and location-dependent propagation knowledge to enable environment-aware wireless network design. This paper investigates the construction and dynamic updating of a particular type of CKM, namely grid-based channel gain maps (CGMs), for large-scale networks using three-dimensional Gaussian splatting (3DGS). First, we formulate a grid-based channel gain model, where each map entry is defined as the locally averaged channel gain over a receiver grid, thereby suppressing phase-sensitive small-scale fluctuations. The resulting channel gain is decomposed into distance-dependent attenuation, path transmittance, and effective scattering contributions. Based on this decomposition, we develop a physics-informed Gaussian-splatting-based channel gain (GS-CG) model, which represents the propagation environment as a set of Gaussian primitives. The proposed model maps Gaussian geometry, opacity, and directional features to propagation-related factors and renders grid-level channel gains through a differentiable process. To accommodate real-time environmental changes, we further propose an incremental learning mechanism that updates a static reference GS-CG representation into a dynamic CGM. Specifically, the reference Gaussian primitives are frozen, while a compact set of tunable Gaussians is introduced to capture newly induced local channel-gain variations from sparse measurements. In addition, we design a low-complexity GS-CG variant based on a multilayer perceptron (MLP), termed MLP-GS, which replaces part of the explicit scattering-path interaction computation with lightweight neural surrogates. Numerical results demonstrate that the proposed GS-CG and MLP-GS methods accurately reconstruct grid-based CGMs, efficiently adapt to dynamic environmental changes, and achieve a favorable accuracy-complexity tradeoff for fast CGM refinement.
\end{abstract}


\section{Introduction}

Channel knowledge maps (CKMs) have attracted growing research interest for sixth-generation (6G) networks, as they enable a new paradigm toward environment-aware wireless communications \cite{zeng2021toward, zeng2024tutorial}. With denser infrastructure deployments, larger antenna arrays, wider bandwidths, and stricter latency requirements, traditional channel acquisition methods based on pilot training and feedback are becoming increasingly costly or even impractical. CKMs address this challenge by mapping transceiver locations to location-specific channel knowledge, enabling networks to proactively infer channel conditions with substantially reduced pilot overhead and improved system performance. Depending on the application scenario, various types of CKMs have been proposed, such as channel state information (CSI) maps, channel gain maps (CGMs), channel angle maps (CAMs), and beam index maps (BIMs) \cite{zeng2024tutorial}. Among these CKM instances, CGMs are the most fundamental and practically useful, as they characterize large-scale propagation quality over space and directly support coverage analysis, deployment planning, beam management, and resource allocation \cite{dall2011channel, zeng2024tutorial}.

Existing CGM construction methods can be broadly categorized into interpolation-based, model-based, and learning-based approaches \cite{ren2026channel, liu2025channel}. Interpolation methods, such as Kriging \cite{dall2011channel}, kernel regression \cite{xu2021efficient}, and matrix completion \cite{sun2022propagation}, are effective under sparse measurements but often struggle to capture sharp propagation transitions caused by blockage and diffraction. Model-based approaches, including empirical path-loss models \cite{seidel1994site, akhpashev2016cost} and ray tracing \cite{yun2015ray}, provide strong physical interpretability but require accurate environmental information and often incur high computational complexity. More recently, deep-learning and generative-AI approaches, such as RadioUNet \cite{levie2021radiounet}, RMTransformer \cite{li2025radiotransformer}, and CKMDiff \cite{fu2025ckmdiff}, have demonstrated impressive map reconstruction capabilities. However, these methods typically rely on discretized scene representations and may be less suitable for scene-specific modeling and efficient map updating. Consequently, developing CGM representations that achieve physical interpretability, data efficiency, and computational affordability remains an important research challenge.

Recent advances in neural scene representations, particularly neural radiance fields (NeRFs) \cite{mildenhall2021nerf} and three-dimensional Gaussian splatting (3DGS) \cite{kerbl3Dgaussians}, provide a new opportunity for CKM construction. Inspired by their success in computer graphics, wireless radiance field (WRF) frameworks have been developed to model radio propagation and predict wireless channels from geometric information \cite{zhao2023nerf2, wen2025wrfgs, zhou20256d, zhang2026rf3dgs, zhang2025rfpgs, liu2026deformable, nukapotula2025gsparc, chen2025gaussian, wang2025radsplatter, li2025wideband}. Compared with NeRF-based approaches \cite{zhao2023nerf2}, which rely on implicit volumetric neural fields and often incur substantial computational overhead, 3DGS provides explicit scene representations, differentiable rendering, and efficient optimization. Existing 3DGS-based wireless studies \cite{wen2025wrfgs, zhou20256d, zhang2026rf3dgs, zhang2025rfpgs, liu2026deformable, nukapotula2025gsparc, chen2025gaussian, wang2025radsplatter, li2025wideband} have demonstrated promising performance in radio field reconstruction, CSI prediction, and spectrum mapping. Nevertheless, these works primarily focus on pointwise channel quantities associated with precise transmitter and receiver locations and have not explicitly considered grid-based CGM construction.

Different from prior works, this paper focuses on the construction of grid-based CGMs, where each map entry represents the locally averaged channel gain over a receiver grid. This formulation is important for practical large-scale wireless deployments, where receiver locations are often subject to spatial uncertainty. In outdoor environments, receiver positions are typically available only with meter-level accuracy, and a nominal receiver location may correspond to a local spatial region rather than an exact point. Under such uncertainty, phase-sensitive channel quantities can fluctuate significantly and become difficult to supervise consistently. By contrast, grid-based CGMs suppress small-scale fluctuations by defining each map entry as the locally averaged channel gain within a receiver grid, thereby providing a more stable and actionable representation for coverage analysis and network optimization. Although virtual-scatterer-based approaches have recently been proposed for grid-based CGM modeling \cite{sun2025channel, sun2026channel}, a physics-informed 3DGS framework specifically tailored to grid-level channel-gain rendering remains largely unexplored.

Beyond the initial construction of CGM, maintaining CGM accuracy under dynamic environments is another critical challenge. Practical wireless environments contain not only permanent structures such as buildings and walls but also dynamic objects like vehicles, pedestrians, and temporary obstacles. These dynamic factors continuously alter blockage, transmission, and scattering conditions, causing the propagation environment to evolve over time. Existing 3DGS-based and virtual-scatterer-based CKM methods are primarily designed for static environments and typically require retraining when the environment changes. Such construction-from-scratch approaches are inefficient because environmental changes often affect only a small subset of propagation paths and local spatial regions. Therefore, an important yet largely unexplored problem is how to incrementally update a previously constructed CGM by efficiently incorporating newly observed channel variations while preserving previously learned propagation knowledge.

In this paper, we investigate the construction of a static/reference grid-averaged CGM and its real-time updating in dynamic environments for a given transmitter using the 3DGS framework. The static/reference CGM represents either a propagation scene without transient dynamic factors or a long-term measured map in which dynamic factors are averaged out or weakened. By contrast, the dynamic CGM characterizes the real-time propagation condition when local blockers or scatterers appear or move. Our objective is to first construct the static CGM from limited measurements and then update it using only a small number of newly collected dynamic measurements.
The main contributions of this paper are summarized as follows.
\begin{itemize}
	\item First, we establish a grid-based average channel gain model for practical CGM construction and updating. By defining each map entry as the locally averaged channel gain over a receiver grid, the proposed model suppresses small-scale phase fluctuations and decomposes the resulting large-scale power quantity into three components: distance-dependent attenuation, path transmittance, and effective scattering responses.
	
	\item Second, based on the proposed model, we develop a physics-informed Gaussian-splatting-based channel gain (GS-CG) model for grid-level channel gain rendering. The proposed model represents the propagation environment as a set of Gaussian primitives and explicitly matches the physical power decomposition of the grid-based channel gain. Specifically, Gaussian geometry, opacity, and directional features are mapped to path transmittance and effective scattering responses, resulting in a unified rendering model for both static and dynamic CGMs.
	
	\item Third, we propose a unified GS-CG framework for static CGM construction and dynamic CGM updating. For static construction, the Gaussian representation is trained from scratch to obtain a reference CGM. For dynamic updating, we develop an incremental learning strategy that freezes the reference representation and introduces a compact set of tunable Gaussian primitives to capture newly induced local channel-gain variations from sparse measurements.
	
	\item Fourth, we develop a low-complexity GS-CG variant based on a multilayer perceptron (MLP), termed MLP-GS, for fast CGM construction and updating. This design preserves the Gaussian spatial support and explicit direct-path rendering, while replacing part of the computationally expensive scattering-branch interaction with lightweight neural surrogates, thereby achieving a practically appealing accuracy-complexity tradeoff for dynamic CGM refinement.
	
	\item Finally, extensive simulations validate the proposed GS-CG and MLP-GS models for both static CGM construction and dynamic updating. The results show that the proposed methods can accurately reconstruct grid-based CGMs, effectively adapt to environmental changes using sparse new measurements, and substantially reduce training time compared with full explicit rendering.
\end{itemize}

The remainder of this paper is organized as follows. Section II presents the grid-based wireless channel model and formulates the static CGM construction and dynamic CGM updating problems. Section III develops the proposed GS-CG model for physics-aware static CGM construction. Section IV presents the incremental GS-CG mechanism for dynamic CGM updating. Section V presents the low-complexity MLP-GS variant. Section VI provides simulation results to validate the proposed methods. Finally, Section VII concludes this paper.

\emph{Notations:} Bold lowercase letters are used for vectors, and bold uppercase letters are used for matrices. Calligraphic letters denote sets. For a finite set \(\mathcal{A}\), \(|\mathcal{A}|\) denotes its cardinality. For a scalar \(a\), \(|a|\) denotes its absolute value or magnitude. For a vector \(\mathbf{a}\), \(\|\mathbf{a}\|\) denotes its Euclidean norm. The transpose operator is denoted by \((\cdot)^T\). \(\mathrm{diag}(\cdot)\) denotes a diagonal matrix. \(\mathbb{R}\) and \(\mathbb{C}\) denote the real and complex fields, respectively. \(\mathbb{E}[\cdot]\) denotes statistical expectation. \(\mathcal{N}(\boldsymbol{\mu},\boldsymbol{\Sigma})\) denotes the Gaussian distribution with mean \(\boldsymbol{\mu}\) and covariance \(\boldsymbol{\Sigma}\). \(\mathcal{U}(a,b)\) denotes the uniform distribution in interval \([a,b]\). The imaginary unit is denoted by \(j=\sqrt{-1}\).
%The notation \(\hat{\cdot}\) denotes rendered or estimated quantities. 

\section{System Model and Problem Formulation}

\begin{figure}[!tb]
	\centering
	\includegraphics[width=0.35\textwidth]{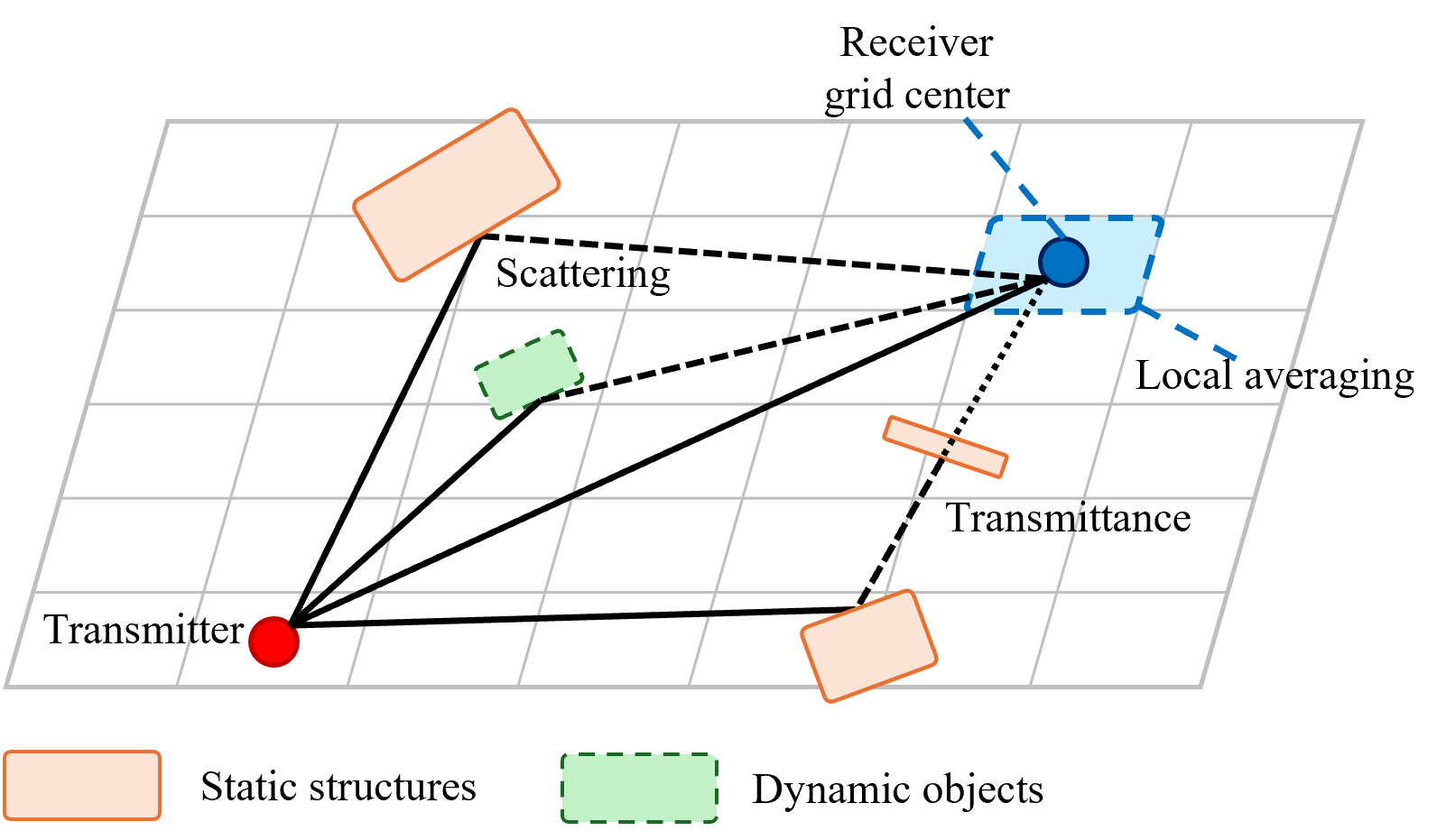}
	\caption{Considered system and grid-based channel gain modeling.}
	\label{fig_system}
\end{figure}

As shown in Fig. \ref{fig_system}, we consider a wireless communication system with a single-antenna base station (BS) located at a fixed position \(\mathbf{p}_{t} \in \mathbb{R}^3\). The receiver region of interest for CGM construction and updating is a three-dimensional (3D) region \(\mathcal{D}\subset\mathbb{R}^3\), which is partitioned into grids. Each grid is indexed by its center \(\mathbf{p}_r\in\mathcal{D}\). Let \(\mathcal{C}(\mathbf{p}_r)\subset\mathcal{D}\) denote the grid centered at \(\mathbf{p}_r\). 
Each CGM entry indexed by \(\mathbf{p}_r\) represents the average channel gain over receiver grid \(\mathcal{C}(\mathbf{p}_r)\). This grid-level representation is relevant to practical outdoor scenarios, where receiver locations are typically reliable at the grid scale but not at wavelength-level precision.

The wireless propagation environment includes static structures, such as buildings and walls, as well as dynamic objects, such as vehicles, pedestrians, or other moving blockers/scatterers. We use \(\mathcal{M}_\text{s}\) to denote the static/reference CGM that is mainly determined by the static structures. It can represent either a scene without transient dynamic objects or a long-term measured map in which the effects of moving objects are averaged out. We use \(\mathcal{M}_\text{d}\) to denote the dynamic/current CGM under real-time environmental changes. Our goal is to first construct \(\mathcal{M}_\text{s}\) from grid-averaged channel gain measurements and then update it to obtain \(\mathcal{M}_\text{d}\) using sparse new measurements.

In the following, we first derive a grid-based channel gain model for a generic CGM state, then formulate the static construction and dynamic updating problems for obtaining \(\mathcal{M}_\text{s}\) and \(\mathcal{M}_\text{d}\), respectively.

\subsection{Grid-Based Channel Gain Model}

We first derive the grid-based channel gain model. Under far-field and local-homogeneity conditions, we assume that each grid size is much smaller than the propagation distances to dominant structures, so large-scale path parameters, such as path losses and angles of departure/arrival, remain approximately invariant within the grid. Meanwhile, the grid size is much larger than the carrier wavelength \(\lambda\), so receiver displacement inside the grid mainly induces small-scale phase variation. Thus, the channel within \(\mathcal{C}(\mathbf{p}_r)\) can be decomposed into a grid-level path response and an intra-grid field response \cite{sun2025channel}.

For an exact receiver location \(\mathbf{x}\in\mathcal{C}(\mathbf{p}_r)\), we adopt the field-response model \cite{sun2025channel} and write the multi-path channel from the transmitter \(\mathbf{p}_t\) as
\begin{equation} \label{h}
	h(\mathbf{x}) = \mathbf{u}^T(\mathbf{p}_r)\mathbf{f}(\mathbf{x};\mathbf{p}_r).
\end{equation}
Here, \(\mathbf{f}(\mathbf{x};\mathbf{p}_r)\) is the field-response vector (FRV) that captures the phase variation caused by the displacement from \(\mathbf{p}_r\) to \(\mathbf{x}\), which is given by
\begin{equation}
	\begin{aligned}
		\mathbf{f}(\mathbf{x};\mathbf{p}_r) &= \big[f_0(\mathbf{x};\mathbf{p}_r), \cdots, f_{N_s}(\mathbf{x};\mathbf{p}_r)\big]^T \\
		&= \Big[e^{-j\frac{2\pi}{\lambda}\Delta_0(\mathbf{x};\mathbf{p}_r)},
		\cdots, e^{-j\frac{2\pi}{\lambda}\Delta_{N_s}(\mathbf{x};\mathbf{p}_r)}\Big]^T,
	\end{aligned}
\end{equation}
where \(\Delta_n(\mathbf{x};\mathbf{p}_r)\) is the excess propagation distance of the \(n\)-th path.
In addition, \(\mathbf{u}(\mathbf{p}_r)\) is the path-response vector (PRV) evaluated at the receiver grid center \(\mathbf{p}_r\). Let \(\mathcal{N}_s\triangleq\{1,\cdots,N_s\}\) denote the set of effective scattering paths. The PRV is given by
\begin{equation}
	\mathbf{u}(\mathbf{p}_r) = \big[u_0(\mathbf{p}_r), u_1(\mathbf{p}_r), \cdots, u_{N_s}(\mathbf{p}_r)\big]^T,
\end{equation}
where the \(0\)-th entry corresponds to the direct path and the remaining \(N_s\) entries correspond to effective one-bounce scattering paths.\footnote{We adopt the effective one-bounce scattering model for tractability. It explicitly models both the incident and outgoing propagation segments. This differs from existing semi-implicit \cite{zhao2023nerf2, wen2025wrfgs} or last-bounce \cite{sun2025channel, sun2026channel} formulations, where only part of the propagation process is explicitly modeled. 
The simulations later show that this model can well represent practical wireless environments involving multi-bounce reflection, diffraction, and scattering. Moreover, the model can be readily extended to six-dimensional (6D) CGM scenarios with varying transceiver locations \cite{zhou20256d}.} Following \cite{goldsmith2005wireless}, the path responses are
\begin{equation}\label{u_0}
	\begin{aligned}
		u_0(\mathbf{p}_r)
		&= \frac{\lambda}{4\pi d_0} e^{-j\frac{2\pi}{\lambda}d_0} \cdot \Phi_0, \\
		u_n(\mathbf{p}_r)
		&= \frac{\lambda}{(4\pi)^{\frac{3}{2}} d_n^\text{in} d_n^\text{out}}  e^{-j \frac{2\pi}{\lambda} (d_n^\text{in} + d_n^\text{out})} \cdot \Omega_n
		\Phi_n^\text{in} \Phi_n^\text{out},
	\end{aligned}
\end{equation}
for all \(n\in\mathcal{N}_s\). Here, \(d_0\) is the direct-path distance, while \(d_n^\text{in}\) and \(d_n^\text{out}\) are the transmitter-to-scatterer and scatterer-to-receiver distances of the \(n\)-th scattering path, respectively.\footnote{The distance-dependent terms in \eqref{u_0} follow free-space spherical spreading for each explicitly modeled propagation segment. General path-loss models may use a tunable exponent to absorb unresolved effects such as ground reflection \cite{sun2025channel}. By contrast, the proposed GS-CG model explicitly represents environmental interactions through Gaussian primitives, so the segment-wise spreading exponent is fixed to its free-space value.} 
\(\Omega_n \in \mathbb{C}\) denotes the anisotropic complex scattering response. The coefficients \(\Phi_0,\Phi_n^\text{in},\Phi_n^\text{out}\in\mathbb{C}\) denote the complex transmittance of the direct path and those of the incident and outgoing segments of the \(n\)-th scattering path, respectively, collecting the blockage or penetration effects along the corresponding propagation paths.

We define each CGM entry by averaging the instantaneous channel power over the receiver grid. The channel gain associated with grid \(\mathcal{C}(\mathbf{p}_r)\) is given by \cite{sun2025channel}
\begin{equation} \label{Gint}
	G(\mathbf{p}_r) = \frac{1}{|\mathcal{C}(\mathbf{p}_r)|} \int_{\mathcal{C}(\mathbf{p}_r)} |h(\mathbf{x})|^2 d\mathbf{x},
\end{equation}
where \(|\mathcal{C}(\mathbf{p}_r)|\) denotes the grid volume and \(d\mathbf{x}\) is the infinitesimal spatial element. The averaged channel gain \(G(\mathbf{p}_r)\) suppresses phase-sensitive small-scale fluctuations and preserves the large-scale propagation quality associated with the grid center \(\mathbf{p}_r\).

\textit{Proposition 1 (Grid-level channel gain decomposition):}
Under the above local-homogeneity and far-field conditions, if distinct propagation paths have sufficiently separated local directions, the grid-based channel gain in \eqref{Gint} can be approximated by the incoherent sum of path powers, i.e.,
\begin{equation} \label{G}
	G(\mathbf{p}_r) \approx \|\mathbf{u}(\mathbf{p}_r)\|^2 
	\triangleq G_0(\mathbf{p}_r) + \sum_{n\in\mathcal{N}_s} G_n(\mathbf{p}_r),
\end{equation}
where the direct and \(n\)-th scattering terms are expressed as
\begin{equation} \label{G_0n}
	\begin{aligned}
		G_0(\mathbf{p}_r) 
		&= \Big(\underbrace{\frac{\lambda}{4\pi d_0}}_{\text{Path loss}} \cdot \underbrace{|\Phi_0|}_{\text{Transmittance}}\Big)^2, \\
		G_n(\mathbf{p}_r) 
		&= \Big(\underbrace{\frac{\lambda}{(4\pi)^{\frac{3}{2}} d_n^\text{in} d_n^\text{out}}}_{\text{Path loss}} \cdot \underbrace{|\Omega_n|}_{\substack{\text{Scattering} \\ \text{response}}} \cdot \underbrace{|\Phi_n^\text{in}| |\Phi_n^\text{out}|}_{\text{Transmittance}}\Big)^2.
	\end{aligned}
\end{equation}

%	The derivation for square grids on a two-dimensional (2D) plane has been given in \cite{sun2025channel}. Here, we consider 3D grid cells with arbitrary shapes. The detailed proof is provided in Appendix~\ref{appendix:proof_prop1}.
\textit{Proof:} Please refer to Appendix A. \hfill \(\blacksquare\)

Proposition 1 shows that a grid-level CGM is a large-scale power map determined by distance-dependent attenuation, path transmittance, and effective scattering strength, rather than a phase-dependent CSI field. We define the generic CGM as the scalar field
\begin{equation}
	\mathcal{M}: \mathbf{p}_r \in \mathcal{D} \mapsto G(\mathbf{p}_r) \in \mathbb{R}_+.
\end{equation}
The static and dynamic CGMs, \(\mathcal{M}_\text{s}\) and \(\mathcal{M}_\text{d}\), are two instances of \(\mathcal{M}\) for long-term/static and real-time environmental states, respectively.

\subsection{CGM Construction and Dynamic Updating Problems}

Let \(\hat{\mathcal{M}}(\varTheta)\) denote a parametric CGM with learnable parameters \(\varTheta\) to represent the CGM \(\mathcal{M}\). This paper instantiates \(\varTheta\) as a 3D Gaussian scene representation and uses it to render the grid-based channel gain over \(\mathcal{D}\). We consider the following two tasks.

\subsubsection{Static CGM Construction}

Let \(\hat{\mathcal{D}}_\text{s}\subset\mathcal{D}\) denote the set of grids where channel gain measurements are available in the long-term/static environment. Given the measurement data \(G(\mathbf{p}_r)\) for \(\mathbf{p}_r\in\hat{\mathcal{D}}_\text{s}\) as the training set, the static construction problem aims to learn a reference model \(\varTheta_\text{s}\) from scratch so that \(\hat{\mathcal{M}}(\varTheta_\text{s})\) reconstructs \(\mathcal{M}_\text{s}\) over both measured and unmeasured grids.

\subsubsection{Dynamic CGM Updating}

Suppose that the static/reference CGM model \(\varTheta_\text{s}\) is already available to the system \emph{a priori}. When dynamic objects appear or move, we assume that only a sparse measurement set \(\hat{\mathcal{D}}_\text{d}\subset\mathcal{D}\) is available. Given the measurement data \(G(\mathbf{p}_r)\) for \(\mathbf{p}_r\in\hat{\mathcal{D}}_\text{d}\) as the training set, the dynamic updating problem aims to learn a refined model \(\varTheta_\text{d}\) such that \(\hat{\mathcal{M}}(\varTheta_\text{d})\) matches \(\mathcal{M}_\text{d}\) while reusing the previously constructed propagation knowledge.

\section{GS-CG Model for Static CGM Construction}

Section II shows that the grid-averaged channel gain can be decomposed into multi-path contributions determined by distance-dependent attenuation, anisotropic scattering, and transmittance. The physical model in \eqref{G} and \eqref{G_0n}, however, depends on unknown scatterer locations, scattering responses \(\{\Omega_n\}\), and transmittance coefficients \(\Phi_0\) and \(\{\Phi_n^\text{in}, \Phi_n^\text{out}\}\). The GS-CG model aims to represent these inaccessible physical quantities with learnable 3D Gaussian primitives that represent latent radio-interaction structures. Since the target is a grid-based CGM, GS-CG uses the transmitter location \(\mathbf{p}_t\) and receiver grid center \(\mathbf{p}_r\) to define propagation paths and renders the grid-averaged channel gain, rather than a phase-sensitive pointwise channel response in prior work \cite{wen2025wrfgs, zhou20256d}.

\subsection{Gaussian Representation and Channel Gain Rendering}

We represent the wireless scene by \(N\) Gaussian primitives with index set \(\mathcal{N}=\{1,\cdots,N\}\). The learnable scene parameters are
\begin{equation} \label{Theta}
	\varTheta \triangleq \{\vartheta_n\}_{n\in\mathcal{N}},
\end{equation}
where the \(n\)-th Gaussian primitive is parameterized as
\begin{equation} \label{Theta_n}
	\vartheta_n = \{\boldsymbol{\mu}_n, \mathbf{s}_n, \mathbf{q}_n, \alpha_n, \mathbf{f}_n\}.
\end{equation}
Here, \(\boldsymbol{\mu}_n\in\mathbb{R}^3\) is the Gaussian center, \(\mathbf{s}_n\in\mathbb{R}_+^3\) gives its anisotropic scale, \(\mathbf{q}_n\) is a quaternion specifying its rotation, \(\alpha_n\in[0,1]\) denotes its opacity, and \(\mathbf{f}_n\) contains learnable directional scattering features. Each primitive can act as an effective interaction point for one-bounce scattering and as a soft volumetric attenuator for propagation paths. Note that the number of Gaussian primitives, or Gaussians for short, \(N\), is an adjustable parameter and need not equal the number of physical effective scatterers \(N_s\) in \eqref{G}. Instead, \(\varTheta\) provides a differentiable surrogate of the unknown propagation environment.

We construct the rendered channel gain as the Gaussian-parameterized counterpart of \eqref{G} and \eqref{G_0n}:
\begin{equation} \label{Gtotal}
	\hat{G}(\mathbf{p}_r) = \hat{G}_0(\mathbf{p}_r) + \sum_{n\in\mathcal{N}} \hat{G}_n(\mathbf{p}_r),
\end{equation}
where the direct term and the scattering term associated with the \(n\)-th Gaussian are given by
\begin{equation} \label{GLoS}
	\begin{aligned}
		\hat{G}_0(\mathbf{p}_r; \varTheta) =& \Big( \underbrace{\frac{\lambda}{4\pi d_0}}_{\text{Path loss}} \cdot \underbrace{\hat{\Phi}_0(\mathbf{p}_t, \mathbf{p}_r)}_{\text{Transmittance}} \Big)^2, \\
		\hat{G}_n(\mathbf{p}_r; \varTheta) =& \Big( \underbrace{\frac{\lambda}{(4\pi)^{\frac{3}{2}} d_n^\text{in} d_n^\text{out}}}_{\text{Path loss}} \cdot \underbrace{\hat{\Omega}_n(\mathbf{p}_r)}_{\substack{\text{Scattering} \\ \text{response}}} \\ 
		&\ \cdot \underbrace{\hat{\Phi}_n(\mathbf{p}_t, \boldsymbol{\mu}_n) \hat{\Phi}_n(\boldsymbol{\mu}_n, \mathbf{p}_r)}_{\text{Transmittance}} \Big)^2.
	\end{aligned}
\end{equation}
In \eqref{GLoS}, \(d_0=\|\mathbf{p}_r-\mathbf{p}_t\|\), \(d_n^\text{in}=\|\boldsymbol{\mu}_n-\mathbf{p}_t\|\), and \(d_n^\text{out}=\|\mathbf{p}_r-\boldsymbol{\mu}_n\|\). The terms \(\hat{\Omega}_n(\mathbf{p}_r)\), \(\hat{\Phi}_0(\mathbf{p}_t,\mathbf{p}_r)\), \(\hat{\Phi}_n(\mathbf{p}_t,\boldsymbol{\mu}_n)\), and \(\hat{\Phi}_n(\boldsymbol{\mu}_n,\mathbf{p}_r)\) are Gaussian-based amplitude surrogates of the unknown scattering response and transmittance coefficients in \eqref{G_0n}, which will be determined later.
Thus, GS-CG preserves the power-decomposition structure of the physical model while replacing inaccessible scatterer parameters with differentiable Gaussian parameters.

\textit{Remark (Compatibility with Grid-Based CGM Learning):}
GS-CG differs from existing wireless 3DGS designs in its prediction target \cite{wen2025wrfgs, zhou20256d}. It does not reconstruct direction-resolved spectra, full CSI, or generic radio fingerprints. Instead, it renders the locally averaged channel gain of each receiver grid. The Gaussian primitives therefore learn spatially stable propagation structures, such as blockage boundaries, penetrable regions, and dominant scattering support, rather than highly oscillatory pointwise channel responses. Such a grid-level representation also supports dynamic CGM updating, since the learned static propagation structure can be reused while local environmental changes are absorbed incrementally.

\begin{figure*}[!tb]
	\centering
	\includegraphics[width=0.65\textwidth]{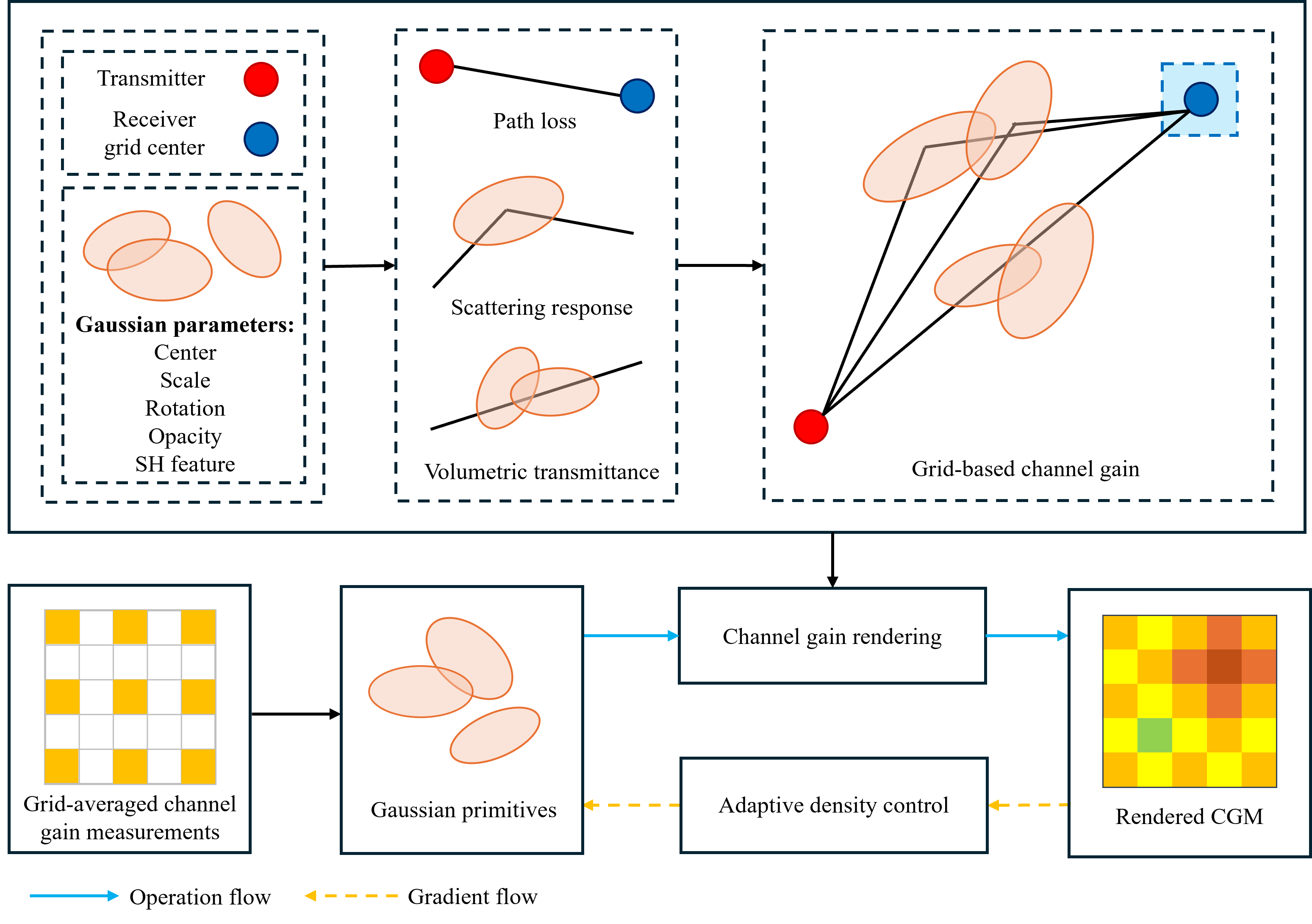}
	\caption{GS-CG pipeline for static CGM construction.}
	\label{fig_CGM_cons}
\end{figure*}

The unknown components in \eqref{GLoS} are the scattering response \(\{\hat{\Omega}_n\}\) and the path transmittance \(\hat{\Phi}_0\) and \(\{\hat{\Phi}_n\}\). The next two subsections will determine these terms. Fig. \ref{fig_CGM_cons} illustrates the overall GS-CG rendering and training pipeline for CGM construction.

\subsection{Effective Scattering Response}

We first model the effective scattering response \(\{\hat{\Omega}_n\}\) as the counterpart of \(\{\Omega_n\}\) in \eqref{G_0n}. It combines three factors: geometric support, opacity-based interaction strength, and scattering directionality.

\subsubsection{Geometric Support}

For a query location \(\mathbf{x}\in\mathcal{D}\), the geometric support of the \(n\)-th primitive is described by an anisotropic 3D Gaussian kernel:
\begin{equation} \label{G_n}
	\Upsilon_n(\mathbf{x}) = \exp \Big(-\frac{1}{2} (\mathbf{x} - \boldsymbol{\mu}_n)^T \boldsymbol{\Sigma}_n^{-1} (\mathbf{x} - \boldsymbol{\mu}_n)\Big),
\end{equation}
where \(\boldsymbol{\Sigma}_n\in\mathbb{R}^{3\times3}\) is the covariance matrix. We parameterize it as
\begin{equation}
	\boldsymbol{\Sigma}_n = \mathbf{R}(\mathbf{q}_n) \mathbf{S}^2(\mathbf{s}_n) \mathbf{R}^T(\mathbf{q}_n),
\end{equation}
where \(\mathbf{S}(\mathbf{s}_n)=\mathrm{diag}(\mathbf{s}_n)\) and \(\mathbf{R}(\mathbf{q}_n)\) is the rotation matrix induced by the quaternion \(\mathbf{q}_n\). This parameterization keeps \(\boldsymbol{\Sigma}_n\) positive definite and allows each primitive to represent object thickness, elongation, and orientation.
As such, \(\Upsilon_n(\mathbf{x})\) measures how strongly \(\mathbf{x}\) lies within the scattering support of the \(n\)-th Gaussian.

\subsubsection{Opacity-Based Interaction Strength}

The opacity \(\alpha_n\in[0,1]\) quantifies the interaction strength between the \(n\)-th Gaussian primitive and the propagating radio wave. It modulates both the scattered response generated by the Gaussian and the transmittance of propagation paths that pass through it. Thus, a larger \(\alpha_n\) represents a stronger scatterer and a more lossy obstacle, whereas a smaller \(\alpha_n\) makes the primitive nearly transparent.

\subsubsection{Scattering Directionality}

To describe anisotropic scattering, we assign each Gaussian a spherical harmonic (SH) feature vector \(\mathbf{f}_n=\{f_{n,l,m}\}_{0\le l\le L_\text{SH},-l\le m\le l}\in\mathbb{R}^{(L_\text{SH}+1)^2}\), where \(L_\text{SH}\) denotes the SH degree. Let
\(\hat{\mathbf{v}}_n(\mathbf{x})=\frac{\mathbf{x}-\boldsymbol{\mu}_n}{\|\mathbf{x}-\boldsymbol{\mu}_n\|}\) denote the outgoing direction from the Gaussian center to \(\mathbf{x}\). The directional scattering factor is
\begin{equation} \label{Gamma}
	\Gamma_n(\mathbf{x}) = \sum_{l=0}^{L_\text{SH}} \sum_{m=-l}^{l} f_{n, l, m} Y_{l,m}\big(\hat{\mathbf{v}}_n(\mathbf{x})\big),
\end{equation}
where \(Y_{l,m}(\cdot)\) denotes the SH basis.
This representation can capture both nearly isotropic scatterers and directional reflecting structures while remaining lightweight and differentiable.

Combining the three factors gives the effective scattering response of Gaussian primitive \(n\in\mathcal{N}\) toward \(\mathbf{x}\) by
\begin{equation} \label{Omega_n}
	\hat{\Omega}_n(\mathbf{x}) = \alpha_n \Upsilon_n(\mathbf{x}) \Gamma_n(\mathbf{x}).
\end{equation}
This formulation captures the reflection and diffusion effects of different Gaussians toward different receiver locations.

\subsection{Volumetric Transmittance Coefficients}

We next model the path transmittance \(\hat{\Phi}_0\) and \(\{\hat{\Phi}_n\}\), which corresponds to \(\Phi_0\) and \(\{\Phi_n^\text{in}, \Phi_n^\text{out}\}\) in \eqref{G_0n}. Transmittance characterizes the attenuation effect along a propagation segment and is determined by the opacity and geometric overlap of Gaussian primitives with that segment.

We define a propagation segment from location \(\mathbf{x}_a \in\mathcal{D}\) to \(\mathbf{x}_b \in\mathcal{D}\) as 
\begin{equation}
	\mathbf{r}(\iota) = \mathbf{x}_a + \iota (\mathbf{x}_b-\mathbf{x}_a), \iota\in[0,1].
\end{equation}
For the \(i\)-th Gaussian, we project its center onto this segment by
\begin{equation}
	\iota_i = \min\bigg(\max\Big(
	\frac{(\boldsymbol{\mu}_i-\mathbf{x}_a)^T(\mathbf{x}_b-\mathbf{x}_a)} {\|\mathbf{x}_b-\mathbf{x}_a\|^2}, 0\Big),1\bigg).
\end{equation}
The perpendicular offset from the Gaussian center to the segment is
\begin{equation}
	\mathbf{v}_{\perp,i} = \mathbf{r}(\iota_i) - \boldsymbol{\mu}_i.
\end{equation}

We then quantify the path overlap through the Mahalanobis distance induced by \(\boldsymbol{\Sigma}_i\):
\begin{equation} \label{Gblock}
	\hat{\Upsilon}_i(\mathbf{x}_a, \mathbf{x}_b) = \exp\big( -\frac{1}{2} \mathbf{v}_{\perp, i}^T \boldsymbol{\Sigma}_i^{-1} \mathbf{v}_{\perp, i} \big).
\end{equation}
Different from the point-support kernel \(\Upsilon_i\) defined in \eqref{G_n}, \(\hat{\Upsilon}_i\) measures path-level overlap. It provides a geometry-aware surrogate for soft blockage, penetration loss, and volumetric attenuation. Unlike projection-domain splatting or virtual projection-plane methods \cite{wen2025wrfgs, zhou20256d}, this formulation evaluates ray-Gaussian interaction directly in 3D space, so that the transmittance rendering is directly coupled with the underlying propagation geometry.

Using this overlap factor, we model the transmittance of a path segment from \(\mathbf{x}_a\) to \(\mathbf{x}_b\) associated with Gaussian index \(n\in\{0\}\cup\mathcal{N}\) as
\begin{equation} \label{trans}
	\hat{\Phi}_n(\mathbf{x}_a,\mathbf{x}_b)
	= \prod_{i\in\mathcal{N}, \ i\ne n}
	\big(1-\alpha_i\hat{\Upsilon}_i(\mathbf{x}_a,\mathbf{x}_b)\big).
\end{equation}
For \(n=0\), \eqref{trans} gives the direct-path transmittance along the transmitter-receiver segment, where only Gaussians with non-negligible path overlap and opacity effectively attenuate the path. For \(n\in\mathcal{N}\), it gives the transmittance of a scattering segment associated with the \(n\)-th Gaussian and excludes its self-obstruction, since that primitive already contributes through \(\hat{\Omega}_n\) in \eqref{Omega_n}. 
%Thus, high-opacity Gaussians with strong path overlap induce strong attenuation, whereas transparent or weakly overlapped Gaussians have limited effect.

\subsection{Construction Optimization and Adaptive Density Control}

Next, we present the loss function for training the CGM. Based on \eqref{Gtotal}, the channel gain rendered by the reference model \(\varTheta_\text{s}\) is \(\hat{G}(\mathbf{p}_r;\varTheta_\text{s})\) in the linear power domain. Since channel gain can span a large dynamic range, a linear-domain loss may overemphasize high-power grids. We therefore fit the model in the logarithmic domain and use the dB-domain mean absolute error (MAE). Given CGM measurements over grids \(\hat{\mathcal{D}}_\text{s}\), the construction loss is given by
\begin{equation} \label{loss_mae}
	\mathcal{L}_{\text{MAE}}^\text{cons}(\varTheta_\text{s}) = \frac{1}{|\hat{\mathcal{D}}_\text{s}|} \sum_{\mathbf{p}_r\in\hat{\mathcal{D}}_\text{s}} \Big|10\log_{10}\hat{G}(\mathbf{p}_r; \varTheta_\text{s}) - 10\log_{10}G(\mathbf{p}_r)\Big|.
\end{equation}
Before training, \(\varTheta_\text{s}\) can be initialized from available scene geometry or environmental structure information.
We then optimize \(\varTheta_\text{s}\) end-to-end by back-propagation. The overall static construction procedure is summarized in Algorithm \ref{alg:cgm_gs_construction}.

GS-CG also adapts the resolution of the scene representation. Instead of placing Gaussian primitives uniformly over the entire domain, it refines the primitive distribution according to local propagation complexity. During training, GS-CG tracks the accumulated positional gradient of each primitive and applies adaptive density control (ADC) at fixed intervals through cloning, splitting, and pruning.
Specifically, a small-scale Gaussian with a large positional gradient is cloned to increase local modeling capacity. A large-scale Gaussian with a large gradient is split into smaller primitives, since it may cover multiple propagation features such as blockage boundaries or sharp transition regions. Gaussians with very low opacity are pruned because they contribute little to the rendered gain. These operations allocate more primitives to electromagnetically complex regions while keeping simple regions sparse.

\begin{algorithm}[!tb]
	\caption{GS-CG Construction}\label{alg:cgm_gs_construction}
	\KwIn{Channel gain measurements \(G(\mathbf{p}_r)\) over grids \(\mathbf{p}_r\in\hat{\mathcal{D}}_\text{s}\), transmitter position \(\mathbf{p}_t\)}
	\KwOut{Reference Gaussian parameters \(\varTheta_\text{s}\)}
	Initialize \(\varTheta_\text{s}\) from available scene geometry\;
	\For{each training iteration}{
		Render \(\hat{G}(\mathbf{p}_r; \varTheta_\text{s})\) for \(\mathbf{p}_r \in \hat{\mathcal{D}}_\text{s}\) using \eqref{Gtotal} and \eqref{GLoS}\;
		Compute \(\mathcal{L}_{\text{MAE}}^\text{cons}(\varTheta_\text{s})\) and update \(\varTheta_\text{s}\) by back-propagation\;
		Apply ADC at fixed intervals\;
	}
	Return \(\varTheta_\text{s}\).
\end{algorithm}

\section{Incremental GS-CG for Dynamic CGM Updating}

\begin{figure*}[!tb]
	\centering
	\includegraphics[width=0.65\textwidth]{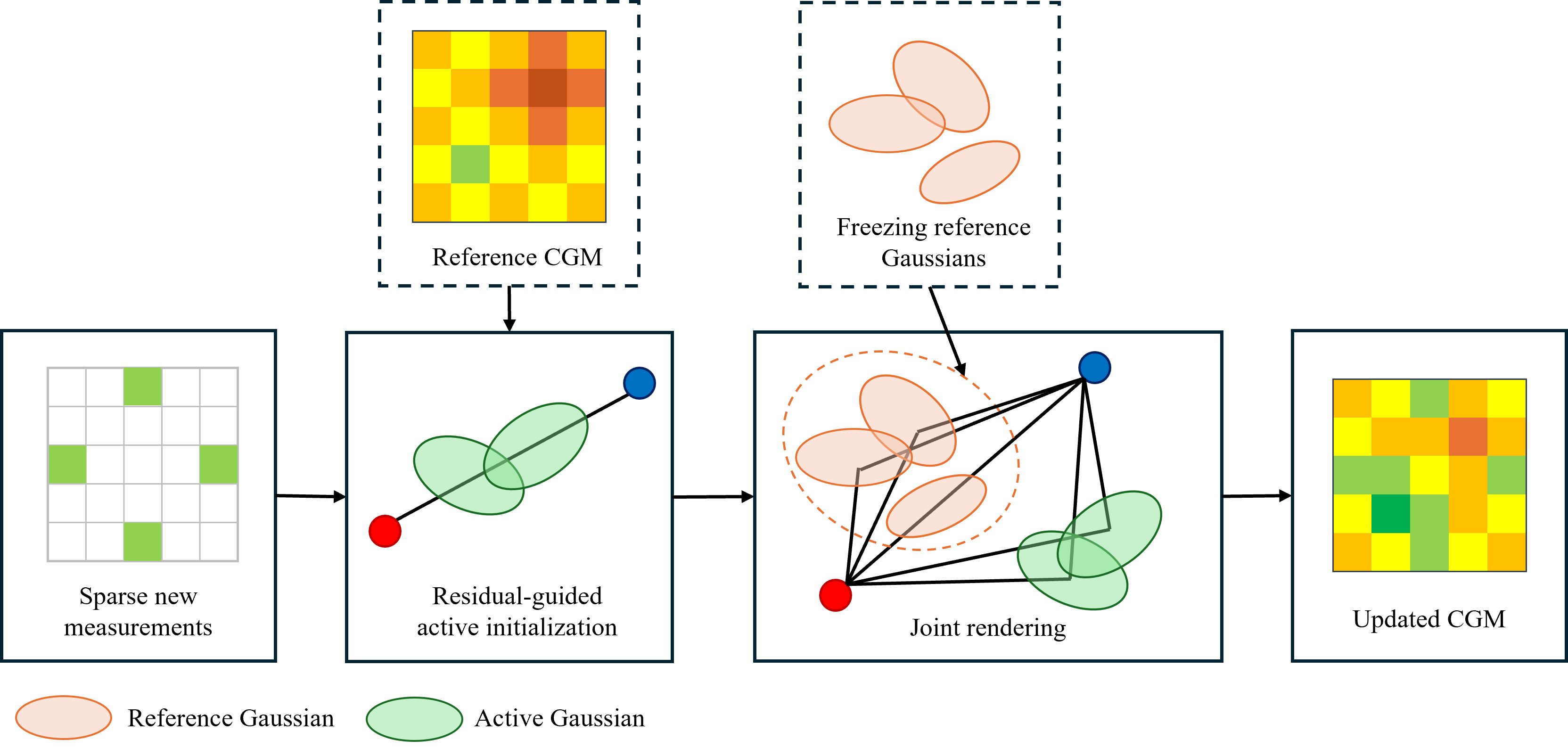}
	\caption{Incremental GS-CG pipeline for dynamic CGM updating.}
	\label{fig_CGM_upd}
\end{figure*}

After constructing the static/reference model \(\varTheta_\text{s}\), we consider updating the CGM from the reference state \(\mathcal{M}_\text{s}\) to the real-time state \(\mathcal{M}_\text{d}\), in dynamic scenarios when objects may appear or move. Such changes usually affect only a local part of the propagation environment. Retraining all Gaussian primitives from sparse dynamic measurements in \(\hat{\mathcal{D}}_\text{d}\) is therefore inefficient and may distort unchanged regions. We instead update the CGM incrementally by freezing the static representation and introducing a compact active Gaussian set whose primitives jointly participate in channel gain rendering. The incremental GS-CG pipeline for CGM updating is illustrated in Fig. \ref{fig_CGM_upd}.

\subsection{Incremental Gaussian Representation}

Recall that \(\varTheta_\text{s}\) denotes the Gaussian parameters learned during static construction. We introduce an active Gaussian set with index set \(\mathcal{N}_\text{act}\), whose parameters are given in the following, similarly to \eqref{Theta} and \eqref{Theta_n}:
\begin{equation} \label{Theta_act}
	\varTheta_\text{act}
	\triangleq \{\vartheta_n\}_{n\in\mathcal{N}_\text{act}}.
\end{equation}
The dynamic representation is then given by
\begin{equation} \label{Theta_dyn}
	\varTheta_\text{d} \triangleq \varTheta_\text{s} \cup \varTheta_\text{act},
\end{equation}
which combines the reference model \(\varTheta_\text{s}\) and active Gaussians \(\varTheta_\text{act}\) for rendering channel gains. 

During the forward pass, all primitives in \(\varTheta_\text{d}\) jointly render the dynamic CGM through \eqref{Gtotal} and \eqref{GLoS}. In the joint rendering, the newly added active primitives may attenuate existing propagation paths or add new effective scattering contributions. During the back-propagation, GS-CG freezes \(\varTheta_\text{s}\) and optimizes only \(\varTheta_\text{act}\). This strategy converts dynamic updating from full-map reconstruction into localized residual learning in the Gaussian domain. 
%If the environment undergoes large-scale reconfiguration or dynamic structures become persistent, the static/reference CGM should be refreshed from denser or long-term measurements.

\subsection{Residual-Guided Active Initialization}

Next, we introduce the initialization of active Gaussian primitives. Since the locations and shapes of newly changed environmental structures are usually unknown, directly placing active Gaussians on these structures is infeasible. Moreover, dynamic changes usually occupy only a small portion of the domain, making random initialization inefficient.
%Since dynamic changes usually occupy only a small portion of the domain, random initialization is inefficient. 
We therefore initialize active Gaussians \(\varTheta_\text{act}\) in nearby regions that likely contain new radio-interaction structures, by the following residual-guided back-projection scheme.\footnote{The active initialization mainly targets newly introduced blockage or attenuation structures, since permanent structures are already captured during static CGM construction. If the environment undergoes large-scale reconfiguration, the reference CGM should be refreshed from denser or long-term measurements. Moreover, the residual-guided back-projection mainly focuses on blockages along direct paths, which is more identifiable from sparse channel gain residuals.}

For each newly measured receiver grid \(\mathbf{p}_r\in\hat{\mathcal{D}}_\text{d}\), define the one-sided channel gain residual for the reference CGM as
\begin{equation} \label{residual}
	e(\mathbf{p}_r) = \max\Big(10\log_{10}\hat{G}(\mathbf{p}_r;\varTheta_\text{s}) - 10\log_{10}G(\mathbf{p}_r),0\Big).
\end{equation}
This residual emphasizes locations where the static model overestimates the channel gain. We then allocate active Gaussians to each measured grid according to the normalized weight
\begin{equation}
	\rho(\mathbf{p}_r)
	= \frac{e(\mathbf{p}_r)}
	{\sum_{\mathbf{p}_r'\in\hat{\mathcal{D}}_\text{d}}e(\mathbf{p}_r')},
\end{equation}
such that grids with larger \(\rho(\mathbf{p}_r)\) receive more Gaussians.

Let \(\mathbf{d}(\mathbf{p}_r)=\mathbf{p}_r-\mathbf{p}_t\) denote the transmitter-to-receiver propagation path. We place active Gaussians along this path, on which the structure causing the gain reduction is likely to lie. Starting from \([\kappa_{\min},\kappa_{\max}]=[0,1]\), we refine the feasible depth interval using nearby measured grids. For another measured grid \(\mathbf{p}_r' \in\hat{\mathcal{D}}_\text{d}\), define the ray-direction similarity as
\begin{equation}
	\eta(\mathbf{p}_r,\mathbf{p}_r')
	= \frac{\mathbf{d}(\mathbf{p}_r)^T\mathbf{d}(\mathbf{p}_r')}
	{\|\mathbf{d}(\mathbf{p}_r)\|\|\mathbf{d}(\mathbf{p}_r')\|}.
\end{equation}
If \(\eta(\mathbf{p}_r,\mathbf{p}_r')\ge\cos\tau_{\theta}\) holds for a small angle threshold \(\tau_{\theta} > 0\), then the two grids are deemed to lie in similar propagation directions. In this case, if \(\mathbf{p}_r'\) is closer to the transmitter and has a smaller residual than a given threshold \(\tau_e^{\min} > 0\), i.e., \(\|\mathbf{d}(\mathbf{p}_r')\|<\|\mathbf{d}(\mathbf{p}_r)\|\) and \(e(\mathbf{p}_r')<\tau_e^{\min}\), the changed structure is unlikely to lie before \(\mathbf{p}_r'\). We then update
\begin{equation}
	\kappa_{\min} \leftarrow \max \Big(\kappa_{\min},
	\frac{\|\mathbf{d}(\mathbf{p}_r')\|}{\|\mathbf{d}(\mathbf{p}_r)\|}\Big).
\end{equation}
Conversely, if a closer grid has a larger residual than a given threshold \(\tau_e^{\max} > \tau_e^{\min}\), i.e., \(\|\mathbf{d}(\mathbf{p}_r')\|<\|\mathbf{d}(\mathbf{p}_r)\|\) and \(e(\mathbf{p}_r')>\tau_e^{\max}\), the changed structure is likely to appear before \(\mathbf{p}_r'\), and we update
\begin{equation}
	\kappa_{\max} \leftarrow \min \Big(\kappa_{\max},
	\frac{\|\mathbf{d}(\mathbf{p}_r')\|}{\|\mathbf{d}(\mathbf{p}_r)\|}\Big).
\end{equation}
In implementation, inconsistent interval constraints are clipped to keep \(\kappa_{\min}\le \kappa_{\max}\). As such, an active Gaussian assigned to \(\mathbf{p}_r\) is uniformly initialized in depth interval \([\kappa_{\min},\kappa_{\max}]\) as
\begin{equation}
	\begin{aligned}
		\boldsymbol{\mu}
		=& \mathbf{p}_t + \kappa \mathbf{d}(\mathbf{p}_r) + \boldsymbol{\zeta}, \\
		&\kappa\sim\mathcal{U}(\kappa_{\min},\kappa_{\max}), \boldsymbol{\zeta}\sim\mathcal{N}(\mathbf{0},\sigma^2\mathbf{I}),
	\end{aligned}
\end{equation}
where \(\boldsymbol{\zeta}\) is a spatial jitter with scale \(\sigma\). This initialization places active primitives in the 3D subregions that best explain the newly observed propagation mismatch.

\subsection{Active Set Optimization}

After initialization, we render the dynamic CGM with \(\varTheta_\text{d} = \varTheta_\text{s} \cup \varTheta_\text{act}\) and back-propagate gradients only to \(\varTheta_\text{act}\). Using the new sparse measurements over grids \(\hat{\mathcal{D}}_\text{d}\), the CGM updating loss is given by
\begin{equation} \label{loss_upd}
	\begin{aligned}
		\mathcal{L}_\text{MAE}^\text{upd}(\varTheta_\text{act}) =
		\frac{1}{|\hat{\mathcal{D}}_\text{d}|} \sum_{\mathbf{p}_r\in\hat{\mathcal{D}}_\text{d}} 
		&\Big|10\log_{10}\hat{G}(\mathbf{p}_r;\varTheta_\text{s} \cup \varTheta_\text{act})\\
		&- 10\log_{10}G(\mathbf{p}_r)\Big|.
	\end{aligned}
\end{equation}

By preserving the reference Gaussian representation and only optimizing parameters for the active set, GS-CG can efficiently adapt to environmental changes from sparse new measurements while maintaining stable predictions in unchanged regions. The overall dynamic updating procedure is summarized in Algorithm \ref{alg:cgm_gs_updating}.

\begin{algorithm}[!tb]
	\caption{Incremental GS-CG Updating}\label{alg:cgm_gs_updating}
	\KwIn{Reference Gaussian parameters \(\varTheta_\text{s}\), new channel gain measurements over grids \(\hat{\mathcal{D}}_\text{d}\), transmitter position \(\mathbf{p}_t\)}
	\KwOut{Updated Gaussian parameters \(\varTheta_\text{d}\)}
	Freeze \(\varTheta_\text{s}\)\;
	Initialize \(\varTheta_\text{act}\) by residual-guided back-projection\; 
%	from \(\mathbf{p}_t\) to \(\mathbf{p}_r\in\hat{\mathcal{D}}_\text{d}\)\;
	\For{each update iteration}{
		Render \(\hat{G}(\mathbf{p}_r;\varTheta_\text{s}\cup\varTheta_\text{act})\) for \(\mathbf{p}_r\in\hat{\mathcal{D}}_\text{d}\)\;
		Compute \(\mathcal{L}_\text{MAE}^\text{upd}(\varTheta_\text{act})\) and update only \(\varTheta_\text{act}\)\;
	}
	Return \(\varTheta_\text{d}=\varTheta_\text{s}\cup\varTheta_\text{act}\).
\end{algorithm}

\section{Low-Complexity MLP-GS Model}

The GS-CG model in Sections III and IV explicitly evaluates the scattering response and the transmittance coefficients of each Gaussian-induced path. This provides physical interpretability, but the scattering branch can be computationally expensive when many Gaussian primitives participate in rendering. To reduce such cost, in this section we further develop a low-complexity MLP-assisted GS-CG model, termed MLP-GS.

MLP-GS uses the following set of Gaussian primitives to represent the scene,
\begin{equation}
	\varTheta^\text{MLP} \triangleq \{\vartheta_n^\text{MLP}\}_{n\in\mathcal{N}},
\end{equation}
where the \(n\)-th primitive is parameterized as 
\begin{equation} \label{Theta_mlp}
	\vartheta_n^\text{MLP} = \{\boldsymbol{\mu}_n, \mathbf{s}_n, \mathbf{q}_n, \alpha_n, \mathbf{F}_n\}.
\end{equation}
Here, \(\boldsymbol{\mu}_n\), \(\mathbf{s}_n\), \(\mathbf{q}_n\), and \(\alpha_n\) retain the same meanings as in GS-CG, and \(\mathbf{F}_n\) is a compact latent feature decoded by an MLP. Following the GS-CG rendering model in \eqref{GLoS}, the direct-path gain and the \(n\)-th scattering gain in MLP-GS are reformulated as follows by incorporating the MLP:
\begin{equation} \label{GLoS_mlp}
	\begin{aligned}
		\hat{G}_0^\text{MLP}(\mathbf{p}_r)
		&= \Big(\frac{\lambda}{4\pi d_0}\hat{\Phi}_0(\mathbf{p}_t,\mathbf{p}_r)\Big)^2,\\
		\hat{G}_n^\text{MLP}(\mathbf{p}_r)
		&= \Big(\alpha_n \Upsilon_n(\mathbf{p}_r)
		\Psi(\mathbf{p}_t,\boldsymbol{\mu}_n;\mathbf{F}_n)
		\Psi(\boldsymbol{\mu}_n,\mathbf{p}_r;\mathbf{F}_n)\Big)^2,
	\end{aligned}
\end{equation}
where \(\Psi(\mathbf{x}_a,\mathbf{x}_b;\mathbf{F}_n)\) denotes a lightweight MLP decoder with two hidden layers that outputs an effective propagation factor for the segment from \(\mathbf{x}_a\) to \(\mathbf{x}_b\).
The rendered channel gain is then obtained in the same additive form as \eqref{Gtotal}:
\begin{equation} \label{Gtotal_mlp}
	\hat{G}^\text{MLP}(\mathbf{p}_r)
	= \hat{G}_0^\text{MLP}(\mathbf{p}_r)
	+ \sum_{n\in\mathcal{N}}\hat{G}_n^\text{MLP}(\mathbf{p}_r).
\end{equation}

The difference between GS-CG and MLP-GS lies mainly in the scattering branch. In \eqref{GLoS}, \(\hat{\Omega}_n(\mathbf{p}_r)\), \(\hat{\Phi}_n(\mathbf{p}_t,\boldsymbol{\mu}_n)\), and \(\hat{\Phi}_n(\boldsymbol{\mu}_n,\mathbf{p}_r)\) are explicitly constructed from Gaussian geometry, opacity, SH-based scattering directionality, and ray-Gaussian overlap. In \eqref{GLoS_mlp}, the Gaussian spatial support \(\alpha_n\Upsilon_n(\mathbf{p}_r)\) is retained, while the remaining scattering-path effects, including distance-dependent attenuation, transmittance, and directional scattering, are absorbed into the MLP decoder with latent feature \(\mathbf{F}_n\), and repeated pairwise scattering/transmittance calculations are avoided.

As a result, MLP-GS is less physically explicit than GS-CG in modeling individual scattering paths, while preserving the geometry-aware inductive bias of GS-CG for structured and data-efficient learning. 
After replacing \(\hat{G}(\mathbf{p}_r)\) with \(\hat{G}^\text{MLP}(\mathbf{p}_r)\), the same static construction and dynamic updating procedures in Algorithms~\ref{alg:cgm_gs_construction} and~\ref{alg:cgm_gs_updating} can be directly applied, with \(\mathbf{F}_n\) and the MLP decoder parameters optimized together with the Gaussian parameters during construction and only the parameters associated with the active set updated during dynamic refinement.
%only the active Gaussian parameters updated during dynamic refinement.

\section{Simulation Results}

In this section, we evaluate the proposed GS-CG and MLP-GS models for grid-based CGM construction and dynamic updating. The simulations aim to verify the construction accuracy, updating efficiency, and the effectiveness of the proposed incremental design.

\subsection{Simulation Setup and Implementation Details}

\begin{figure}[!tb]
	\centering 
	\subfigure[Campus scene for CGM construction.]
	{\includegraphics[width=0.2\textwidth]{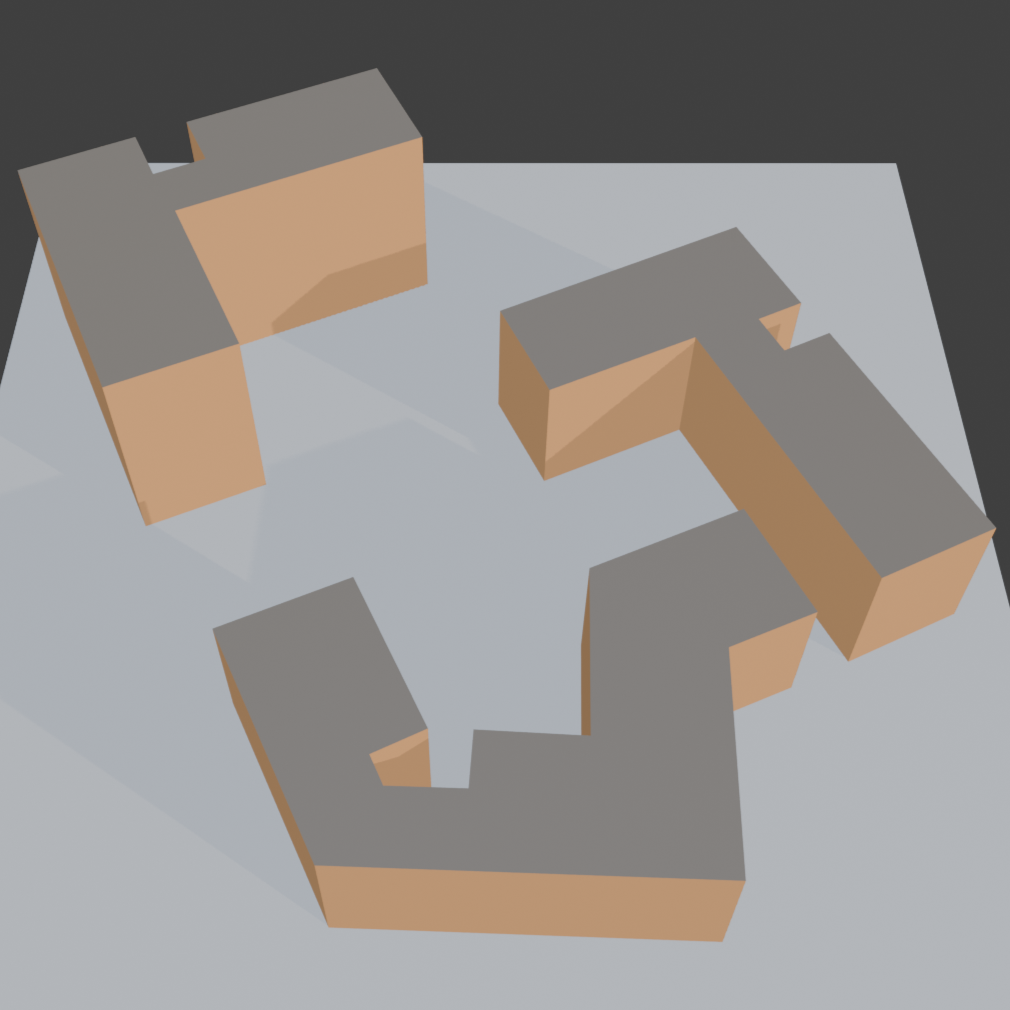}}
	\subfigure[Campus scene for CGM updating.]
	{\includegraphics[width=0.2\textwidth]{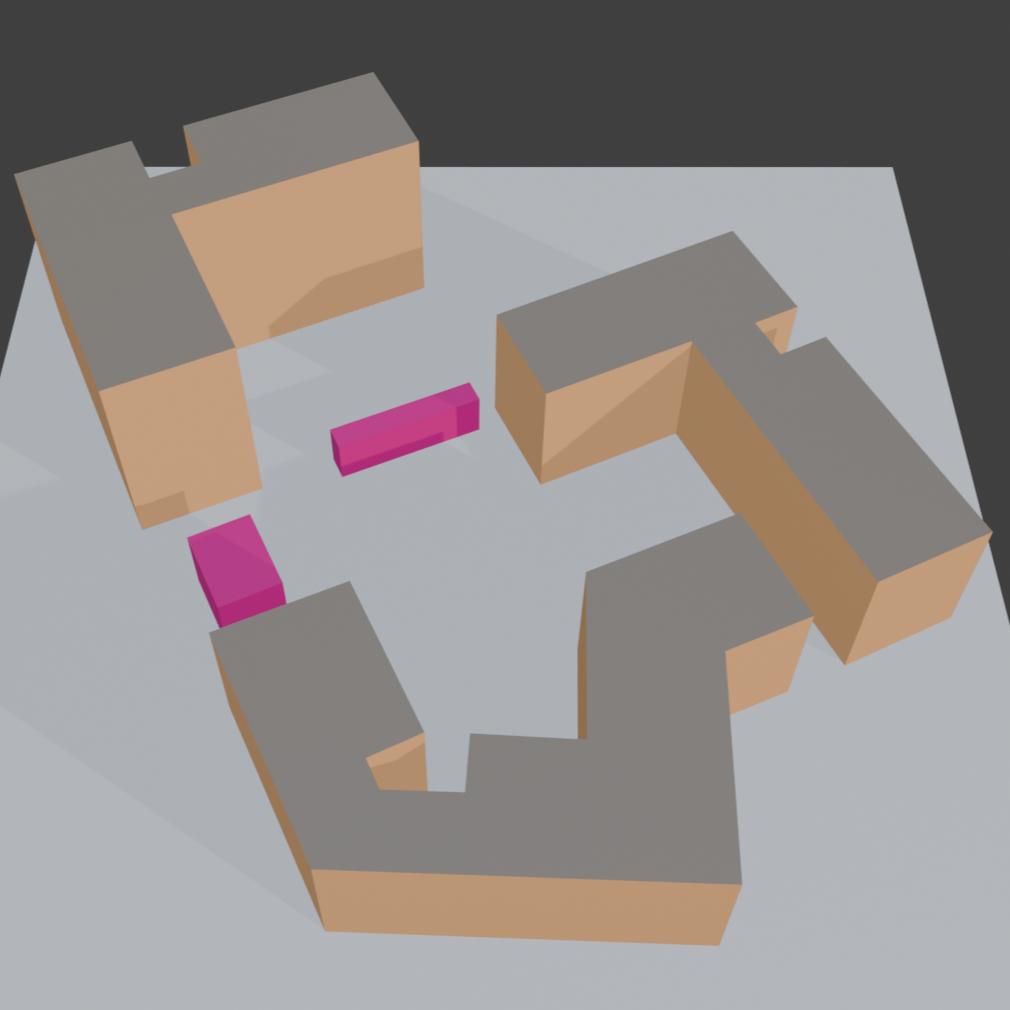}}
	\subfigure[Road scene for CGM construction.]
	{\includegraphics[width=0.2\textwidth]{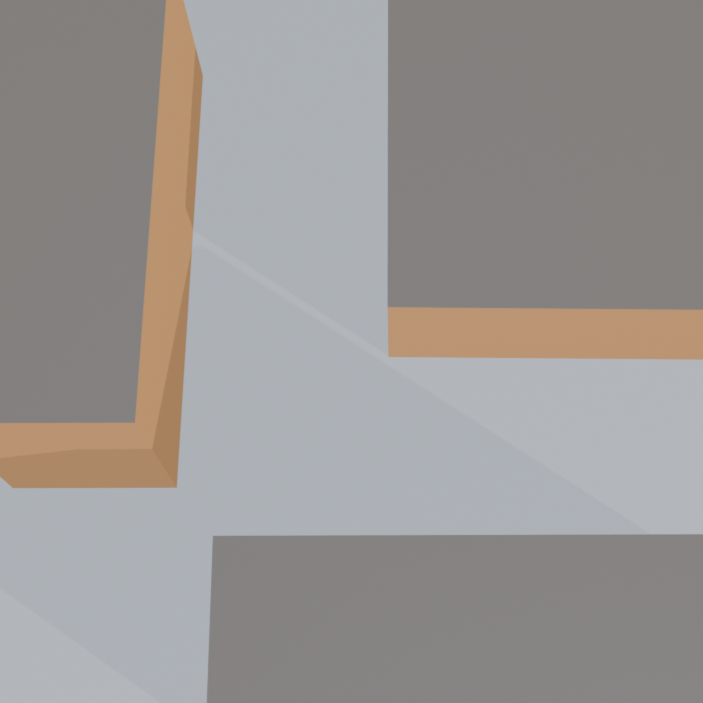}}
	\subfigure[Road scene for CGM updating.]
	{\includegraphics[width=0.2\textwidth]{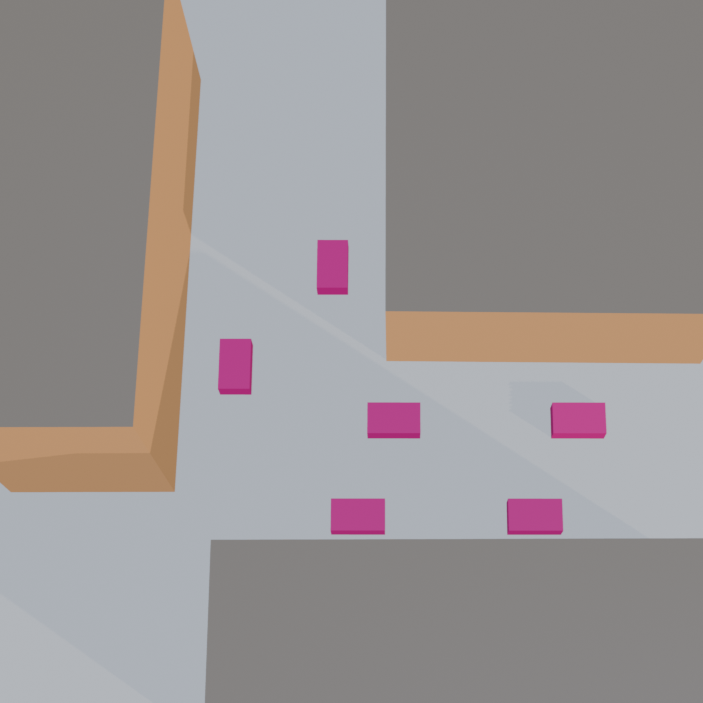}}
	\subfigure[Low-altitude scene for CGM construction.]
	{\includegraphics[width=0.2\textwidth]{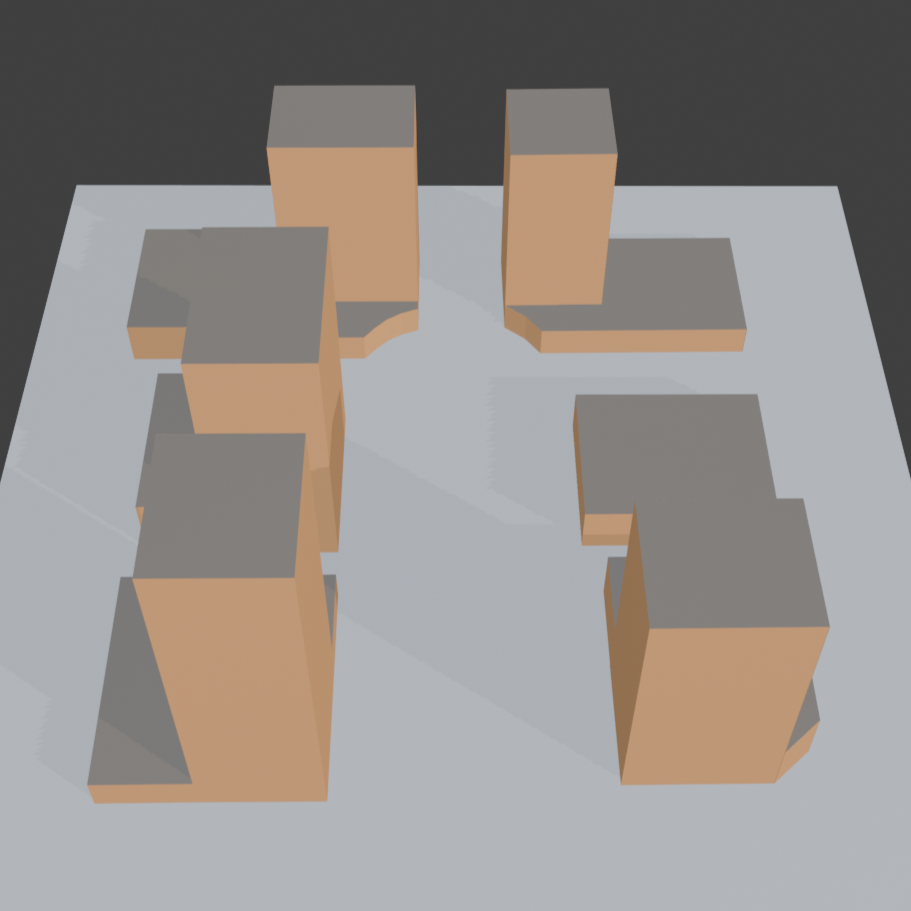}}
	\subfigure[Low-altitude scene for CGM updating.]
	{\includegraphics[width=0.2\textwidth]{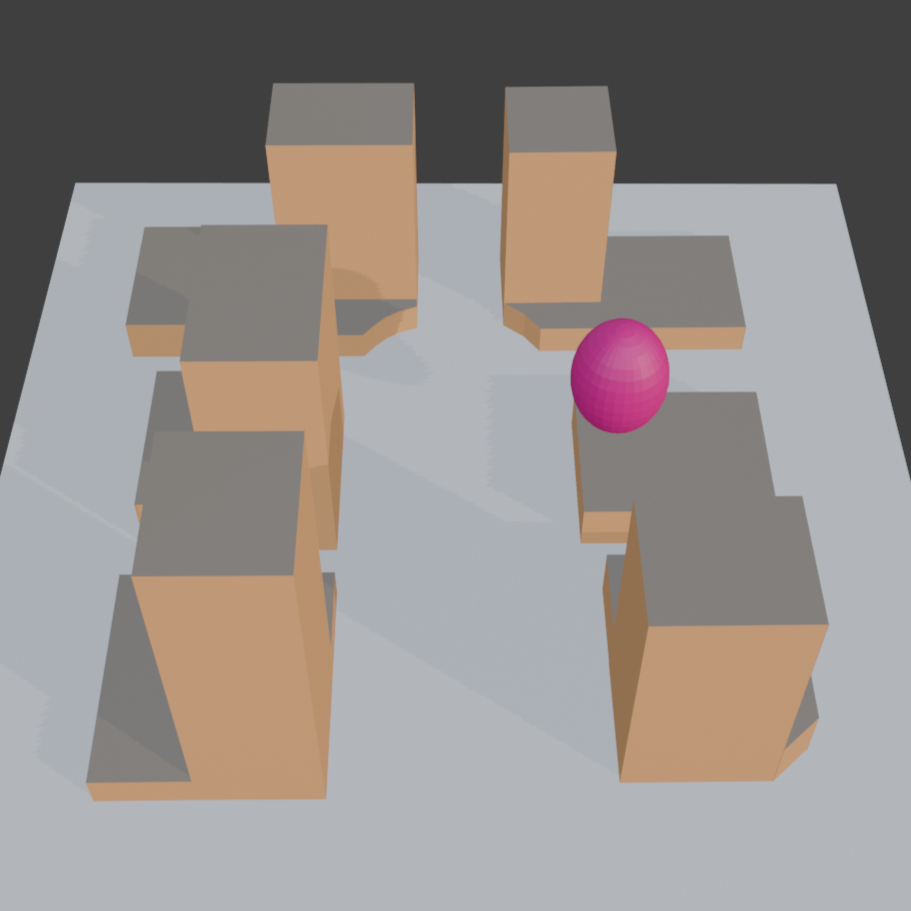}}
	\caption{Test scenes for CGM construction and dynamic updating.}
	\label{fig:test_scenes}
\end{figure}

As shown in Fig. \ref{fig:test_scenes}, we consider three representative test scenes, including a campus scene, a road scene, and a low-altitude scene. The campus scene has a map size of $100\times100$ m\textsuperscript{2} and contains dense building blockage. The road scene has a map size of $50\times50$ m\textsuperscript{2} and is used to evaluate CGM updating under continuously changing blockers. The low-altitude scene has a map size of $300\times300$ m\textsuperscript{2} and is used to examine the scalability of the proposed design in a larger area with relatively sparse blockage. The buildings are modeled according to OpenStreetMap \cite{OpenStreetMap}. For all scenes, the CGM is partitioned into $50\times50=2500$ grids in the horizontal plane, with transmitter altitude of $5$ m and receiver altitudes of $1$ m, $1$ m, and $100$ m, respectively.\footnote{Although the proposed GS-CG model supports a general 3D receiver region, the simulations consider horizontal receiver-grid slices at fixed altitudes, which are common CGM instances for ground and low-altitude coverage analysis.} The carrier frequency is $f_c=3.5$ GHz.

The ground-truth CGM is generated by the radio map solver in Sionna-RT \cite{sionna}. Specifically, each CGM entry is computed over multiple sampling points within each grid according to \eqref{Gint} as a grid-averaged channel gain rather than a pointwise channel response. In the training process, we use the center of each grid as the reference receiver coordinate. Accordingly, the model learns a mapping from the grid-center location to the corresponding average channel gain. These implementations are consistent with the grid-based channel model and CGM formulation in Section II.

The GS-CG model is implemented in PyTorch and trained on an NVIDIA RTX 4090 GPU. For each Gaussian primitive \(n\in\mathcal{N}\), the trainable parameters include its center $\boldsymbol{\mu}_n$, scale $\mathbf{s}_n$, rotation $\mathbf{q}_n$, opacity $\alpha_n$, and SH coefficients $\mathbf{f}_n$ of degree $3$. For MLP-GS, the trainable parameters include $\boldsymbol{\mu}_n$, $\mathbf{s}_n$, $\mathbf{q}_n$, $\alpha_n$, and a latent feature vector $\mathbf{F}_n$ of dimension $16$. The Adam optimizer is adopted with an initial learning rate of $0.01$ and a cosine annealing scheduler.
For static CGM construction, $25\times25=625$ grids, corresponding to $25\%$ of all grids, are uniformly selected as training samples. The model is initialized with $5000$ Gaussians derived from the scene geometry and trained for $2000$ epochs. For dynamic updating, only $10\times10=100$ grids, corresponding to $4\%$ of all grids, are uniformly selected as new measurements. In this stage, the previously learned representation is reused, and $1000$ additional active Gaussians are introduced and trained for $1000$ epochs.

For comparison, we consider the following benchmark schemes.\footnote{NeRF\textsuperscript{2} \cite{zhao2023nerf2}, WRF-GS \cite{wen2025wrfgs}, and generative CKM predictors \cite{li2025radiotransformer} are not included as direct quantitative baselines. NeRF\textsuperscript{2} learns an RF radiance field and WRF-GS reconstructs spatial spectra, whereas GS-CG is built on a grid-based channel gain model. Adapting NeRF\textsuperscript{2} or WRF-GS to this task would require redesigning their rendering process. Generative predictors usually rely on large cross-scene training datasets, while our method constructs and updates a newly encountered scene from its own measurements. In addition, these methods do not directly address dynamic CGM updating from sparse new measurements.}
\begin{itemize}
	\item \emph{MLP-based radiance field (MLP-RF) model:} Motivated by NeRF-based modeling \cite{zhao2023nerf2}, this baseline uses MLPs with inputs of transceiver and scatterer positions to represent the radiance fields and render the direct and scattering channel gain. Without explicitly modeling scatterers' geometry and opacity, its trainable variables are scatterer positions and MLP parameters.
%	It adopts the same incremental updating logic and residual-guided initialization as GS-CG.
	\item \emph{Virtual scatterer (VS) model:} This baseline follows the scatterer-based CGM construction framework in \cite{sun2025channel}, which also targets grid-averaged channel gain. It estimates large-scale multi-path parameters including scattering responses and path-loss-related coefficients, while the scatterer locations are not optimized.
	\item \emph{Kriging interpolation:} This baseline constructs the CGM by spatial interpolation from measured grid samples \cite{OLIVER01071990}.
\end{itemize}
All methods are evaluated by the MAE in the dB domain over all grids.

\subsection{CGM Construction and Dynamic Updating Performances}

\begin{table}[!tb]
	\centering
	\small
	\caption{MAE performance of different schemes (in dB).}
	\begin{tabular}{c|c|c|c|c|c|c} 
		\hline
		\multirow{2}{*}{Scene} & \multicolumn{2}{c|}{Campus} & \multicolumn{2}{c|}{Road} & \multicolumn{2}{c}{Low-altitude} \\
		\cline{2-7}
		& Cons. & Upd. & Cons. & Upd. & Cons. & Upd. \\
		\hline
		GS-CG & \textbf{0.60} & \textbf{1.46} & \textbf{0.54} & \textbf{2.26} & \textbf{0.58} & \textbf{1.07} \\
		\hline
		MLP-GS & 0.76 & 1.76 & 0.64 & 2.73 & 0.67 & 1.31 \\
		\hline
		MLP-RF & 1.48 & 3.99 & 0.85 & 4.26 & 0.93 & 2.07 \\
		\hline
		VS & 1.64 & 3.74 & 1.15 & 3.88 & 1.32 & 1.90 \\
		\hline
		Kriging & 1.87 & 4.34 & 1.38 & 5.44 & 1.38 & 2.95 \\
		\hline
		\multicolumn{7}{l}{\footnotesize Note: Cons. = Construction, Upd. = Updating.}
	\end{tabular}
\end{table}

Table I summarizes the MAE performance in the three test scenes. GS-CG achieves the best accuracy in all cases. For static construction, it obtains sub-dB MAEs in the three scenes. This verifies that the proposed Gaussian-parameterized rendering can accurately reconstruct grid-based channel gain from $25\%$ training samples. MLP-GS achieves slightly higher but still competitive MAEs, showing that replacing part of the explicit scattering interaction with neural surrogates causes only moderate accuracy loss. In comparison, the benchmark schemes of MLP-RF, VS, and Kriging show larger errors, because they are less effective in jointly capturing smooth path-loss variation and sharp blockage-induced discontinuities.

In Table I, the advantage of GS-CG becomes more evident in dynamic updating. With only $4\%$ new measurements, GS-CG achieves the lowest MAEs in the three scenes. By preserving the previously learned Gaussian structures and adding active Gaussians for local residual changes, GS-CG converts the underdetermined full-map reconstruction problem into a localized residual learning problem. By contrast, although MLP-RF uses the same incremental update and residual initialization strategy, its updating errors remain much larger. This indicates that the performance gain comes not only from the incremental training protocol, but also from the Gaussian-based physical representation.

\begin{figure}[!tb]
	\centering 
	\subfigure[Ground truth.]
	{\includegraphics[width=0.22\textwidth]{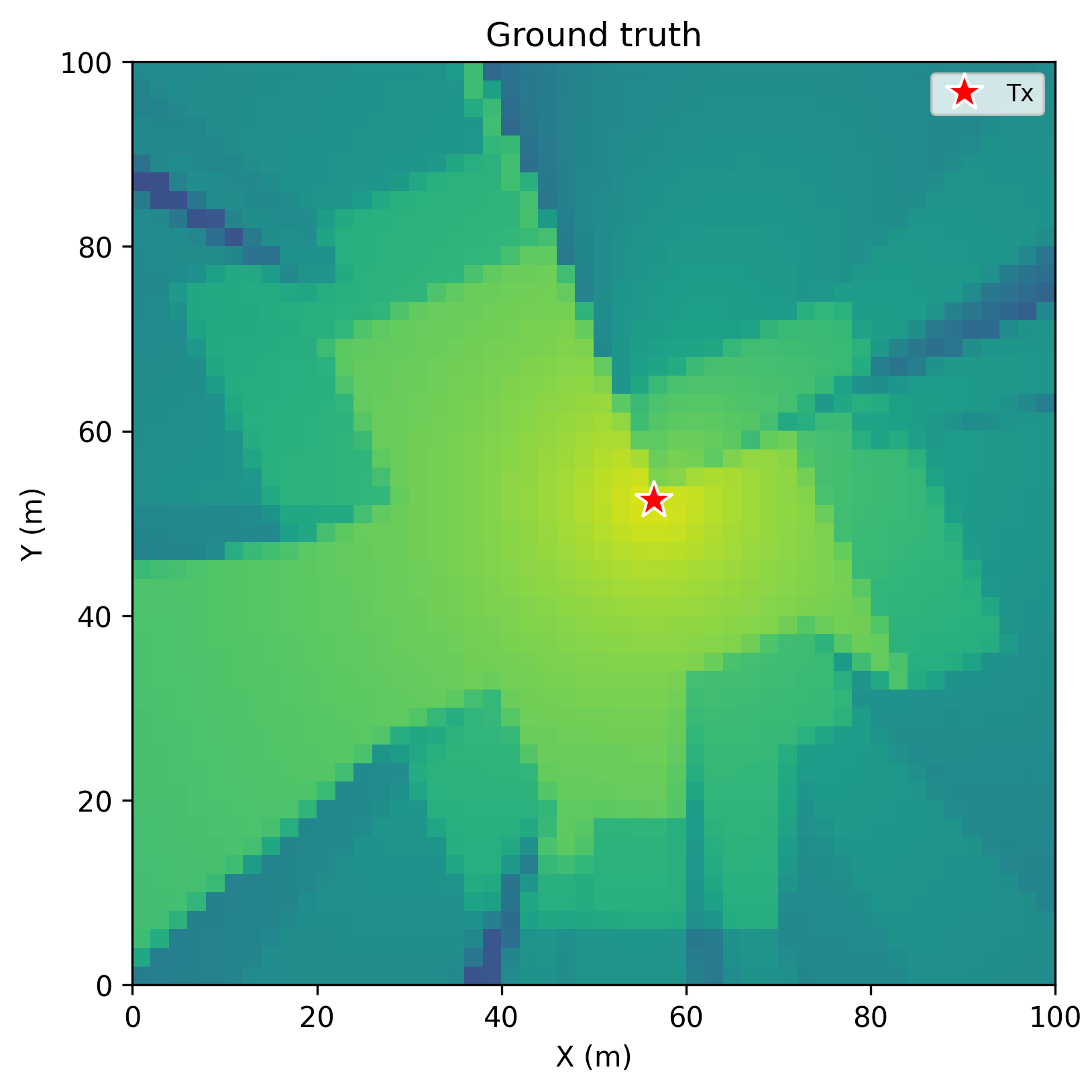}}
	\subfigure[GS-CG.]
	{\includegraphics[width=0.22\textwidth]{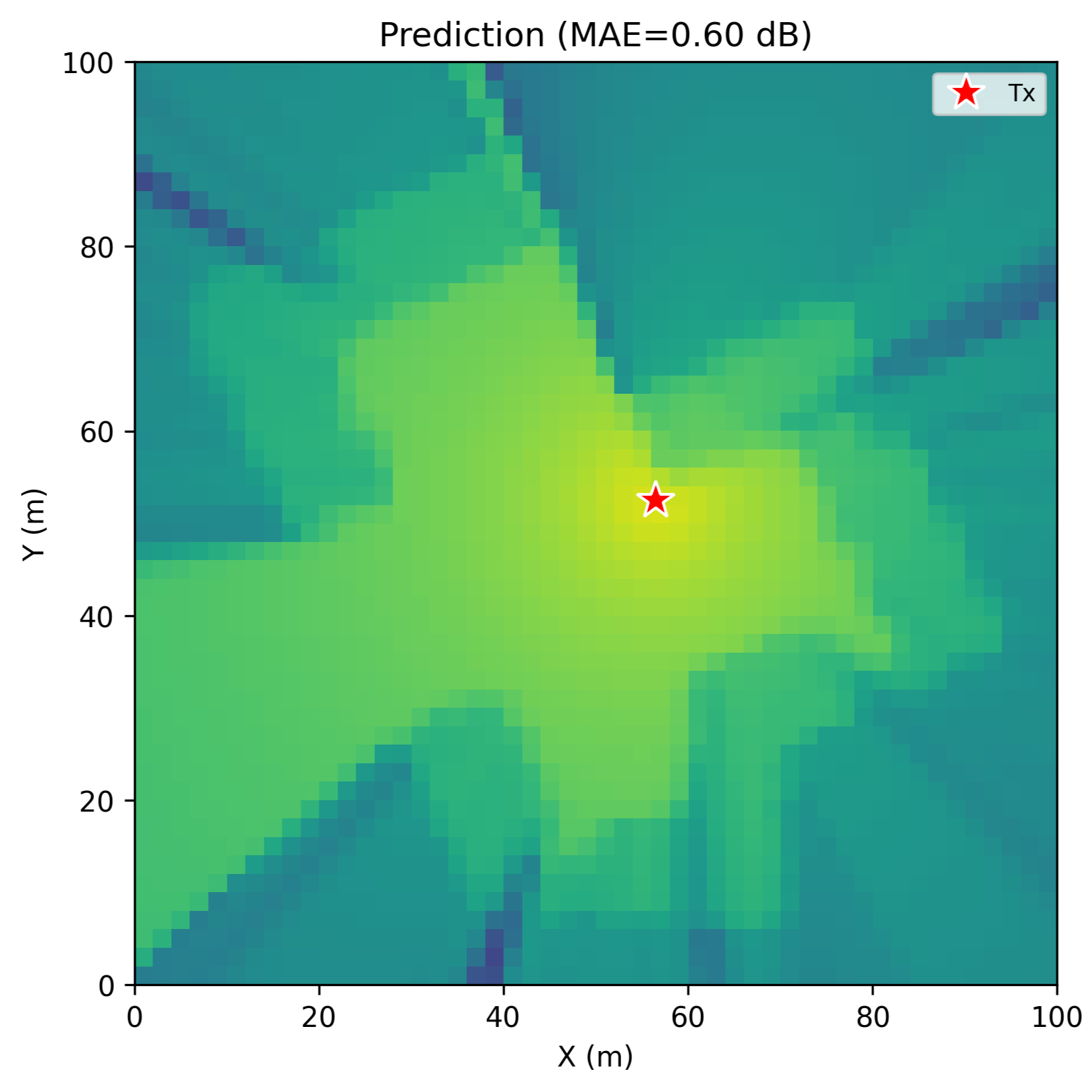}}
	\subfigure[MLP-GS.]
	{\includegraphics[width=0.22\textwidth]{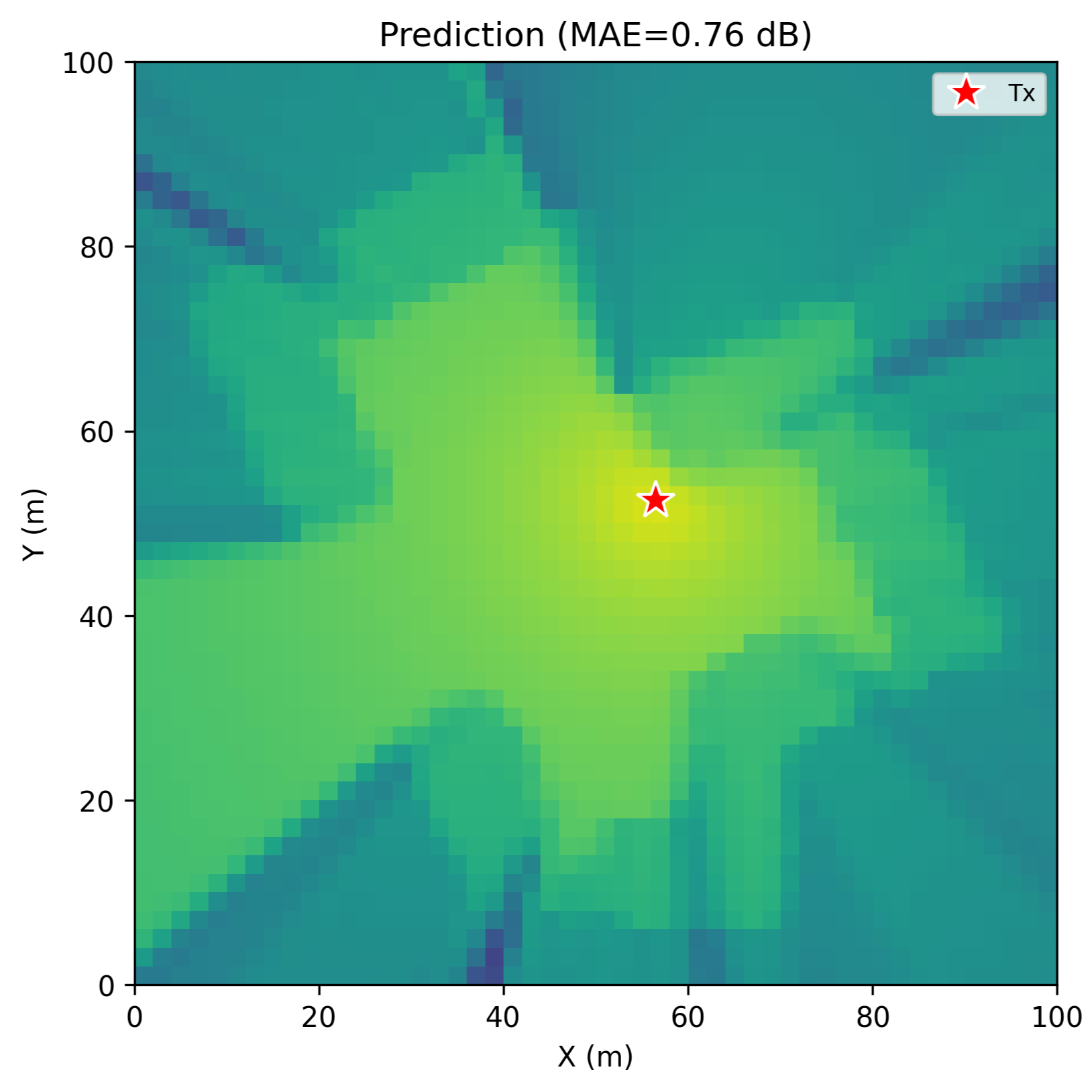}}
	\subfigure[MLP-RF.]
	{\includegraphics[width=0.22\textwidth]{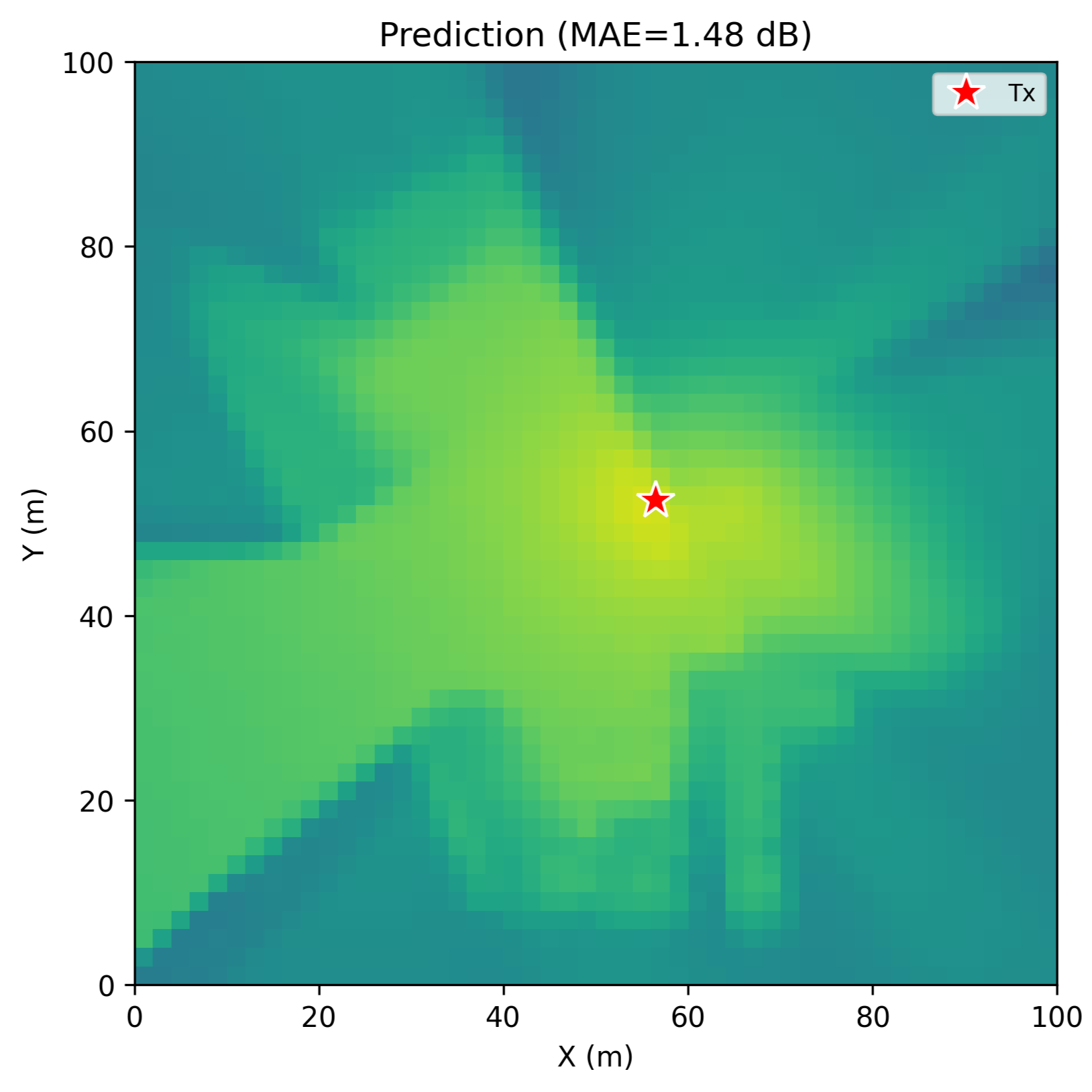}}
	\subfigure[VS.]
	{\includegraphics[width=0.22\textwidth]{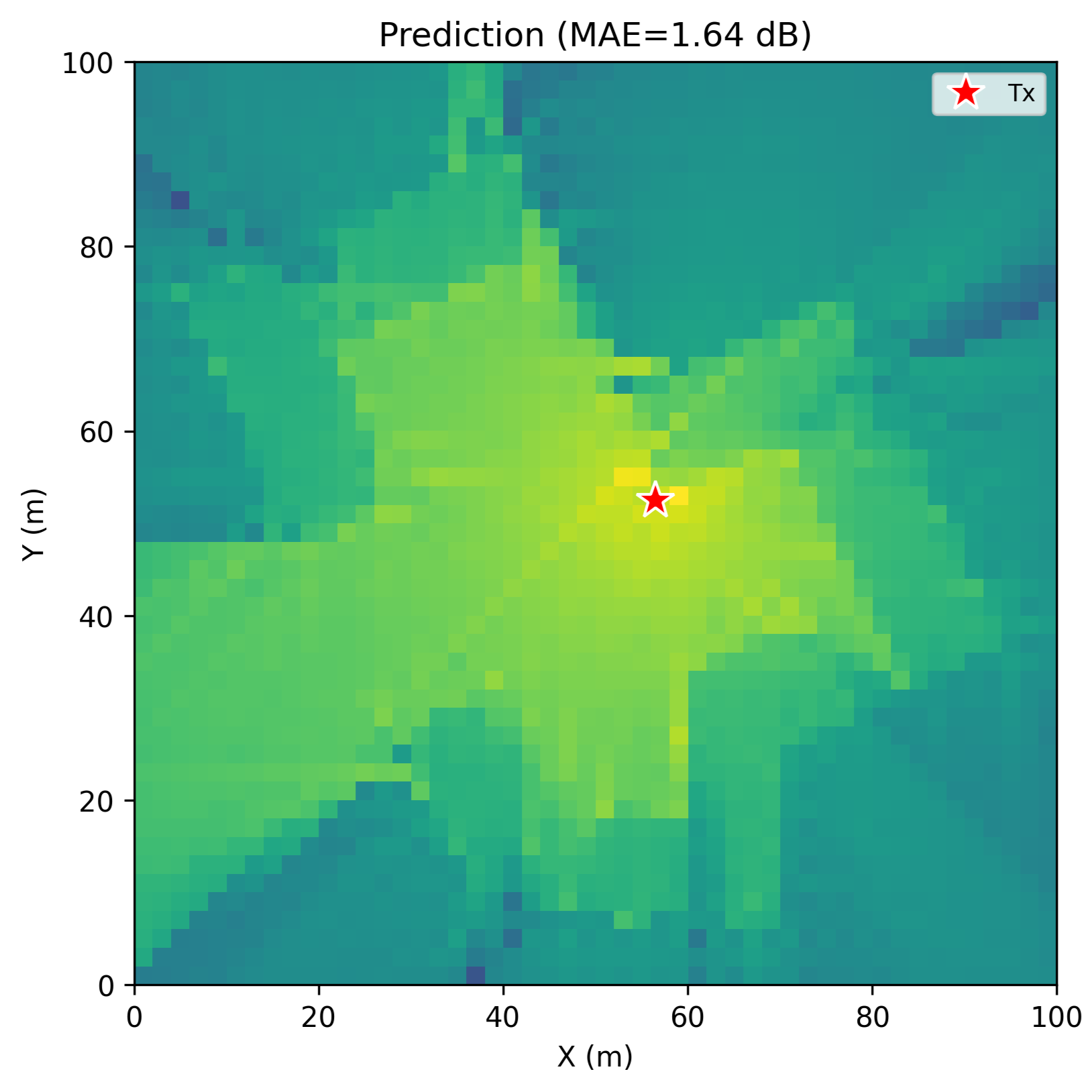}}
	\subfigure[Kriging interpolation.]
	{\includegraphics[width=0.22\textwidth]{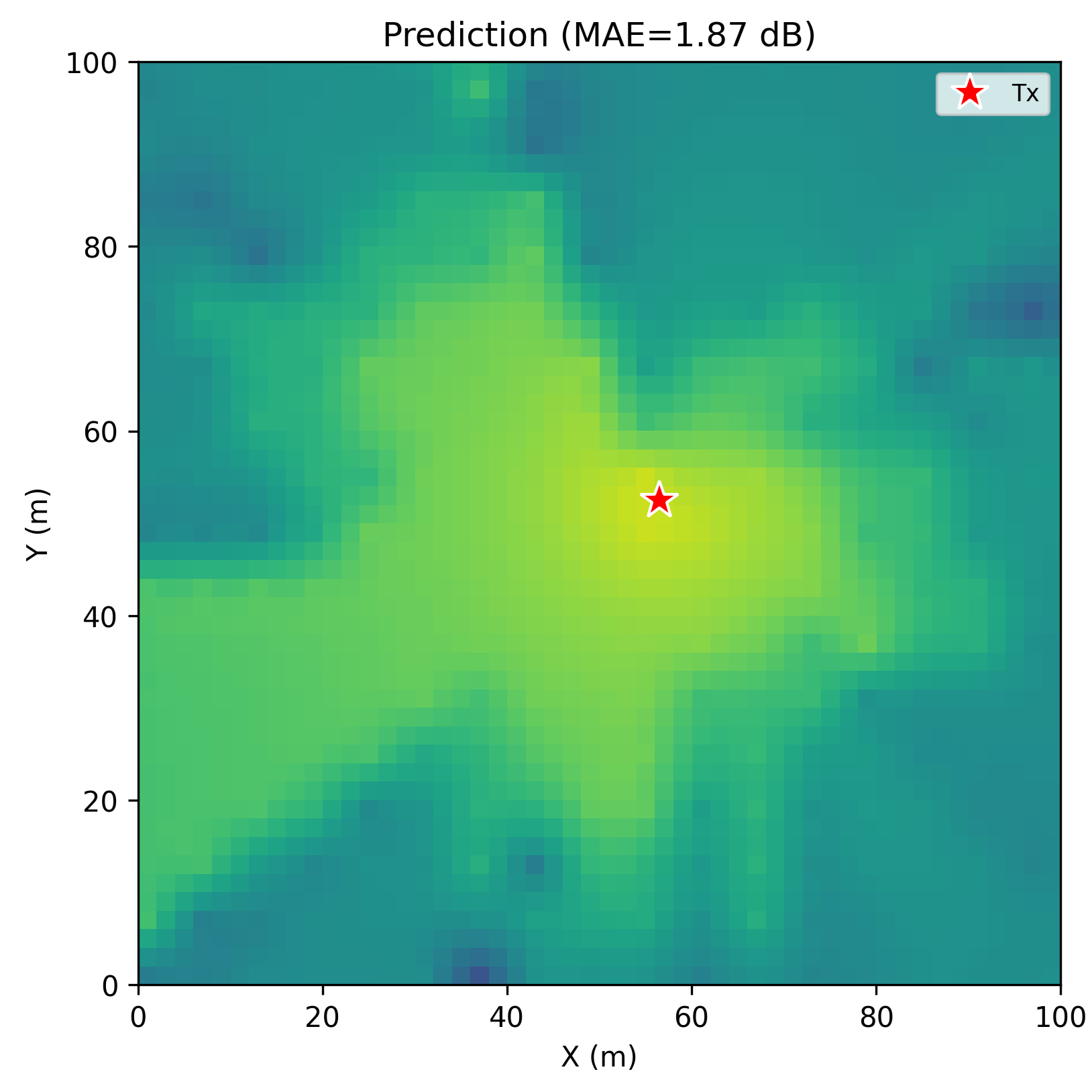}}
	\caption{CGM construction results in the campus scene.}
	\label{fig:campus_construction}
\end{figure}

\begin{figure}[!tb]
	\centering 
	\subfigure[Ground truth.]
	{\includegraphics[width=0.22\textwidth]{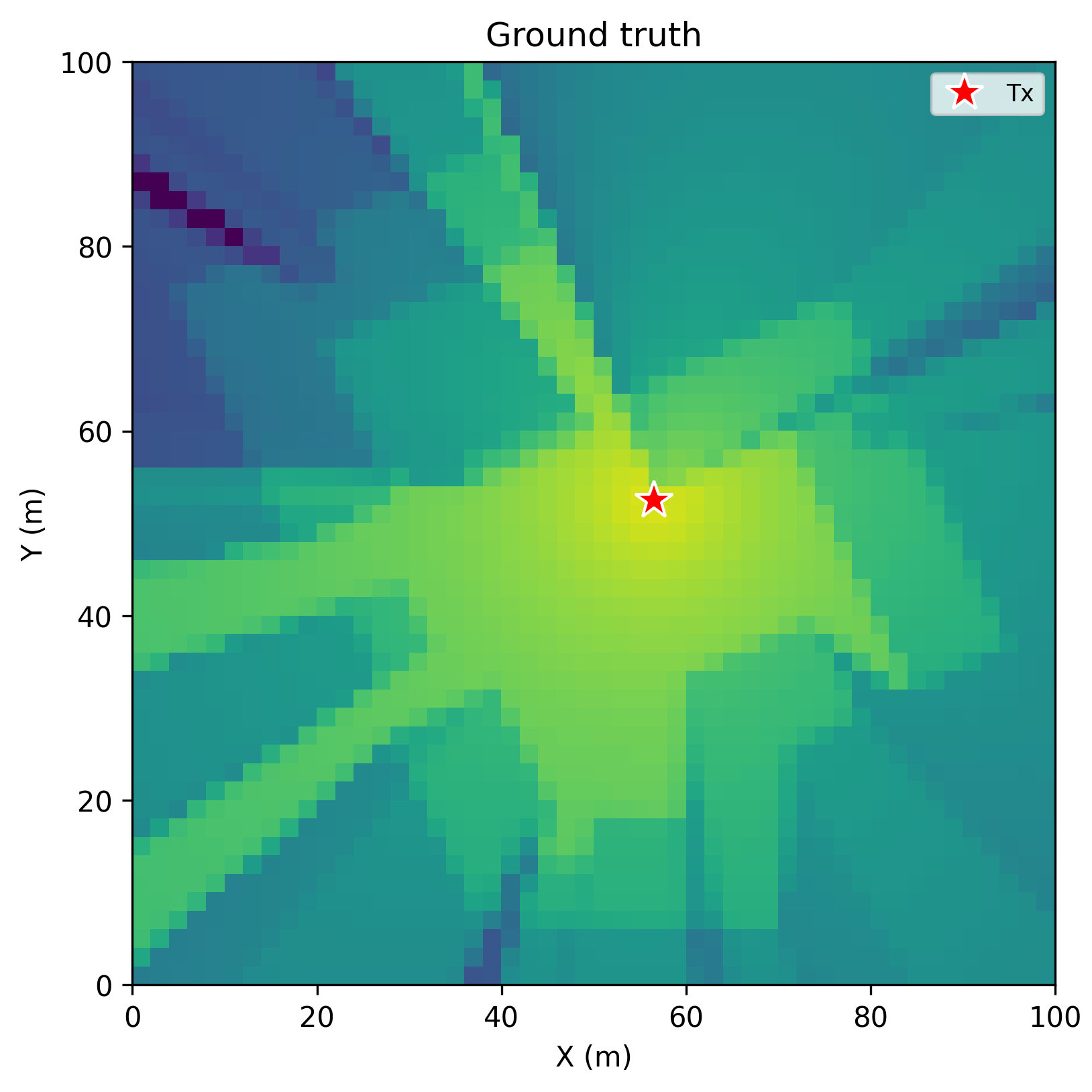}}
	\subfigure[GS-CG.]
	{\includegraphics[width=0.22\textwidth]{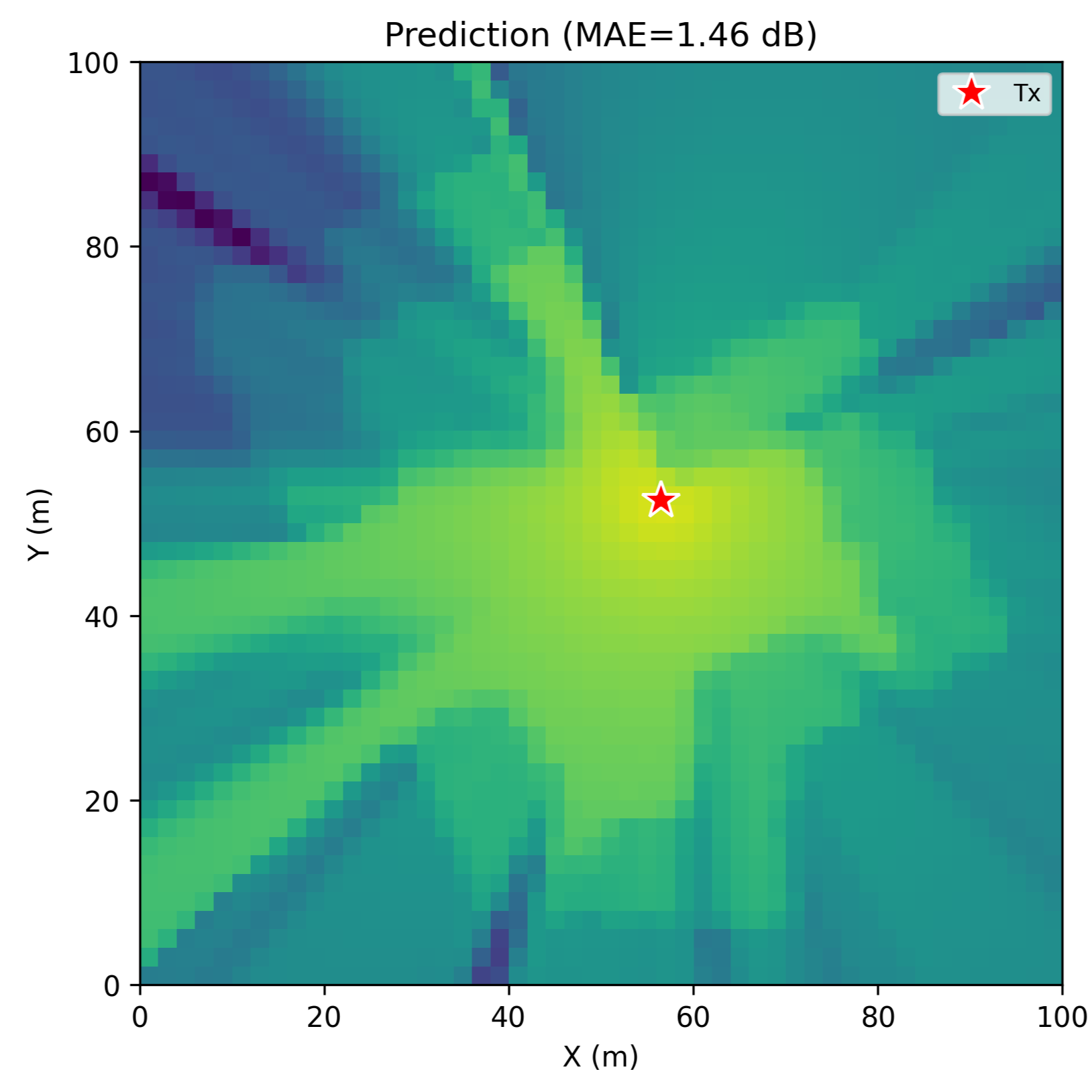}}
	\subfigure[MLP-GS.]
	{\includegraphics[width=0.22\textwidth]{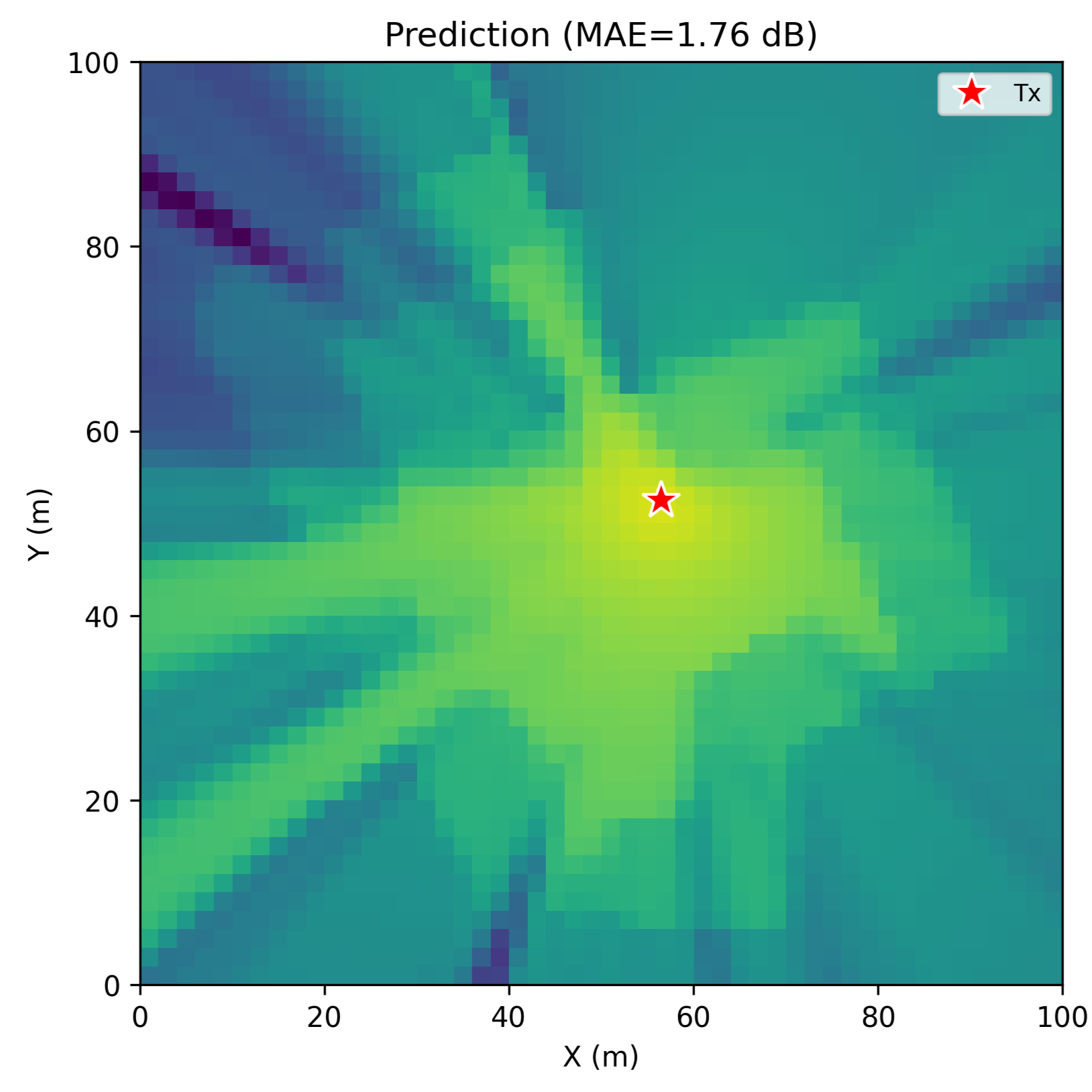}}
	\subfigure[MLP-RF.]
	{\includegraphics[width=0.22\textwidth]{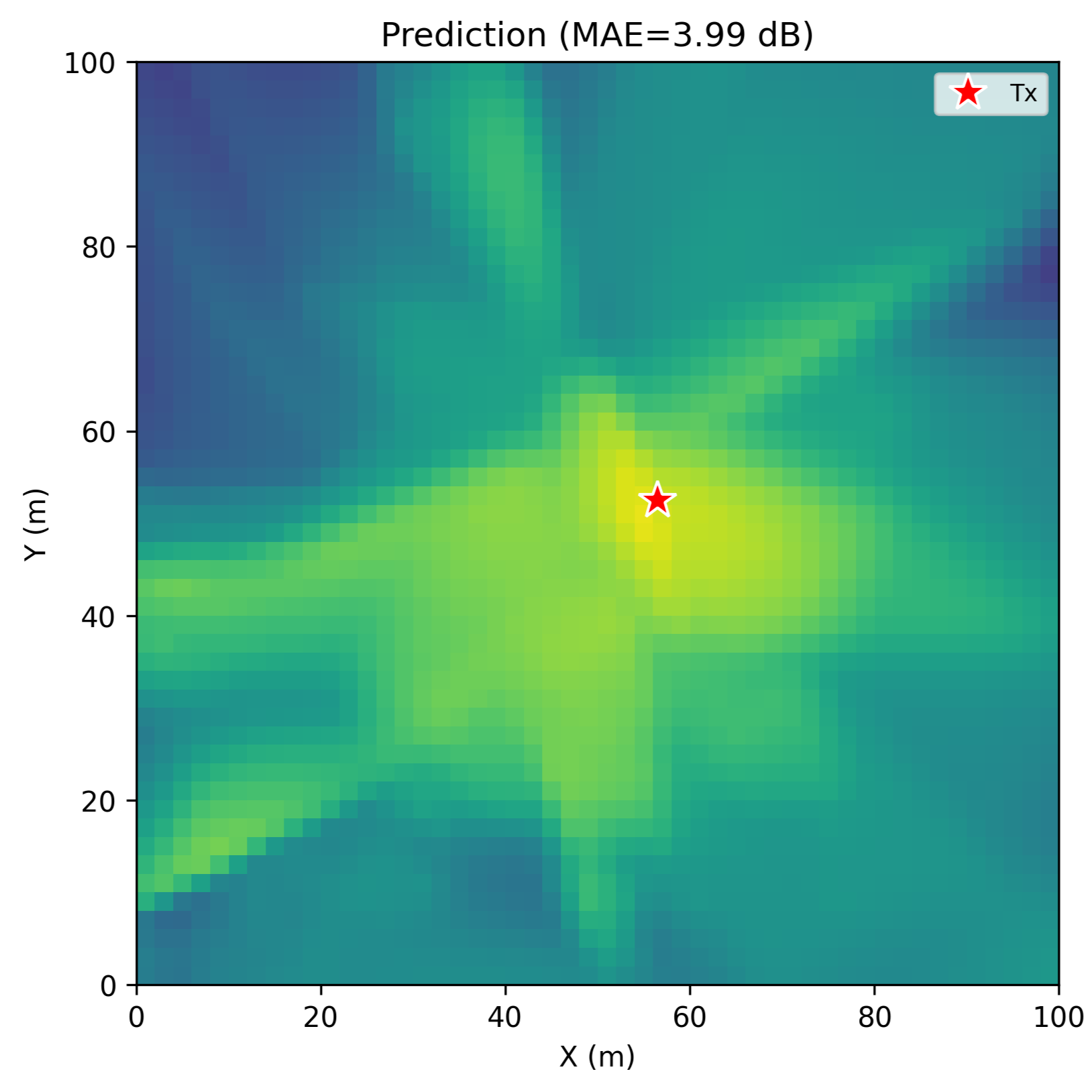}}
	\subfigure[VS.]
	{\includegraphics[width=0.22\textwidth]{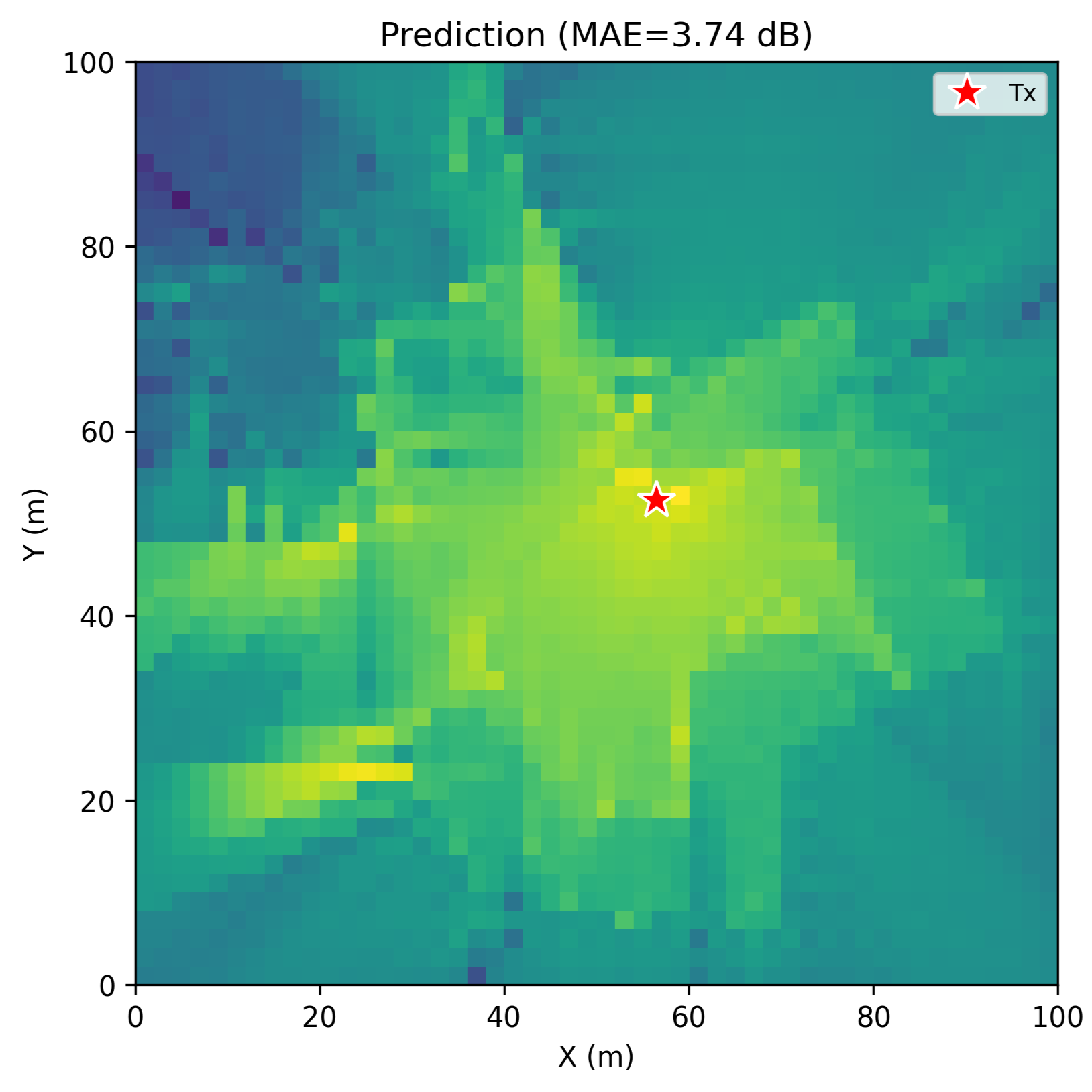}}
	\subfigure[Kriging interpolation.]
	{\includegraphics[width=0.22\textwidth]{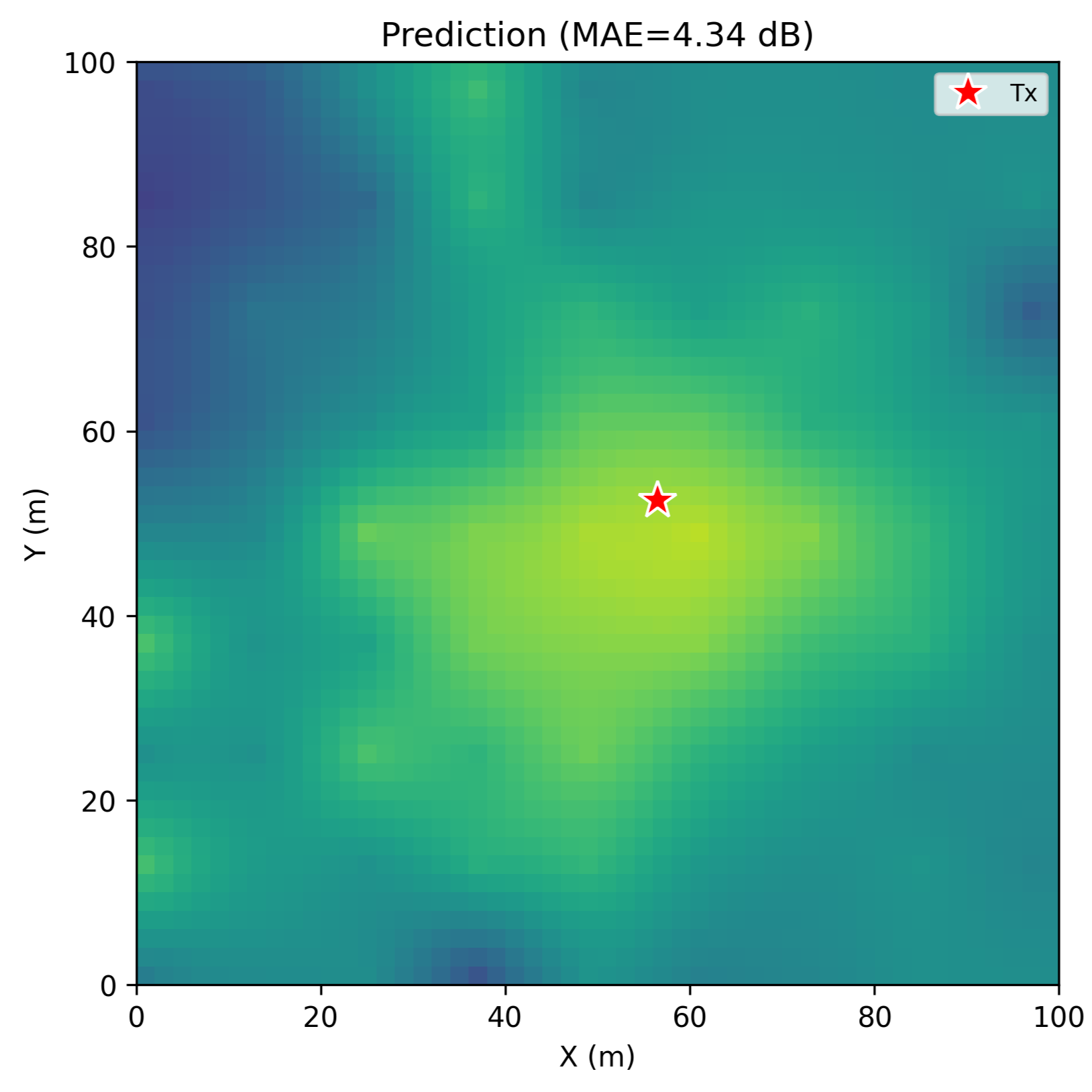}}
	\caption{CGM updating results in the campus scene.}
	\label{fig:campus_update}
\end{figure}

Figs. \ref{fig:campus_construction} and \ref{fig:campus_update} show the construction and updating results in the campus scene. Brighter colors indicate larger channel gain. In Fig. \ref{fig:campus_construction}, the ground-truth CGM contains both smooth power decay around the transmitter and sharp shadow regions caused by buildings. GS-CG accurately reconstructs these features, while MLP-GS slightly smooths some blockage boundaries due to its simplified scattering rendering. By contrast, MLP-RF misses several sharp attenuation transitions, VS introduces noticeable artifacts, and Kriging interpolation blurs the shadow regions. In Fig. \ref{fig:campus_update}, newly introduced structures create additional directional shadow regions. GS-CG and MLP-GS correctly capture their orientation and spatial extent, which validates the proposed incremental updating with residual-guided active initialization. By contrast, under sparse measurements, MLP-RF fails to recover building-shaped attenuation regions and directional shadows, VS produces fragmented and misaligned shadow patterns around changed structures, and Kriging interpolation over-smooths the updated map.

\begin{figure}[!tb]
	\centering 
	%	\subfigure[Initialization for construction.]
	%	{\includegraphics[width=0.24\textwidth]{figs/gs_init1.png}}
	\subfigure[Representation for reference CGM.]
	{\includegraphics[width=0.22\textwidth]{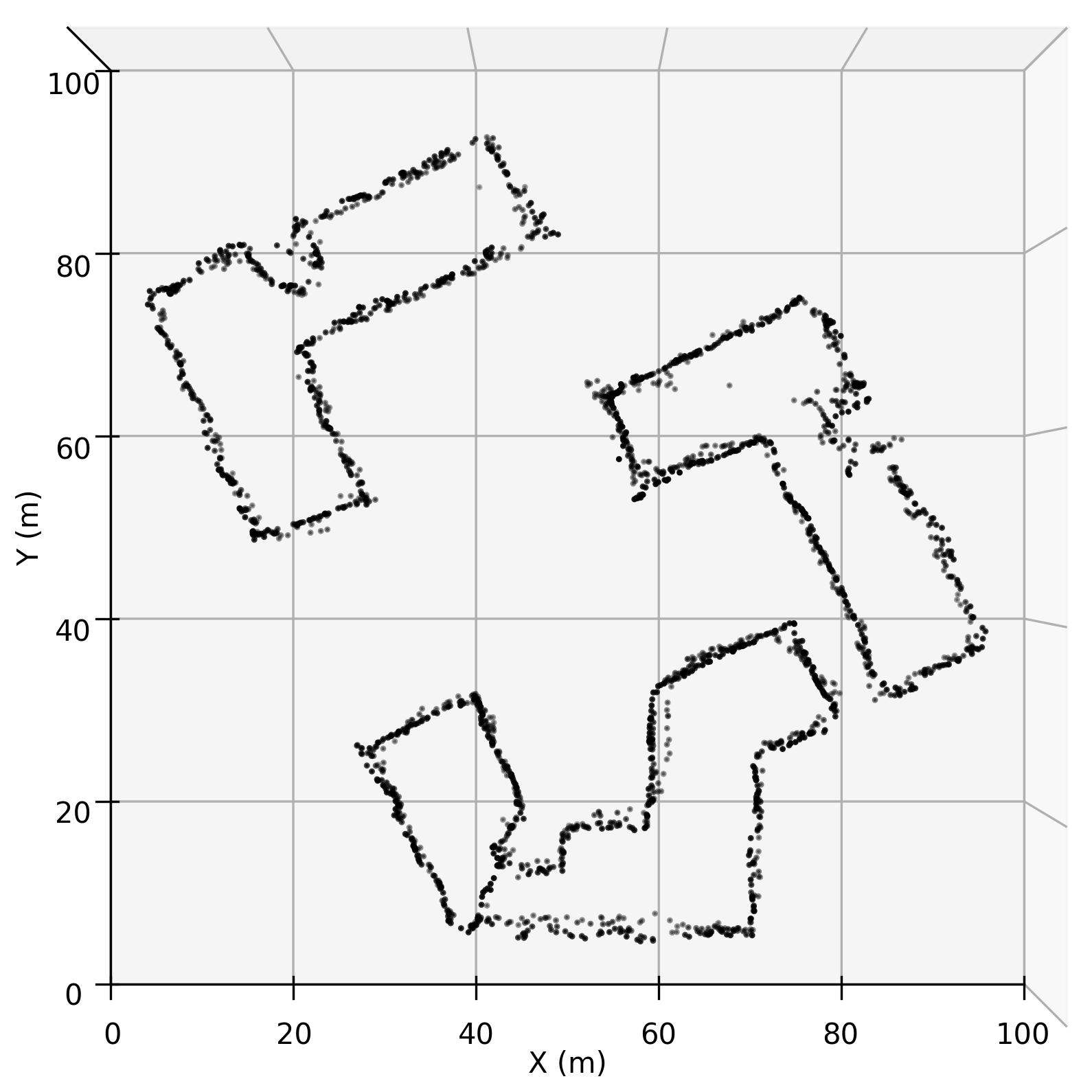}}
	%	\subfigure[Initialization for updating.]
	%	{\includegraphics[width=0.24\textwidth]{figs/gs_init2.png}}
	\subfigure[Representation for updated CGM.]
	{\includegraphics[width=0.22\textwidth]{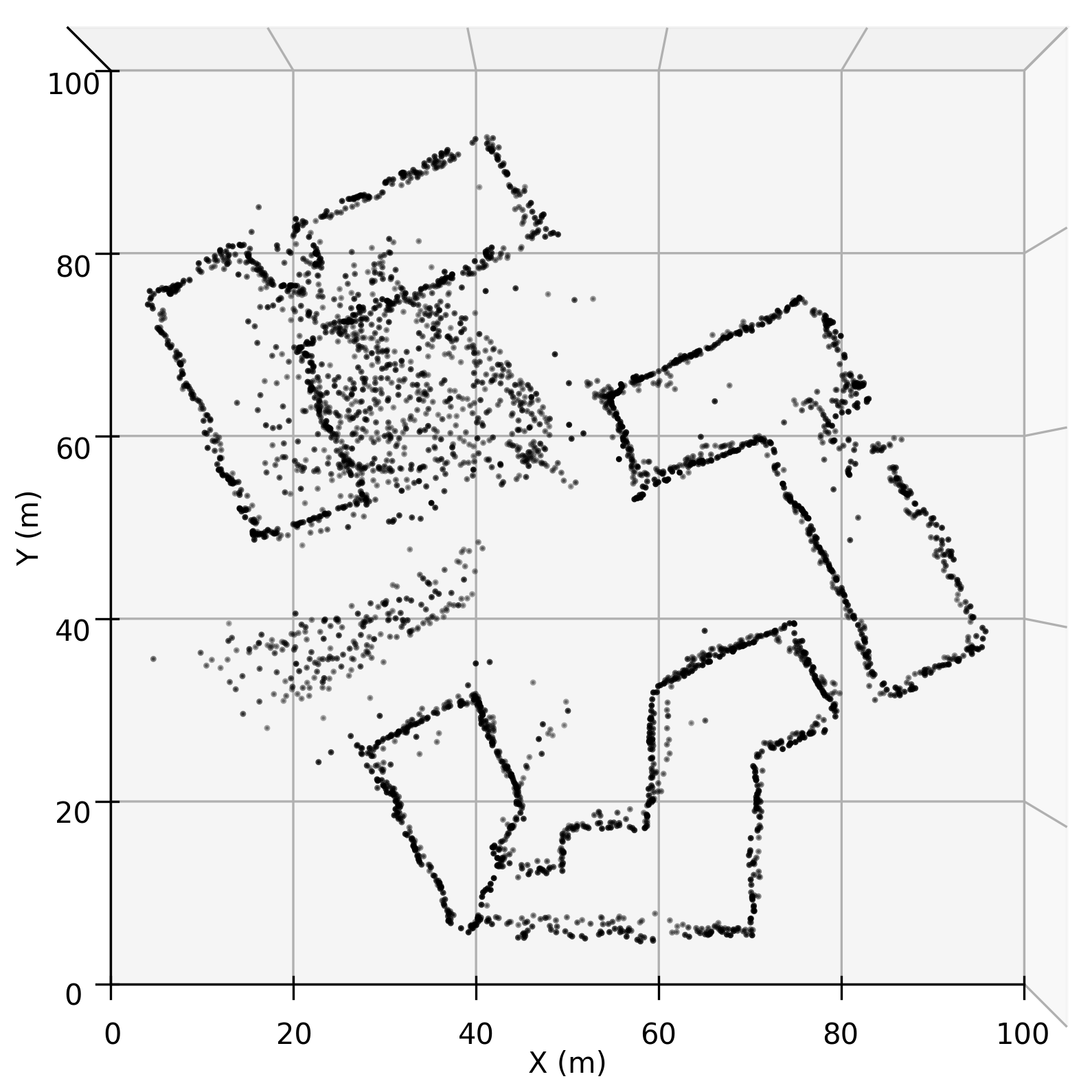}}
	\caption{Learned Gaussian representation for the campus scene.}
	\label{fig:gaussian_representation}
\end{figure}

Fig. \ref{fig:gaussian_representation} further illustrates the learned Gaussian representation in the campus scene. For visual clarity, we display only the Gaussians associated with building structures, while the ground-surface Gaussians used by GS-CG are omitted from the visualization. During construction, the initialization places Gaussians on building structures, and subsequent optimization further adapts their locations and shapes toward propagation-relevant regions. This indicates that the model refines a geometry-aware representation for channel gain rendering rather than merely fitting the CGM as an image. During updating, the active Gaussians initialized from residual-guided back-projection are refined toward the newly changed regions, while the previously learned Gaussians remain unchanged. This observation supports the physical intuition of the proposed initialization and shows that the learned representation is interpretable.

\begin{figure*}[!tb]
	\centering 
	\subfigure[Static: Ground truth.]
	{\includegraphics[width=0.22\textwidth]{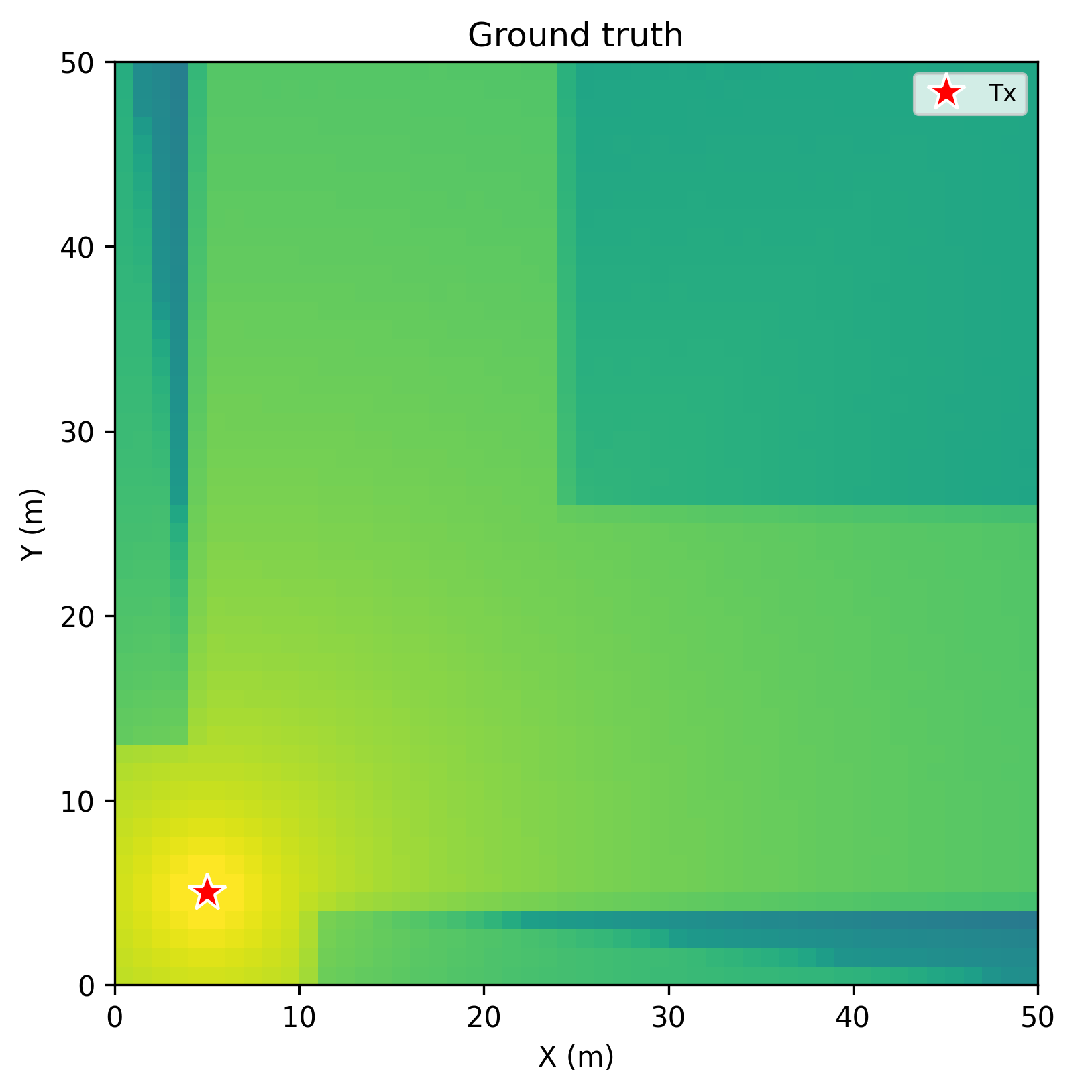}}
	\subfigure[Stage 1: Ground truth.]
	{\includegraphics[width=0.22\textwidth]{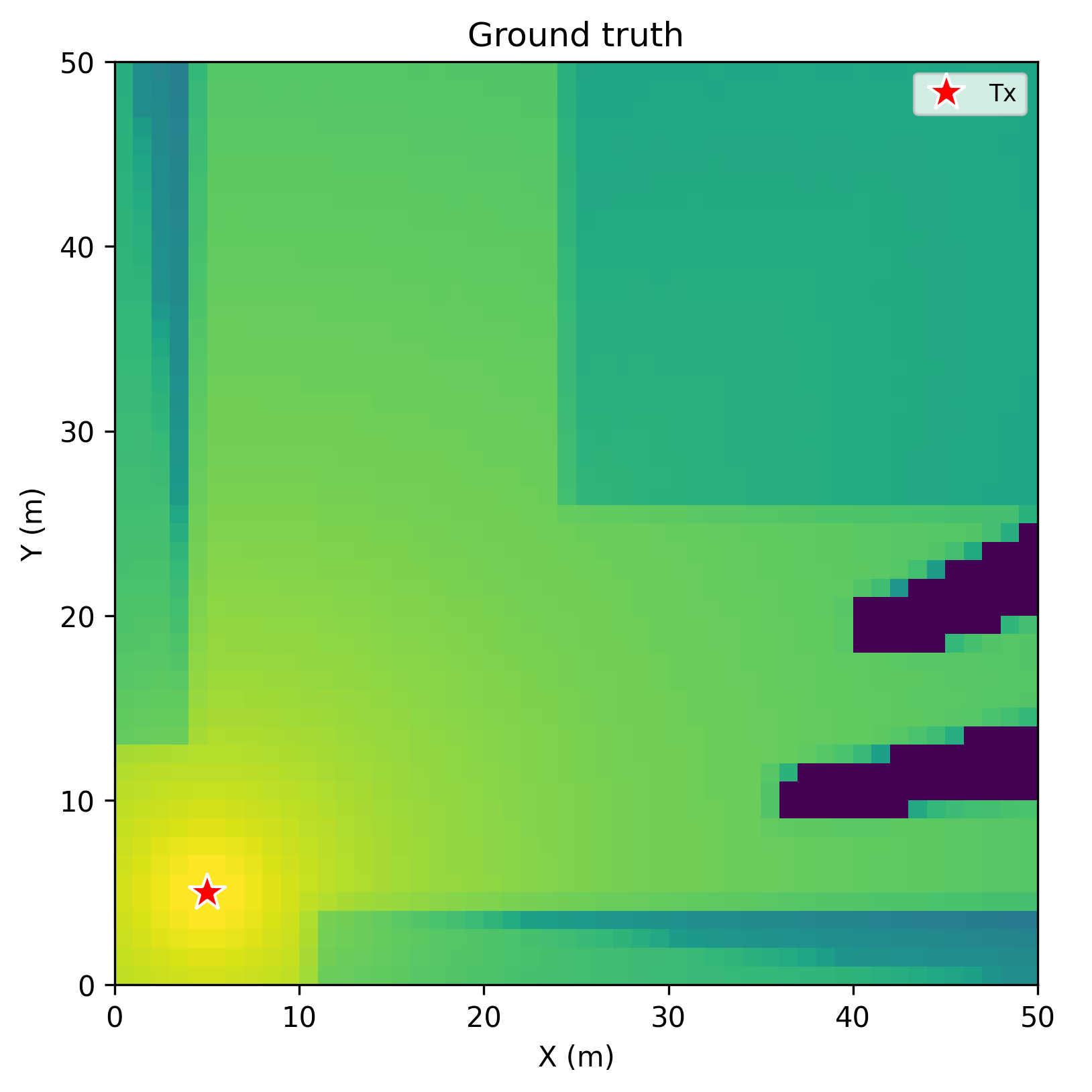}}
	\subfigure[Stage 2: Ground truth.]
	{\includegraphics[width=0.22\textwidth]{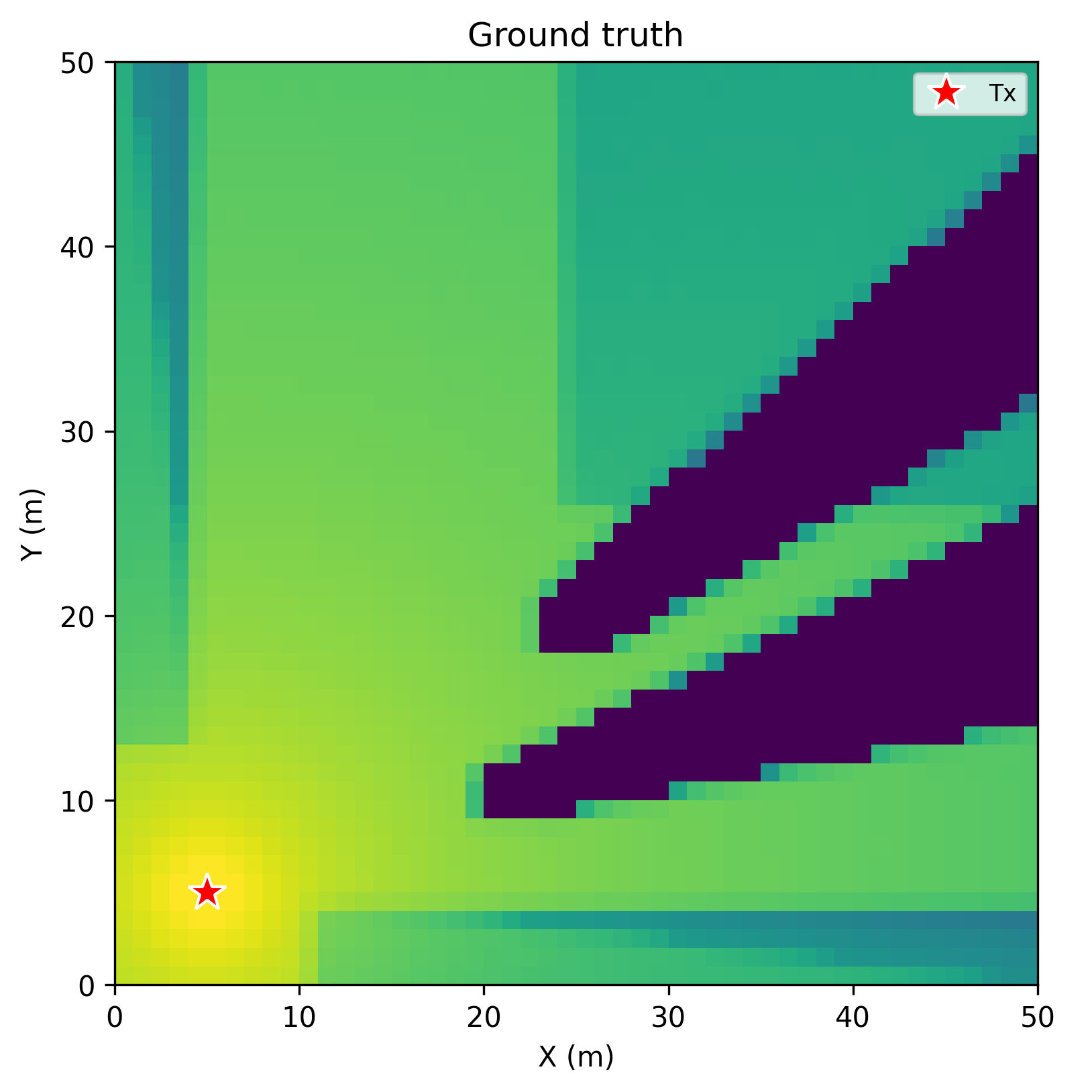}}
	\subfigure[Stage 3: Ground truth.]
	{\includegraphics[width=0.22\textwidth]{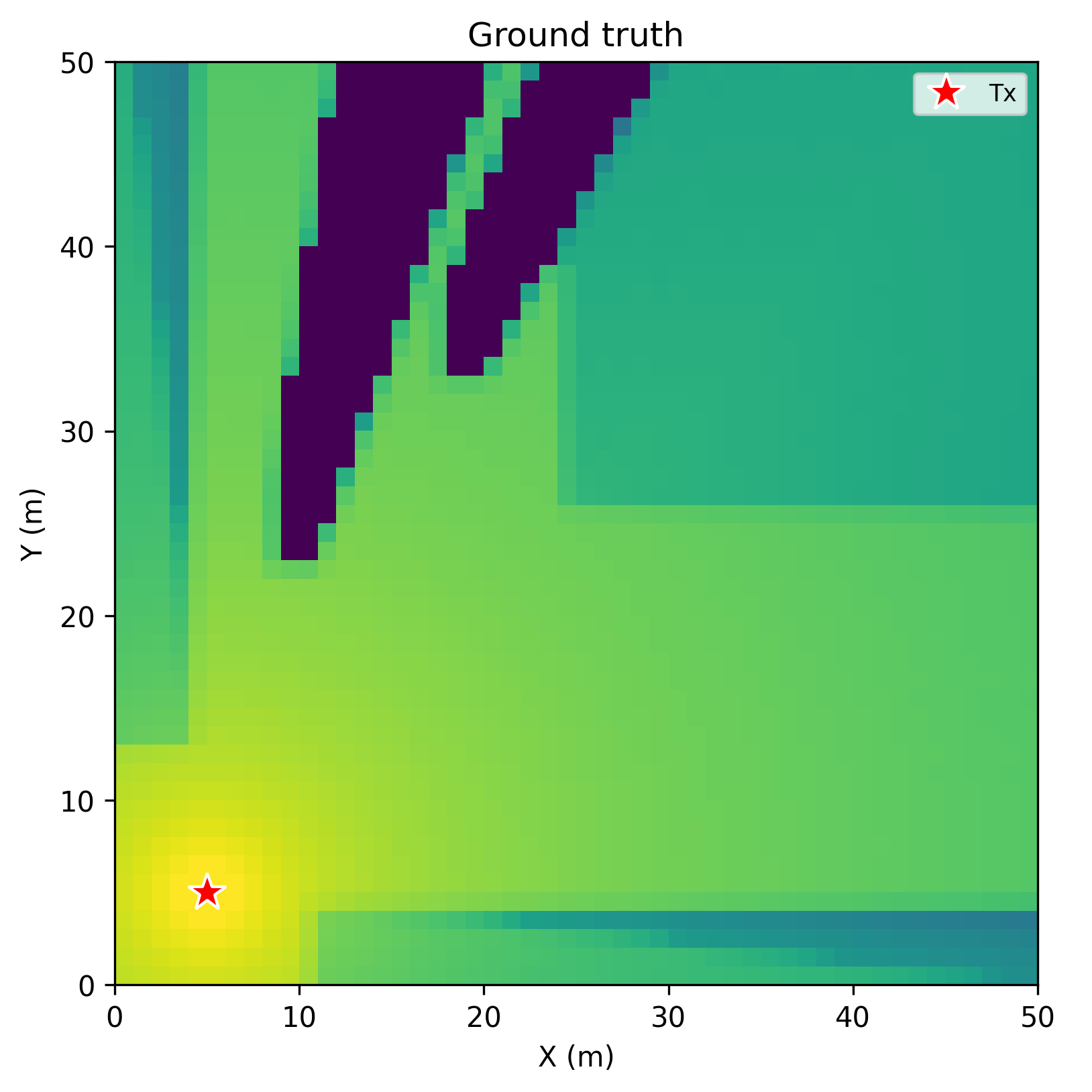}}
	\subfigure[Static: Prediction.]
	{\includegraphics[width=0.22\textwidth]{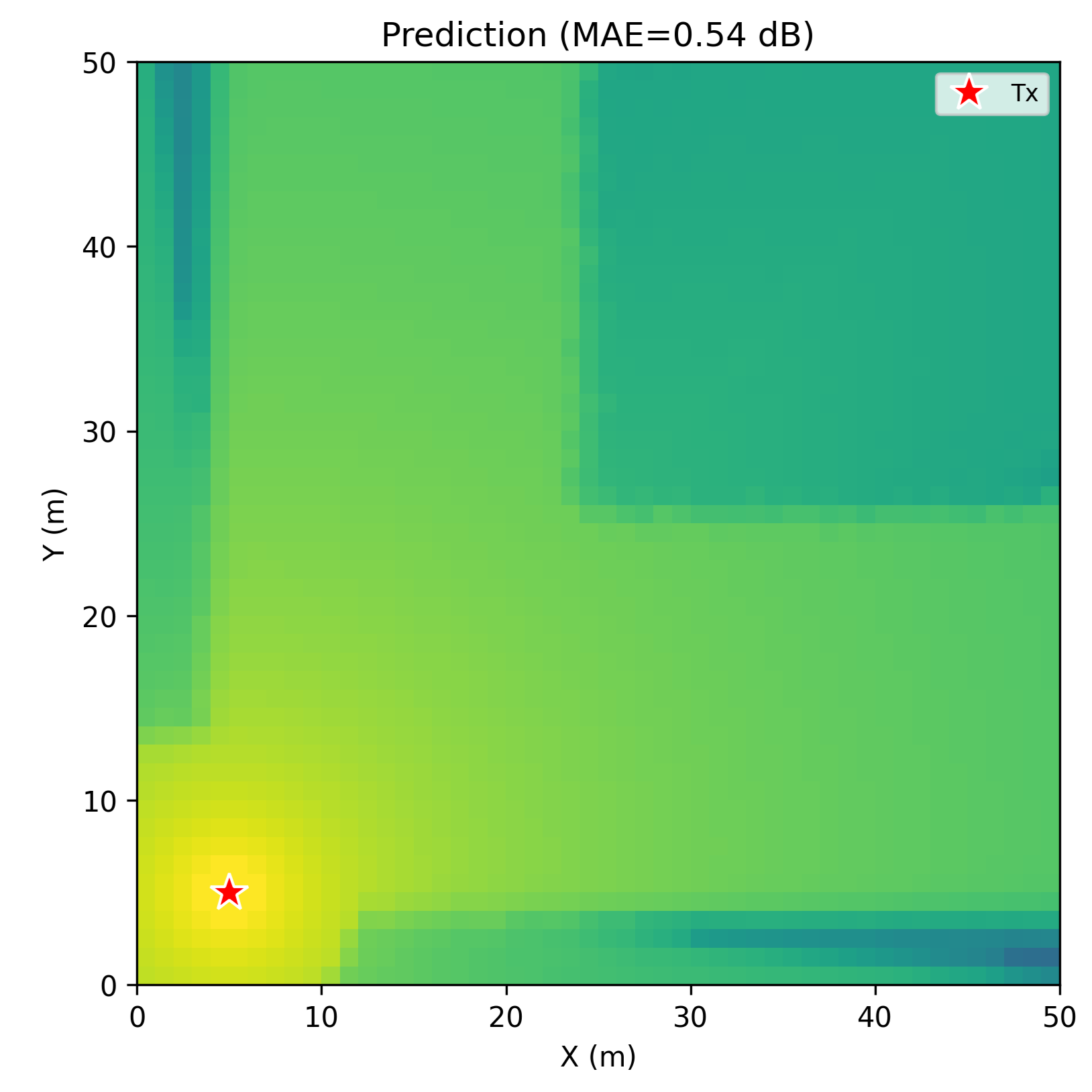}}
	\subfigure[Stage 1: Prediction.]
	{\includegraphics[width=0.22\textwidth]{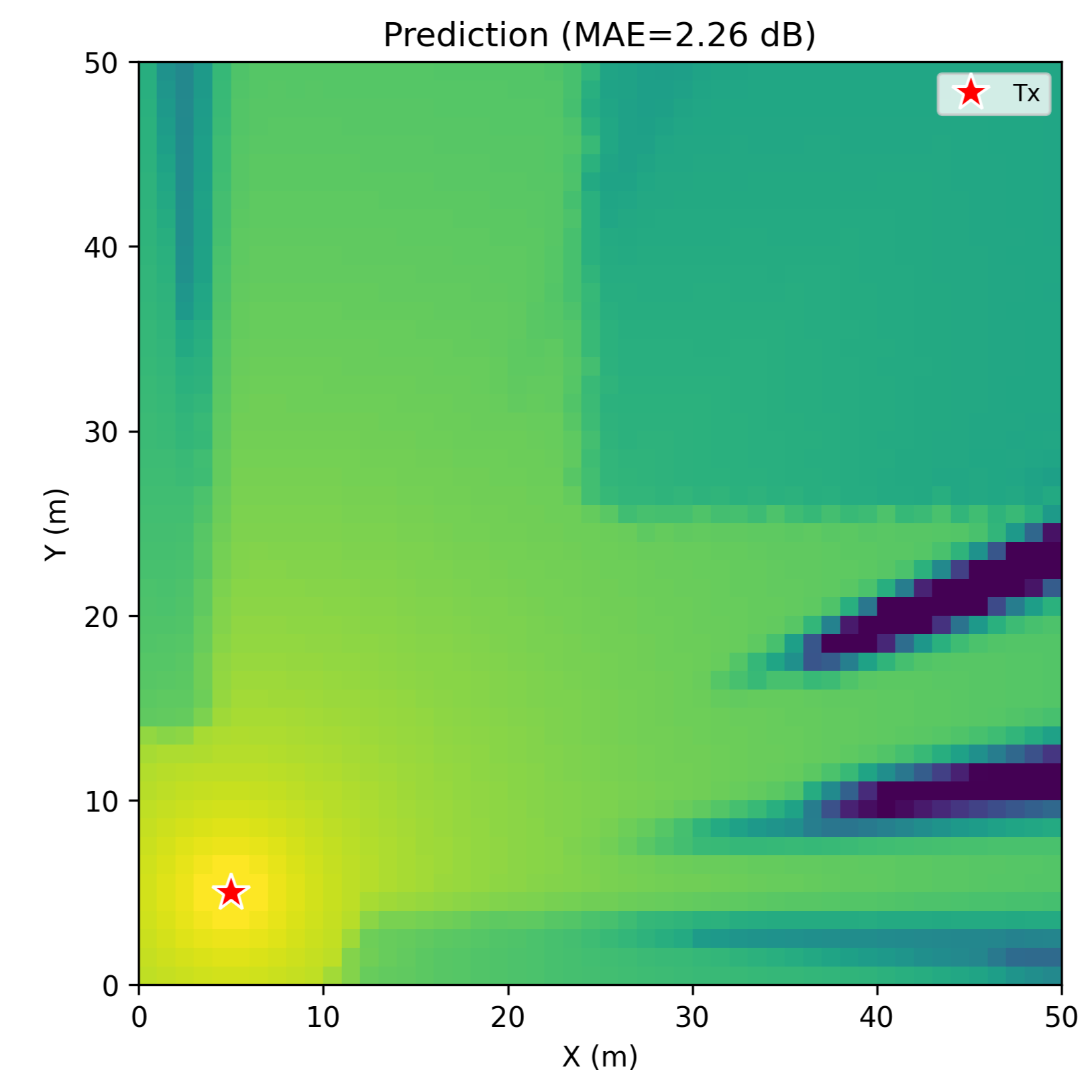}}
	\subfigure[Stage 2: Prediction.]
	{\includegraphics[width=0.22\textwidth]{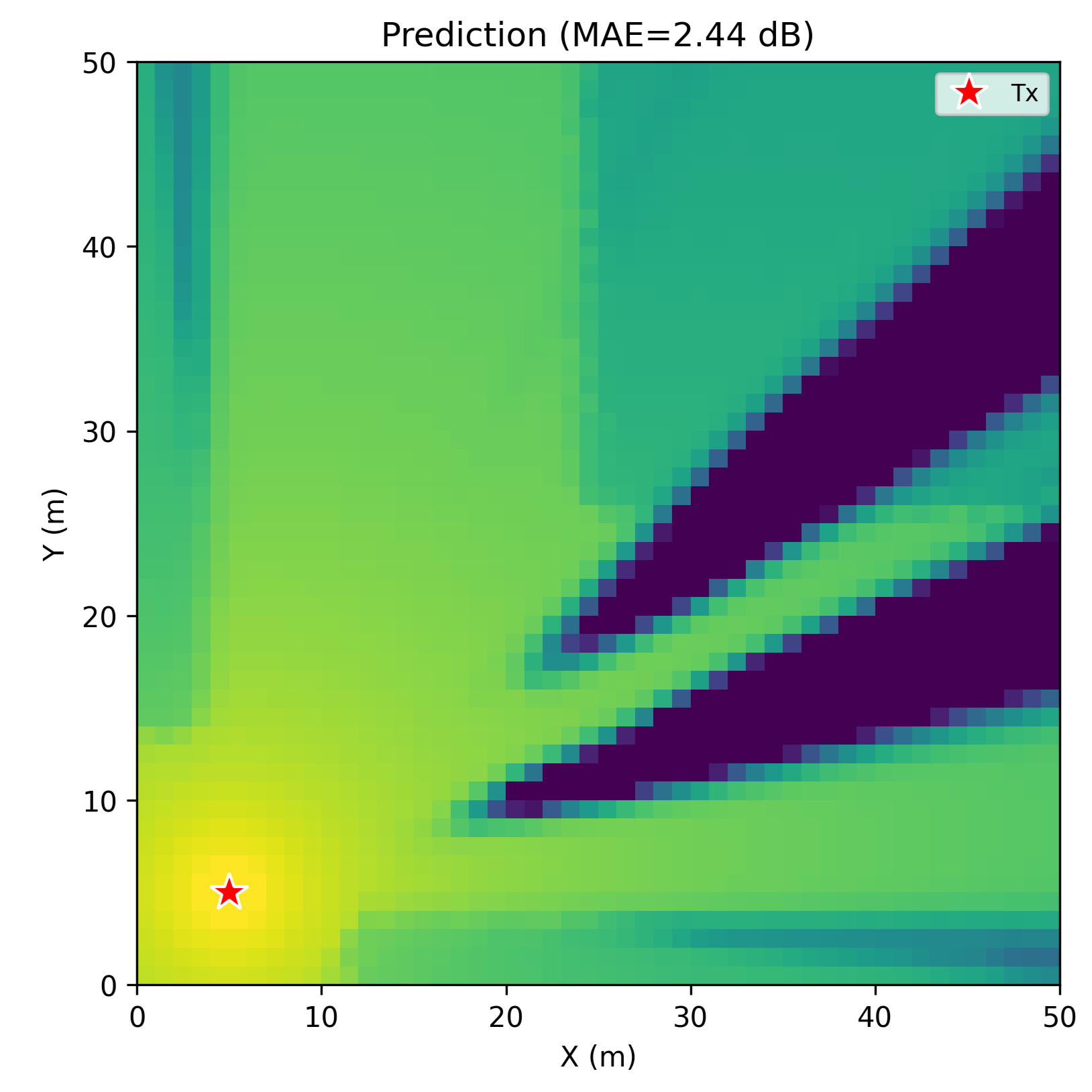}}
	\subfigure[Stage 3: Prediction.]
	{\includegraphics[width=0.22\textwidth]{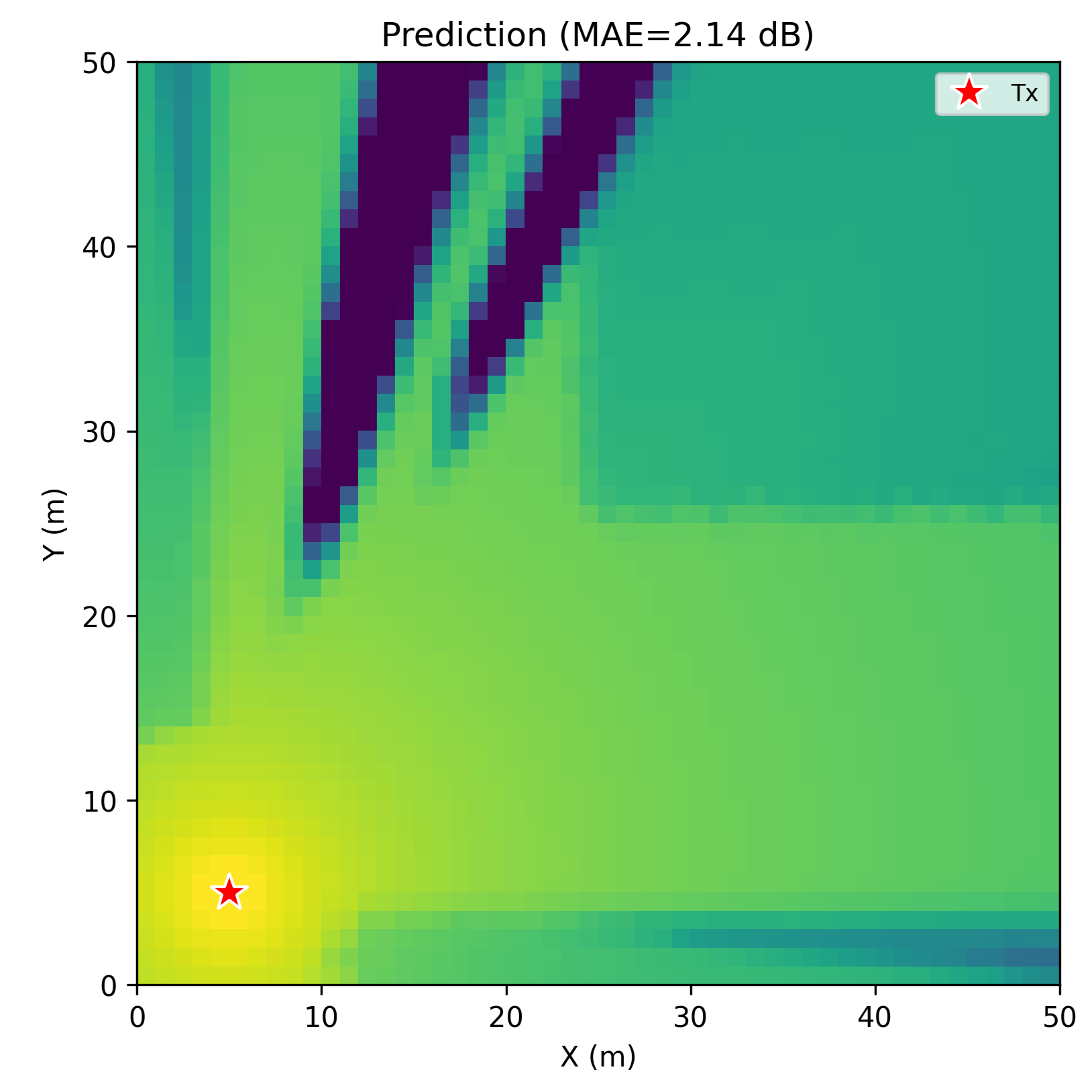}}
	\caption{CGM prediction results by GS-CG in the road scene.}
	\label{fig:road_update}
\end{figure*}

Fig. \ref{fig:road_update} evaluates the road scene, where the environment changes continuously across multiple stages. The ground-truth CGMs show that different newly introduced blockers generate different wedge-shaped shadow regions. The predicted CGMs demonstrate that GS-CG can track these evolving shadow patterns over consecutive updates. Although some sharp shadow boundaries are slightly smoothed due to the limited new measurements, the dominant attenuation regions and their directional structures are well reconstructed. This result indicates that the proposed method is not limited to a one-time update, but can support repeated CGM refinement as the environment evolves.

\begin{figure}[!tb]
	\centering 
	\subfigure[Static: Ground truth.]
	{\includegraphics[width=0.22\textwidth]{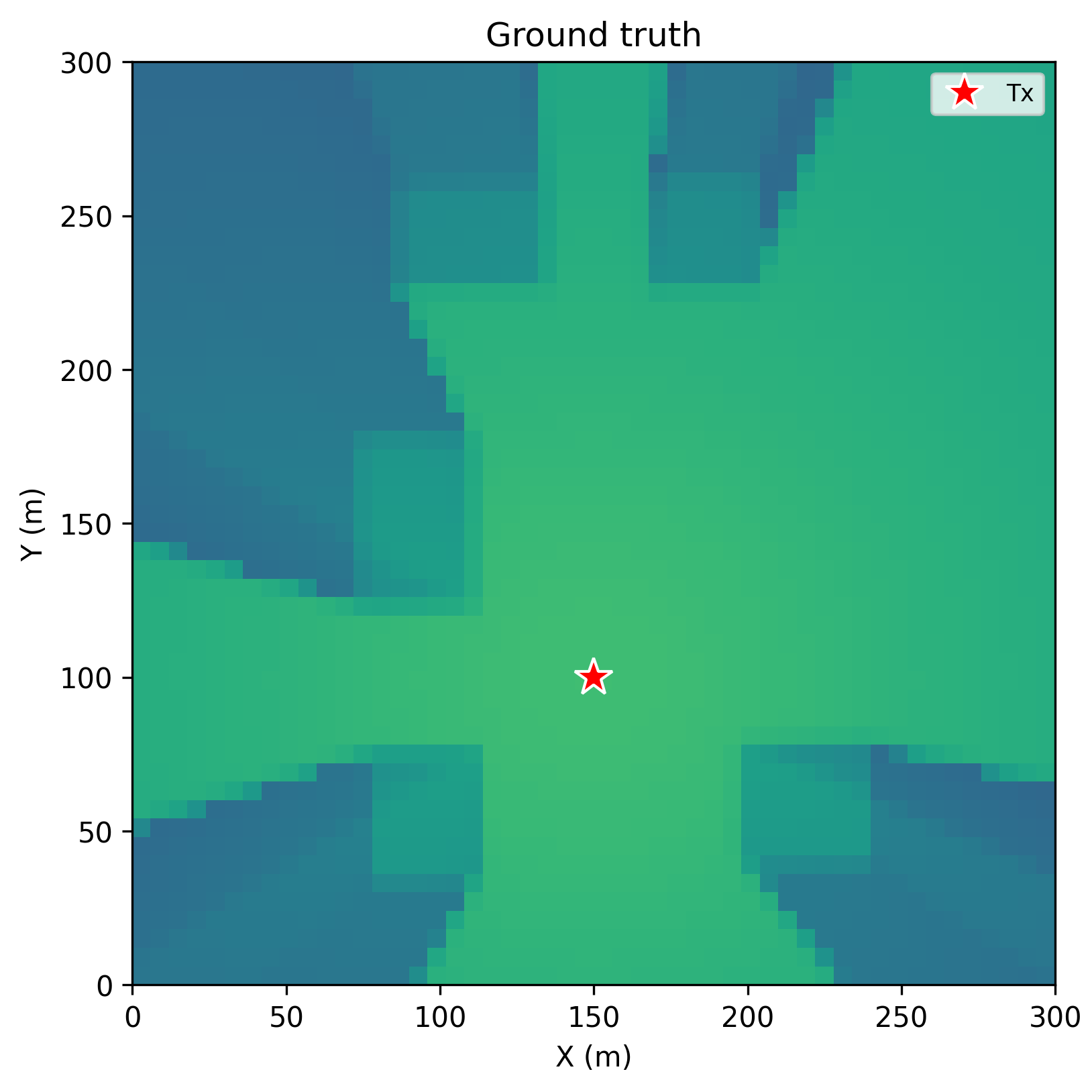}}
	\subfigure[Updated: Ground truth.]
	{\includegraphics[width=0.22\textwidth]{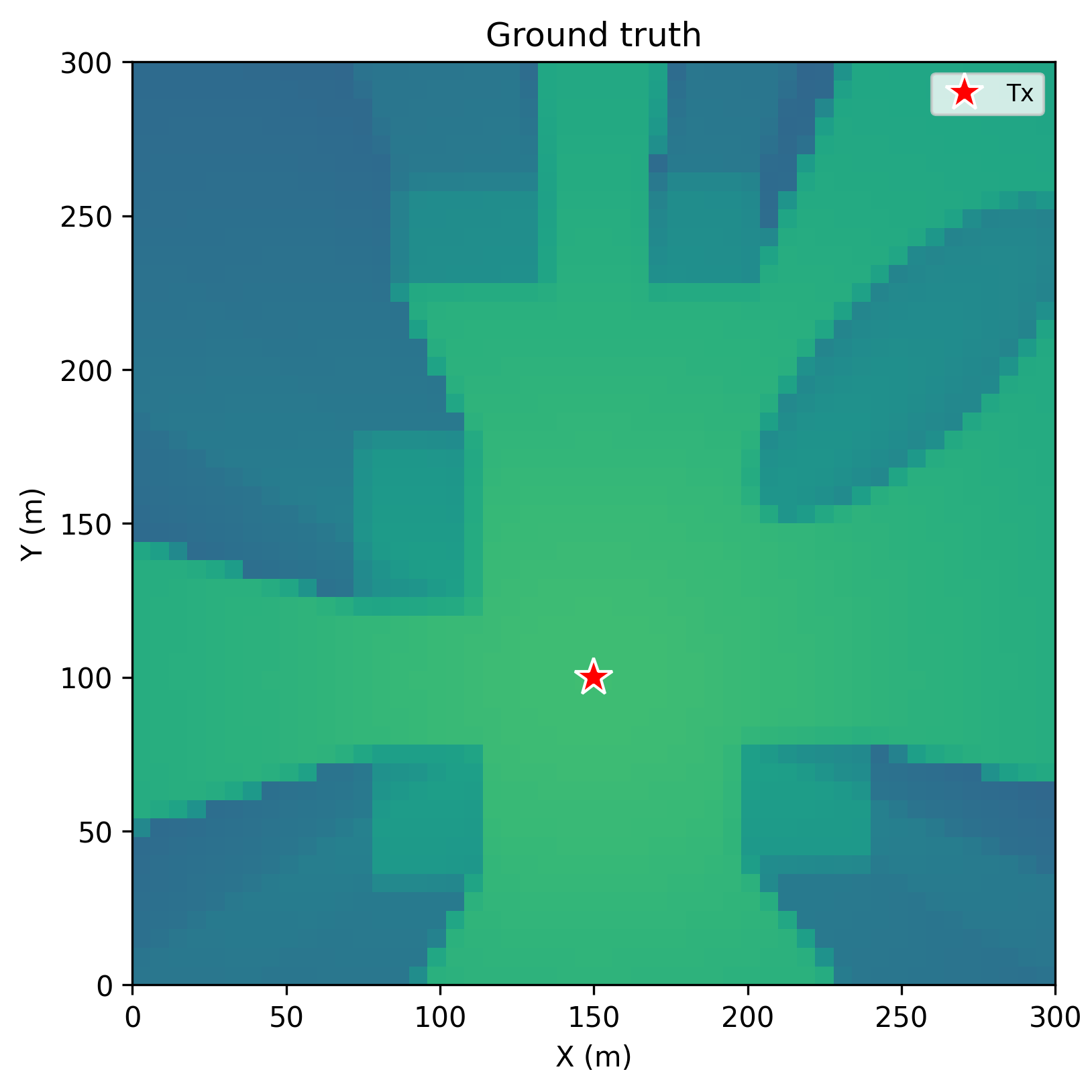}}
	\subfigure[Static: Prediction.]
	{\includegraphics[width=0.22\textwidth]{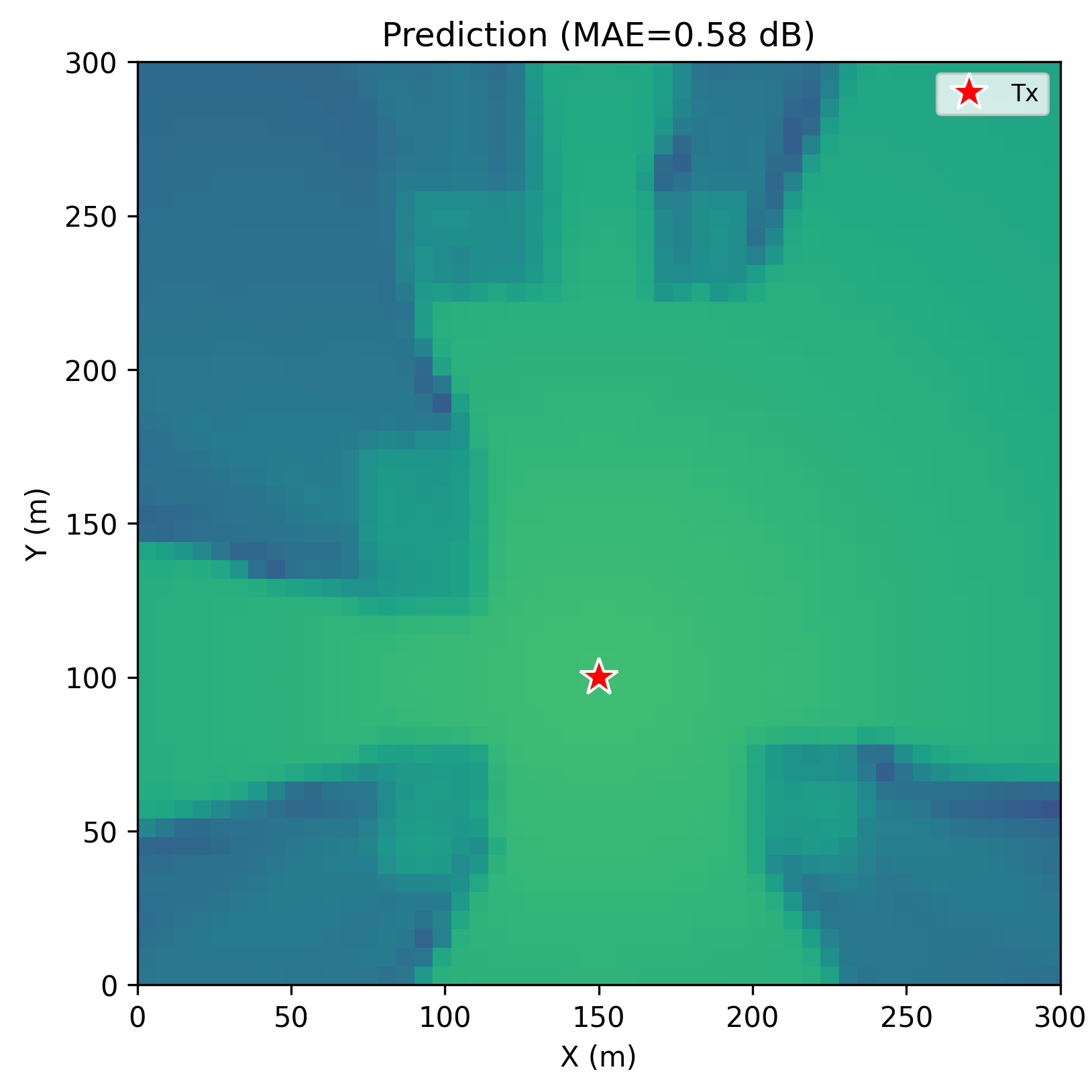}}
	\subfigure[Updated: Prediction.]
	{\includegraphics[width=0.22\textwidth]{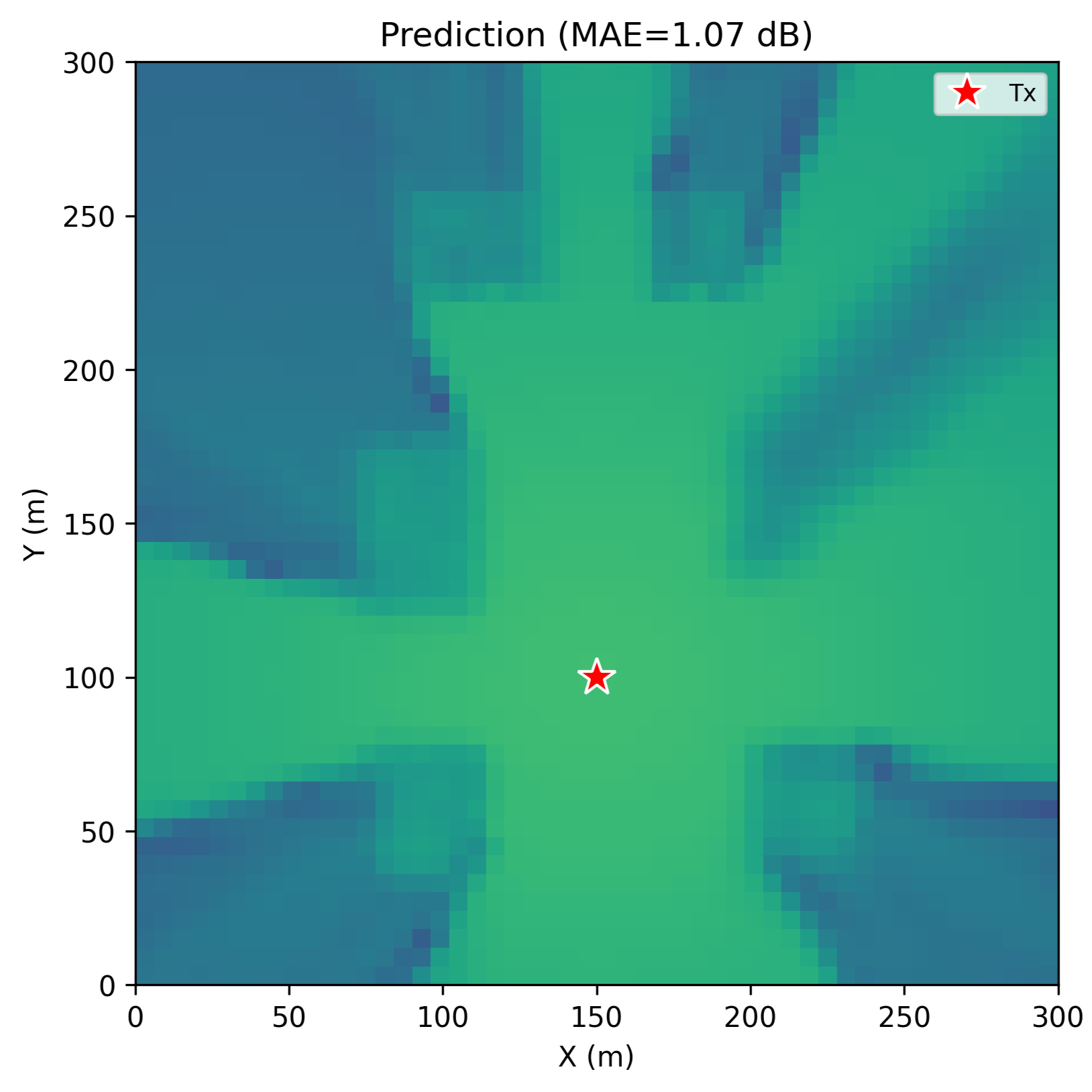}}
	\caption{CGM prediction results by GS-CG in the low-altitude scene.}
	\label{fig:low_altitude}
\end{figure}

Fig. \ref{fig:low_altitude} shows the low-altitude scene, which has a much larger map size and the channel gain is mainly determined by sparse building-induced shadows. In this case, GS-CG accurately preserves the global power distribution while still recovering the main blockage effects. This confirms that the proposed grid-based physical rendering model remains effective in a larger-scale scenario, where the absolute propagation distance and channel gain dynamic range are both different from those in the ground-level scenes.

\begin{table}[!tb]
	\centering
	\small
	\caption{Average training time of different schemes (in s).}
	\begin{tabular}{c | c | c} 
		\hline
		& Construction & Updating \\
		\hline
		GS-CG & \(279.7\) & \(46.5\) \\
		\hline
		MLP-GS & \(57.8\) & \(22.4\) \\
		\hline
		MLP-RF & \(51.6\) & \(9.2\) \\
		\hline
		VS & \(9.8\) & \(4.7\) \\
		\hline
		Kriging & \(0.2\) & \(0.1\) \\
		\hline
	\end{tabular}
\end{table}

Table II compares the average training time. GS-CG requires $279.7$ s for construction and $46.5$ s for updating. The much shorter update time confirms the efficiency of reusing the previously learned Gaussian representation and optimizing only the active set with sparse measurements. MLP-GS further reduces the training time to $57.8$ s and $22.4$ s, respectively, with only moderate accuracy degradation. Therefore, GS-CG is preferred when reconstruction fidelity and physical interpretability are important, whereas MLP-GS provides a favorable accuracy-complexity tradeoff for fast CGM construction and frequent updating. Although MLP-RF is faster, especially during updating, its accuracy is much lower, confirming the importance of retaining the geometry-aware Gaussian representation.

\subsection{Ablation and Grid-Resolution Study}

\begin{figure}[!tb]
	\centering
	\subfigure[Direct reconstruction without prior knowledge.]
	{\includegraphics[width=0.22\textwidth]{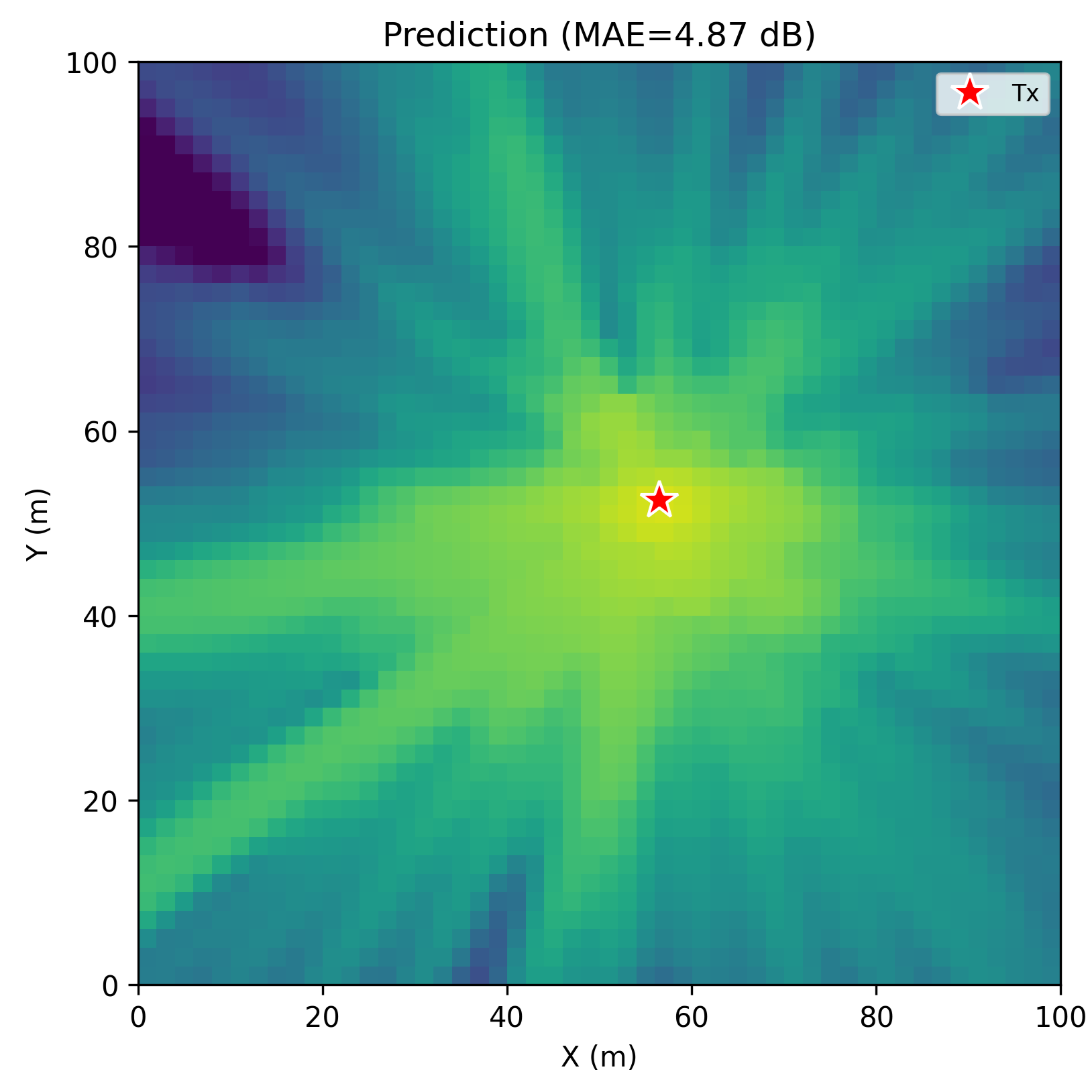}}
	\subfigure[Full-model updating without parameter freezing.]
	{\includegraphics[width=0.22\textwidth]{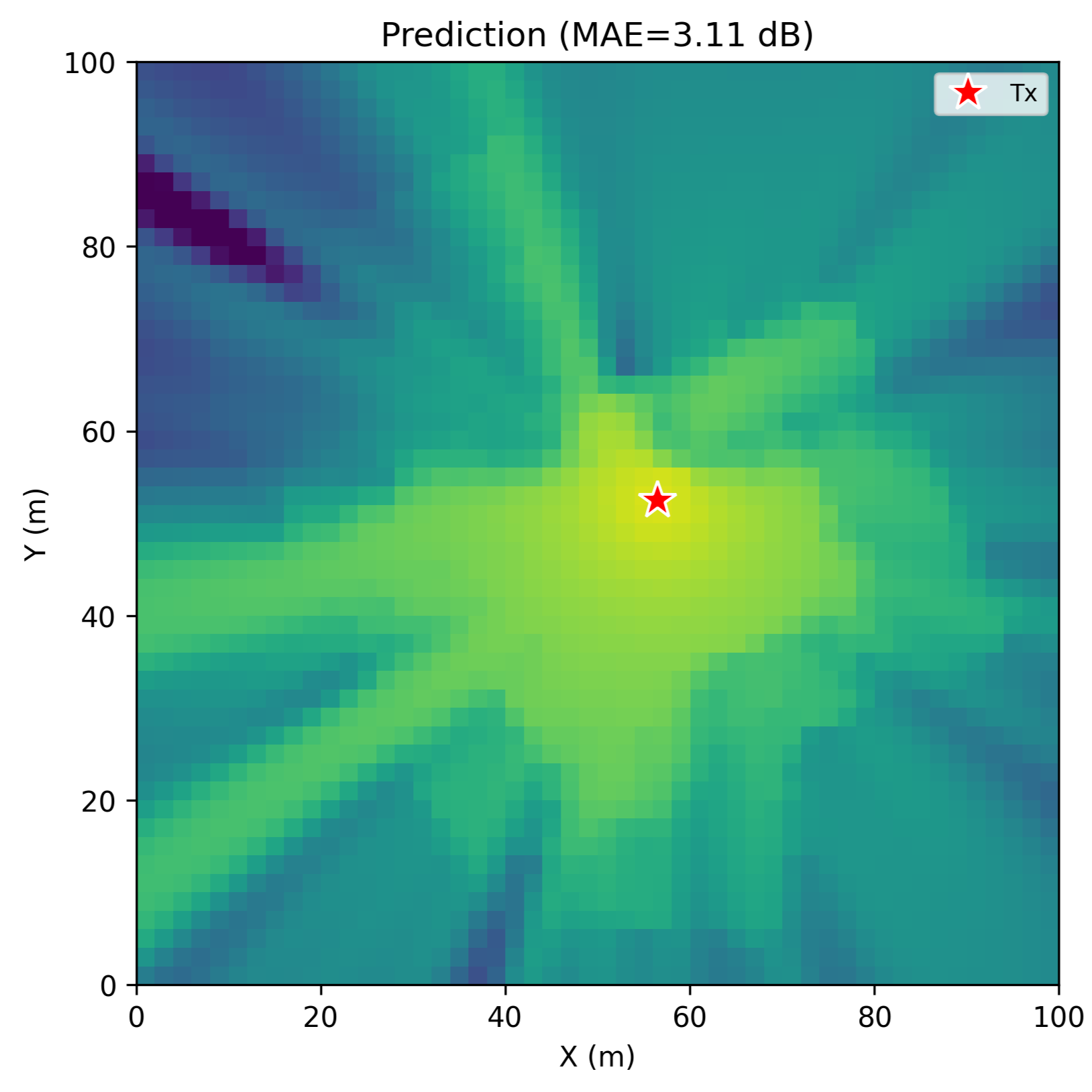}}
	\subfigure[Random active initialization without residual guidance.]
	{\includegraphics[width=0.22\textwidth]{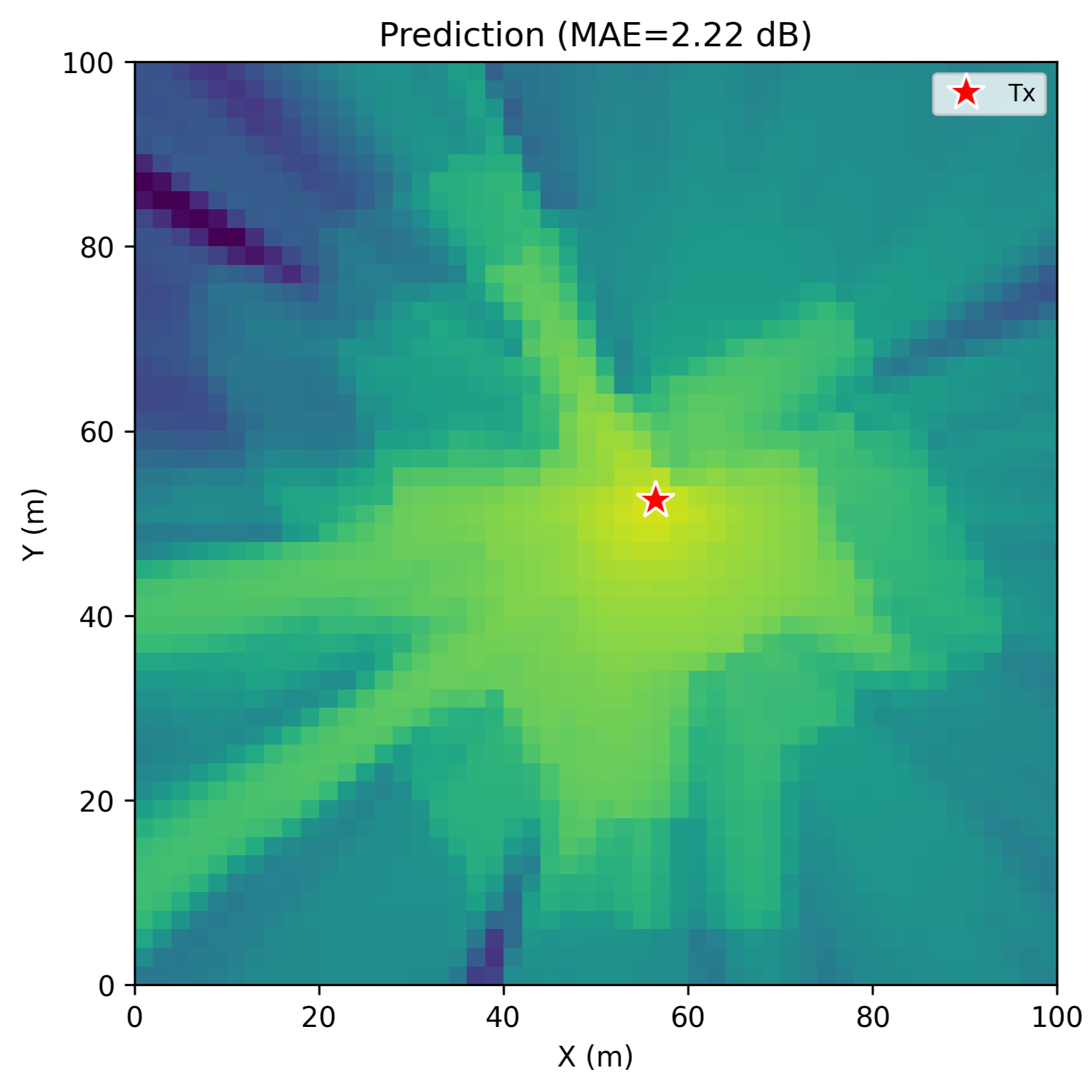}}
	\caption{Ablation study on incremental GS-CG for CGM updating.}
	\label{fig:update_ablation}
\end{figure}

Fig. \ref{fig:update_ablation} evaluates the main components of the incremental GS-CG for dynamic CGM update in the campus scene. As compared to Fig. \ref{fig:campus_update}(b), Fig. \ref{fig:update_ablation}(a) shows the result of directly reconstructing the updated CGM without using prior propagation knowledge. The resulting prediction exhibits strong distortion and a much larger MAE, since the sparse update measurements are insufficient for reconstructing the whole CGM from scratch. 
Fig. \ref{fig:update_ablation}(b) shows the full-retraining baseline, where all parameters are re-optimized during updating. Although it uses the previous model as initialization, updating the entire representation with sparse new data disturbs the learned structures in unchanged regions and yields a larger error. This verifies the necessity of parameter freezing.
Fig. \ref{fig:update_ablation}(c) replaces the proposed residual-guided active initialization with random initialization. The error increases in this case because randomly initialized active Gaussians are less likely to lie near the true changed structures, making it harder to explain the newly formed shadow regions. These results confirm that prior reuse, parameter freezing, and residual-guided active initialization are all important for stable and data-efficient CGM updating.

\begin{figure}[!tb]
	\centering
	\subfigure[Construction with \(25 \times 25\) grids.]
	{\includegraphics[width=0.22\textwidth]{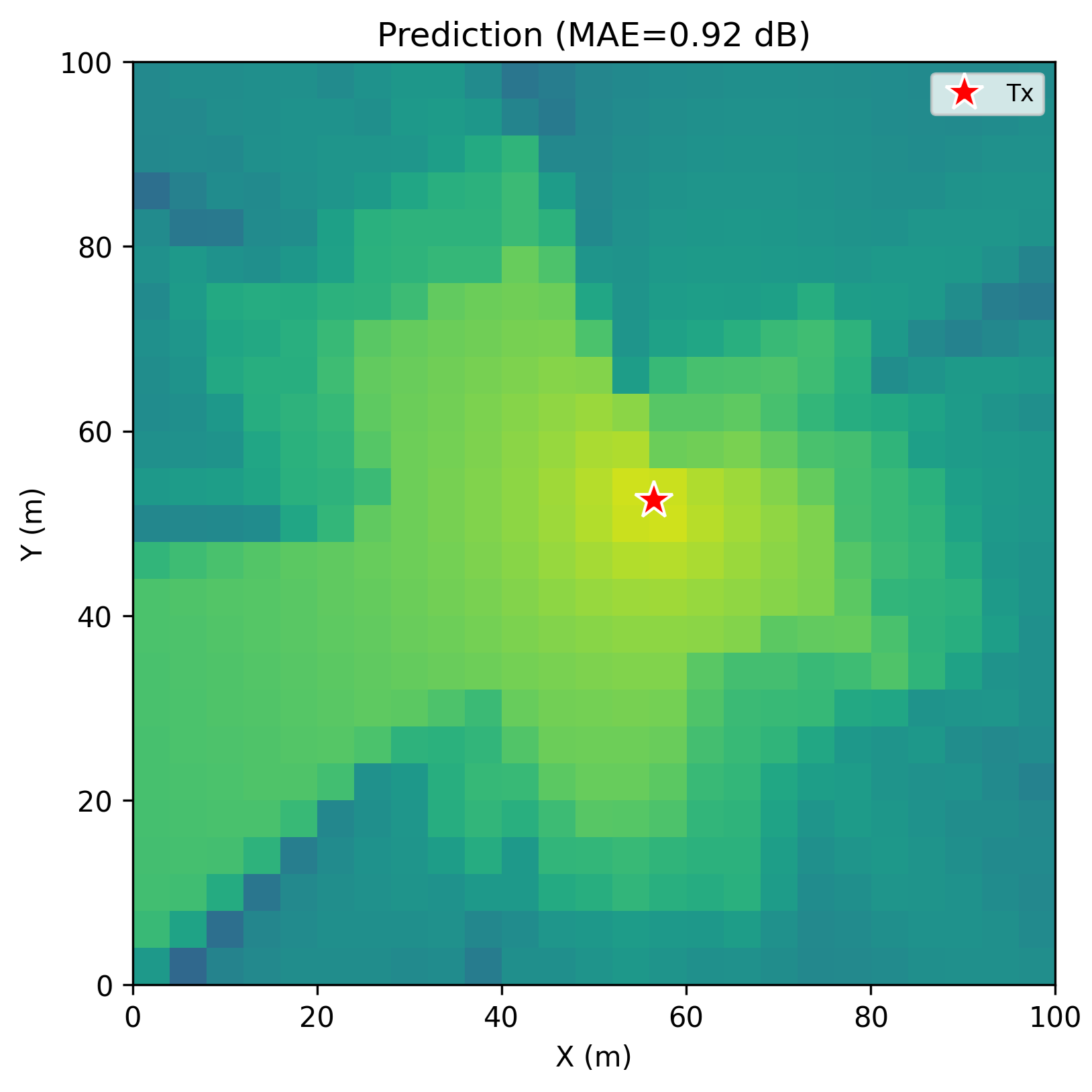}}
	\subfigure[Updating with \(25 \times 25\) grids.]
	{\includegraphics[width=0.22\textwidth]{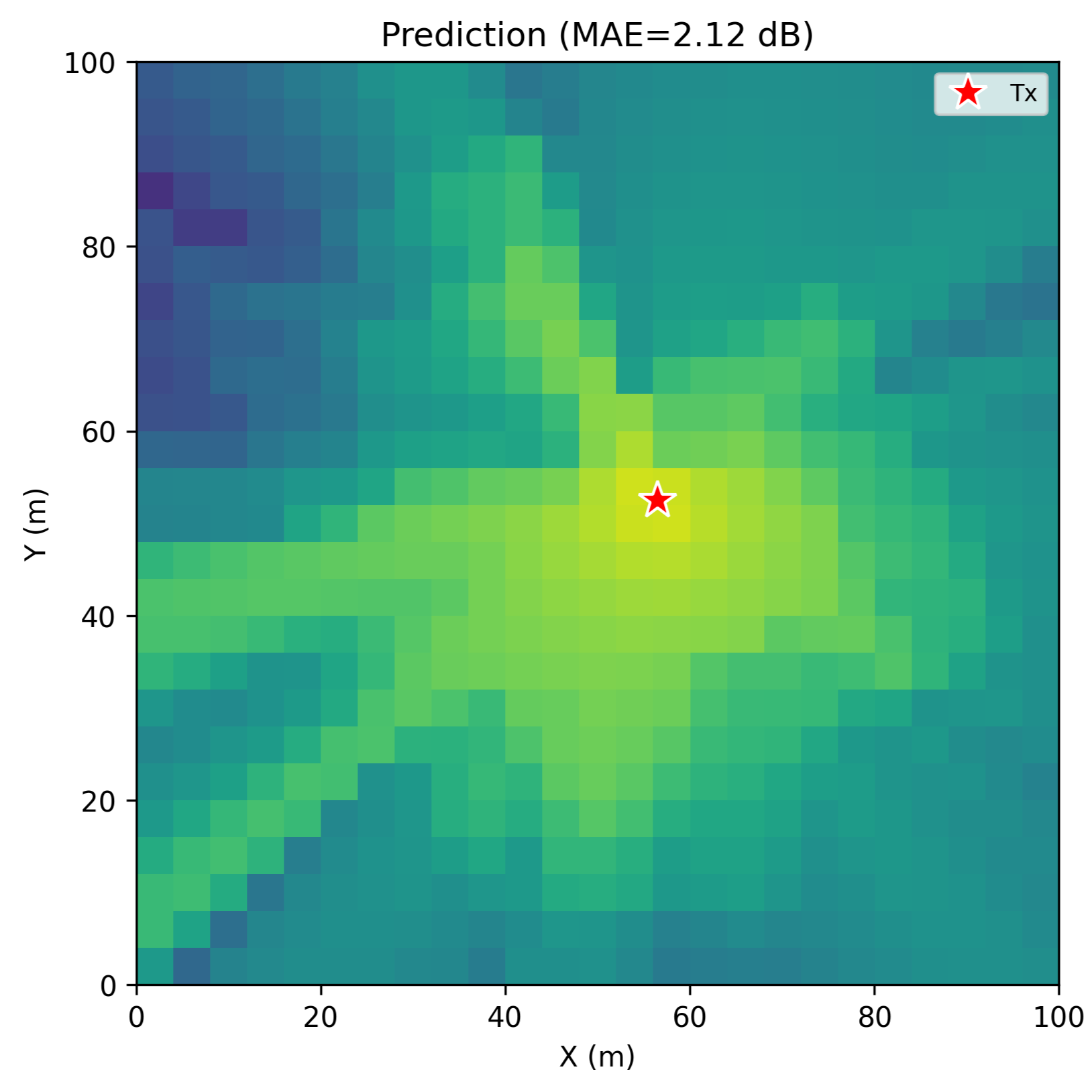}}
	\subfigure[Construction with \(100 \times 100\) grids.]
	{\includegraphics[width=0.22\textwidth]{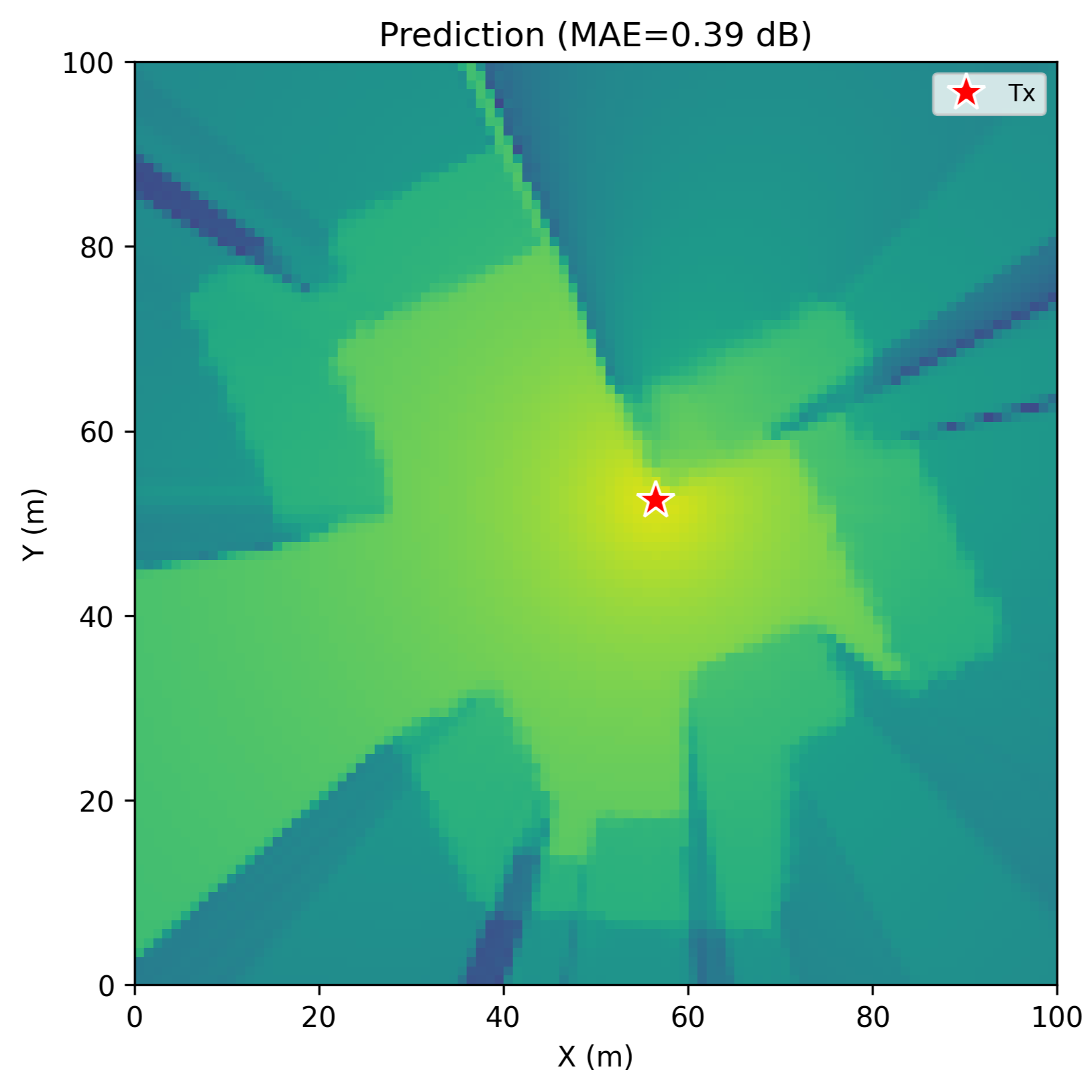}}
	\subfigure[Updating with \(100 \times 100\) grids.]
	{\includegraphics[width=0.22\textwidth]{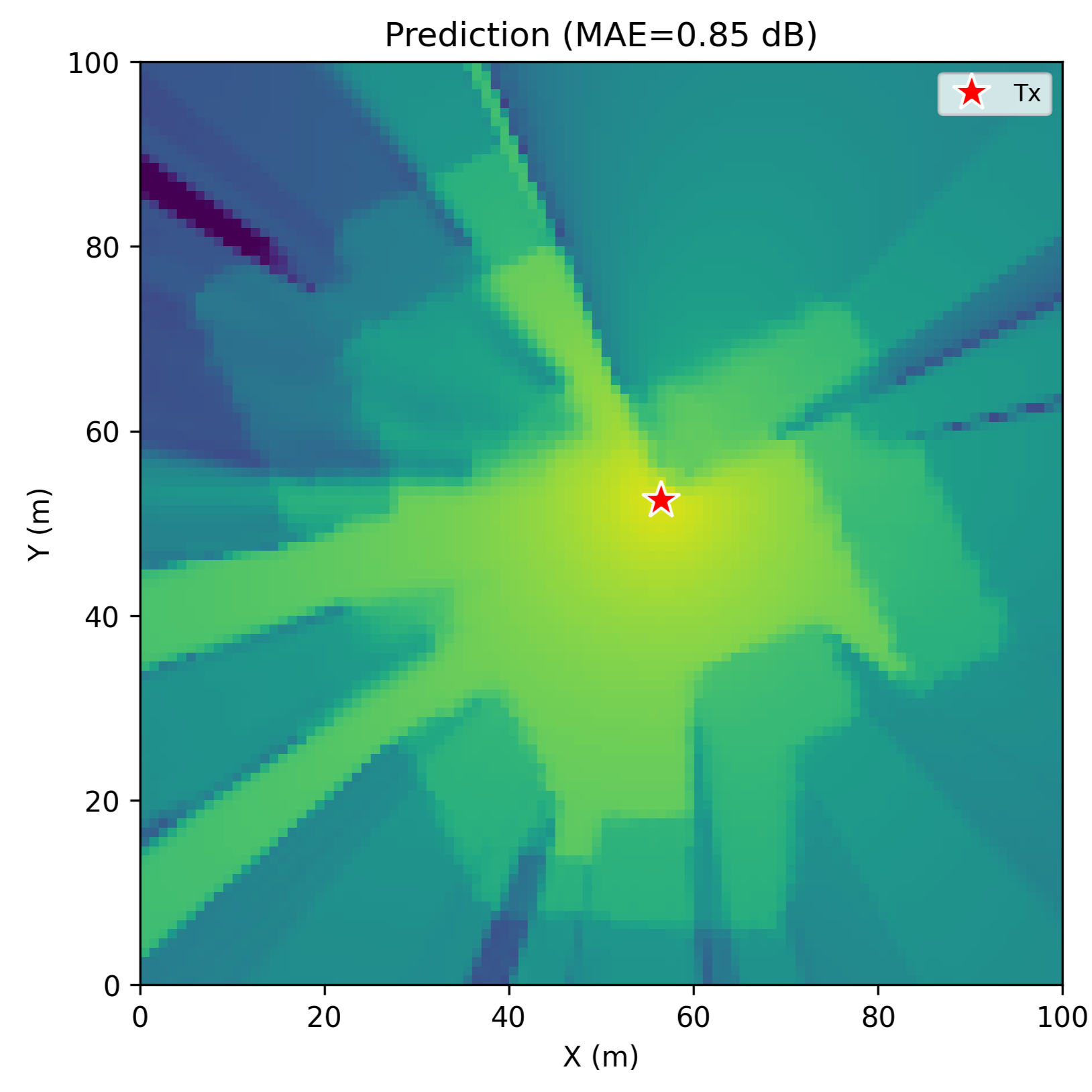}}
	\caption{GS-CG prediction results under different grid resolutions.}
	\label{fig:grid_resolution}
\end{figure}

Fig. \ref{fig:grid_resolution} studies the impact of grid resolution in the campus scene. The same $100\times100$ m\textsuperscript{2} area is divided into $25\times25$ and $100\times100$ grids. For both resolutions, $25\%$ samples are used for construction and $4\%$ samples are used for updating. The proposed framework works well under both settings. With the coarse $25\times25$ grids, the model captures the overall power distribution and major blockage regions. With the finer $100\times100$ grids, the predicted CGM contains sharper shadow boundaries and more detailed spatial variation. This is because the Gaussian representation is continuous in space, while the grid only defines the supervision and evaluation resolution. Therefore, GS-CG can adapt to different CGM resolutions according to the required planning granularity and measurement density.

\section{Conclusion}

This paper investigated 3DGS-enabled construction and dynamic updating of grid-based CGMs for environment-aware wireless networks. We first formulated a grid-averaged channel gain model that suppresses phase-sensitive small-scale fluctuations and decomposes the gain into distance-dependent attenuation, path transmittance, and effective scattering components. Based on this model, we developed GS-CG to represent latent radio-interaction structures with Gaussian primitives and render grid-level channel gains through propagation-aware scattering and transmittance modeling. We then proposed an incremental updating mechanism that freezes the static/reference Gaussian representation and optimizes only a compact active Gaussian set for dynamic CGM refinement, and further introduced MLP-GS to reduce the cost of explicit scattering-branch rendering with a lightweight decoder. Simulation results in campus, road, and low-altitude scenarios showed that GS-CG accurately reconstructs static CGMs from limited measurements and effectively tracks local environmental changes from sparse new measurements. The dynamic updating results confirmed that prior reuse, parameter freezing, and residual-guided active initialization are essential for stable refinement. The comparisons also showed that MLP-GS substantially reduces training time with only moderate accuracy loss, providing a practical accuracy-complexity tradeoff for fast and frequent CGM updating.

\appendices

\section{Proof of Proposition 1}

The proof is motivated by \cite{sun2025channel}, which gives the derivation for square grids on a two-dimensional (2D) plane. Here, we extend to 3D grid cells with a generic shape.
Let \(\mathbf{x}\in\mathcal{C}(\mathbf{p}_r)\) be expressed as \(\mathbf{x}=\mathbf{p}_r+\boldsymbol{\xi}\), where \(\boldsymbol{\xi}\in\mathcal{C}(\mathbf{0})\) is the local displacement and \(\mathcal{C}(\mathbf{0})\) denotes the grid cell translated to the origin. Under the local plane-wave approximation, the excess distance of the \(n\)-th path varies approximately linearly with \(\boldsymbol{\xi}\), i.e., \(\Delta_n(\mathbf{x};\mathbf{p}_r)\approx \boldsymbol{\nu}_n^T\boldsymbol{\xi}\),
%\begin{equation} \label{Delta_n}
%	\Delta_n(\mathbf{x};\mathbf{p}_r)\approx \boldsymbol{\nu}_n^T\boldsymbol{\xi},
%\end{equation}
where \(\boldsymbol{\nu}_n\) is the unit vector associated with the direction of arrival (DoA) of the \(n\)-th path.\footnote{We assume that distinct paths have sufficiently separated local directions. If two paths have very similar DoAs, their cross-correlation may not vanish after spatial averaging. In that case, they should be treated as one effective path or scattering cluster rather than two separable paths.} Thus,
\begin{equation} \label{f_n}
	f_n(\mathbf{x};\mathbf{p}_r)
	=\exp\Big(-j\frac{2\pi}{\lambda}\Delta_n(\mathbf{x};\mathbf{p}_r)\Big)
	\approx \exp(-j\mathbf{k}_n^T\boldsymbol{\xi}),
\end{equation}
where \(\mathbf{k}_n \triangleq \frac{2\pi}{\lambda}\boldsymbol{\nu}_n\) is the effective spatial wave vector of the \(n\)-th path.

Substituting \eqref{h} into \eqref{Gint} gives
\begin{equation}
	\begin{aligned}
		G(\mathbf{p}_r)
		&= \frac{1}{|\mathcal{C}(\mathbf{p}_r)|} \int_{\mathcal{C}(\mathbf{p}_r)}
		\Big|\sum_{n=0}^{N_s} u_n(\mathbf{p}_r) f_n(\mathbf{x};\mathbf{p}_r)\Big|^2 d\mathbf{x} \\
		&= \sum_{n=0}^{N_s}\sum_{n'=0}^{N_s}
		u_n(\mathbf{p}_r)u_{n'}^*(\mathbf{p}_r) R_{n,n'}(\mathbf{p}_r),
	\end{aligned}
\end{equation}
where
\begin{equation}
	R_{n,n'}(\mathbf{p}_r)
	\triangleq
	\frac{1}{|\mathcal{C}(\mathbf{p}_r)|}
	\int_{\mathcal{C}(\mathbf{p}_r)}
	f_n(\mathbf{x};\mathbf{p}_r) f_{n'}^*(\mathbf{x};\mathbf{p}_r)
	d\mathbf{x}
\end{equation}
is the spatial cross-correlation between the \(n\)-th and \(n'\)-th field responses over the grid. Using \eqref{f_n} and changing variables from \(\mathbf{x}\) to \(\boldsymbol{\xi}\), we obtain
\begin{equation}
	R_{n,n'}(\mathbf{p}_r)
	\approx
	\frac{1}{|\mathcal{C}(\mathbf{0})|}
	\int_{\mathcal{C}(\mathbf{0})}
	e^{-j(\mathbf{k}_n-\mathbf{k}_{n'})^T\boldsymbol{\xi}} d\boldsymbol{\xi}.
\end{equation}
The last expression is the normalized Fourier transform of the cell indicator evaluated at the differential wave vector \(\mathbf{k}_n-\mathbf{k}_{n'}\). For \(n=n'\), it gives \(R_{n,n}(\mathbf{p}_r)=1\). For \(n\ne n'\), sufficiently separated local directions make \(\mathbf{k}_n-\mathbf{k}_{n'}\) non-negligible. Since the grid cell spans many wavelengths, the phase term oscillates rapidly over \(\mathcal{C}(\mathbf{0})\), and the integral becomes negligible. Hence, \(R_{n,n'}(\mathbf{p}_r)\approx 0\).
%\begin{equation}
%	R_{n,n'}(\mathbf{p}_r)\approx 0,\ n\neq n'.
%\end{equation}
Substituting these correlations into the quadratic expansion of \(G(\mathbf{p}_r)\) yields \eqref{G}. The expressions in \eqref{G_0n} follow by taking the squared magnitudes of the path responses.

\bibliographystyle{IEEEtran}
\bibliography{myref}

\end{document}